\documentclass{aa}  
\usepackage{graphicx}
\usepackage[varg]{txfonts}
\usepackage{lscape}
\usepackage{natbib}
\usepackage{rotating}
\usepackage{amsmath}

\def\deg{\ifmmode^\circ\else$^\circ$\fi}
\def\alphaTF{\ifmmode{\alpha_{\mathrm{\,{\small TF}}}}\else{$\alpha_{\mathrm{\,{\small TF}}}$}\fi}

\newcommand{\epse}{\ensuremath{\epsilon_\mathrm{e}}}
\newcommand{\lambdae}{\ensuremath{\lambda_\mathrm{e}}}

\begin{document}
\title{Evolution induced by dry minor mergers onto\\ fast-rotator S0 galaxies}

\author{Trinidad Tapia\inst{1,2,7}, M.~Carmen Eliche-Moral\inst{3}, Miguel Querejeta\inst{4}, Marc Balcells\inst{5,1,2}, A.~C\'esar Gonz\'{a}lez-Garc\'{\i}a\inst{6,1,2}, Mercedes Prieto\inst{1,2}, J.~Alfonso L.~Aguerri\inst{1,2}, Jes\'us Gallego\inst{3},  Jaime Zamorano\inst{3}, Cristina Rodr\'{\i}guez-P\'{e}rez\inst{3}, \and Alejandro Borlaff\inst{3}}

\institute{Instituto de Astrof\'{\i}sica de Canarias, C/ V\'{\i}a L\'actea, E-38200 La Laguna, Tenerife, Spain
\and 
Departamento de Astrof\'{\i}sica, Universidad de La Laguna, E-38200 La Laguna, Tenerife, Spain
\and
Departamento de Astrof\'{\i}sica y CC.~ de la Atm\'osfera, Universidad Complutense de Madrid, E-28040 Madrid, Spain  
\and
Max-Planck-Institut f\"{u}r Astronomie, K\"{o}nigstuhl, 17, 69117 Heidelberg, Germany
\and
Isaac Newton Group of Telescopes, Apartado 321, E-38700 Santa Cruz de La Palma, Canary Islands, Spain
\and
Present address: Instituto de Ciencias del Patrimonio, CSIC, R\'{u}a San Roque 2, Santiago de Compostela, E-15704 A Coru\~{n}a,  Spain 
\and
Present address: Instituto de Astronom\'{\i}a, Universidad Nacional Aut\'onoma de M\'exico, Apdo.~877, Ensenada BC 22800, Mexico, \email{ttapia@astro.unam.mx}
}

   \date{Accepted march 6 of 2014}

\abstract{
Numerical studies have shown that the properties of the S0 galaxies with 	kinematics intermediate  between fast and slow rotators are difficult to explain by a scenario of major mergers.
}
{
We investigate whether the smoother perturbation induced by minor mergers can reproduce these systems.
}
{
We analysed collisionless $N$-body simulations of intermediate and minor dry mergers onto S0s to determine the structural and kinematic evolution induced by the encounters. The original primary galaxies represent gas-poor fast-rotator S0b and S0c galaxies with high intrinsic ellipticities. The original bulges are intrinsically spherical and have low rotation. Different mass ratios, parent bulges, density ratios, and orbits were studied.
}
{
Minor mergers induce a lower decrease of the global rotational support (as provided by \lambdae) than encounters of lower mass ratios, which results in S0s with properties intermediate between fast and slow rotators. The resulting remnants are intrinsically more triaxial, less flattened, and span the whole range of apparent ellipticities up to $\epse \sim 0.8$. They do not show lower apparent ellipticities in random projections than initially; on the contrary, the formation of oval distortions and the disc thickening increase the percentage of projections at $0.4<\epse< 0.7$. In the experiments with S0b progenitor galaxies, minor mergers tend to spin up the bulge and to slightly decrease its intrinsic ellipticity, whereas in the cases of primary S0c galaxies they keep the rotational support of the bulge nearly constant and significantly decrease its intrinsic ellipticity. The remnant bulges remain nearly spherical ($B/A \sim C/A > 0.9$), but exhibit a wide range of triaxialities ($0.20 < T < 1.00$). In the plane of global anisotropy of velocities ($\delta$) vs.\,intrinsic ellipticity ($\epsilon_\mathrm{e,intr}$), some of our models extend the linear trend found in previous major merger simulations towards higher $\epsilon_\mathrm{e,intr}$ values, while others clearly depart from it (depending on the progenitor S0). This is consistent with the wide dispersion exhibited by real S0s in this diagram compared with ellipticals, which follow the linear trend drawn by major merger simulations.
}
{
The smoother changes induced by minor mergers can explain the existence of S0s with intermediate kinematic properties between fast and slow rotators that are difficult to explain with major mergers. The different trends exhibited by ellipticals and S0 galaxies in the $\delta$ -- $\epsilon_\mathrm{e}$ diagram may be pointing to the different role played by major mergers in the build-up of each morphological type.
}

\titlerunning{Evolution induced by dry minor mergers on fast-rotator S0 galaxies}	
\authorrunning{Tapia et al.}

   \maketitle

\section{Introduction}
\label{sec:introduction}
Recent studies have shown that classifying early-type galaxies (ETGs) into fast and slow rotators provides a more consistent distinction in terms of their physical properties than the traditional morphological classification into ellipticals and S0s \citep[E11 hereafter]{Emsellem2007,Emsellem2011}. This is because this criterion is almost independent of the viewing angle (E11), whereas S0 galaxies can be morphologically confused with ellipticals in face-on views \citep[][B11 hereafter]{Bois2011}. According to this classification, the vast majority of ETGs are fast rotators when considered as single-component systems, meaning that they have a noticeable regular rotation pattern, with aligned photometric and kinematic axes, they host inner discs and often bars, and span a wide range of apparent ellipticities ($0<\epse<0.85$). Only a small fraction of ETGs are slow rotators ($\sim$ 15\%), and usually have complex stellar velocity fields and kinematically decoupled cores (E11).  Approximately 10-20\% of lenticular galaxies (S0s) in the ATLAS$^\mathrm{3D}$ sample exhibit hybrid properties between fast and slow rotators, lying in the limiting region defined to isolate these two families of objects, with ellipticities spanning the whole range up to $\epse \sim 0.7$ (E11).

Fast rotators may be the result of the rebuilding of a stellar disc around a central spheroid, which in turn may come from the destruction of a pre-existing disc through mergers, or by gas exhaustion in spirals \citep[see][]{Khochfar2011}. Observations indicate that these evolutionary mechanisms might depend on environment. Gas stripping and strangulation seem to have been responsible of transforming spirals into S0s in clusters since $z\sim 0.8$ \citep{Barr2007,Desai2007}, while minor mergers, galaxy harassment, and tidal interactions may have triggered an even more dramatic evolution in groups during the same period of time \citep[][]{Moran2007,Bekki2011}. This variety of formation processes agrees pretty well with the diversity of properties exhibited by S0s depending on the environment and the stellar mass \citep{Laurikainen2010,Roche2010,Wei2010,Silchenko2012,Barway2013}. A  description of the processes that may have been relevant for the formation of S0s can be found in \citet{Aguerri2012}.

Numerical studies indicate that simple fading by itself is not sufficient to produce a fast rotator \citep{Khochfar2011}. However, simulations of major mergers have succeeded in producing both slow- and fast-rotator remnants \citep[][B11]{Gonzalez-Garcia2006a,Jesseit2009}. The fast rotators formed in this way have intermediate apparent flattening ($0.4< \epse <0.6$) and high rotational support ($\lambdae > 0.4$), whereas the resulting slow rotators span the whole range of ellipticities and usually host kinematically decoupled components (B11). Mergers of disc galaxies with higher mass ratios (3:1 and 6:1) basically give rise to fast rotators with intermediate-to-high ellipticities and high rotational support, too (B11). This means that the remnants resulting from mergers with mass ratios lower than 6:1 cannot properly reproduce the region in the $\lambdae - \epse$ parameter space populated by slow rotators with low apparent ellipticities and by galaxies with intermediate properties between fast and slow rotators   \citep[with $0.1\lesssim \lambdae \lesssim0.3$, see][B11]{Burkert2008,Khochfar2011}. 

Intermediate and minor mergers are expected to be much more frequent than major ones in standard hierarchical scenarios \citep{Naab2009,Bezanson2009,Hopkins2010} and more likely to produce S0-like remnants \citep[][]{Bournaud2004,Bournaud2005}.  Therefore, it is straightforward to question whether mergers of mass ratios higher than 6:1 are a feasible evolutionary channel for giving rise to S0s with intermediate kinematic properties or not. 

\citet{Naab2013} have recently shown that subsequent minor mergers in a cosmological context can explain the rare class of slow rotators with low ellipticities. However, their simulations hardly reproduce the location in the \lambdae\ -- \epse\ diagram of the S0s with hybrid kinematic properties ($0.15<\lambdae <0.25$) and $\epse > 0.3$ (see their Fig.\,11). This does not necessarily mean that mergers must be discarded as a feasible mechanism to explain the properties of these galaxies. Cosmological N-body simulations have the advantage (over idealized binary merger simulations) of analysing more realistic pathways to form galaxies, but they are also more limited in numerical resolution. This problem directly affects the way  baryons accumulate in the centre of the potential wells, the physics of star formation, the rotational support of the gas component, and the formation of substructures \citep[see][]{Bournaud2008,Piontek2011,Regan2013}.

To complement the numerical studies cited above, we have investigated whether dry mergers with mass ratios ranging from 6:1 to 18:1 can explain the formation of S0s with intermediate kinematic properties. We used N-body simulations of binary mergers, starting from gas-poor progenitors with high initial intrinsic ellipticities and rotational support. In this paper, we describe the results obtained for S0s that have spherical original bulges. The effects of considering non-axisymmetric primary bulges will be explored in a forthcoming paper, as the vertical buckling of an original bar (induced by the encounter or by simple natural secular evolution) implies changes in both the velocity ellipsoid and the structure of the bulge \citep{Mihos1995,Martinez-Valpuesta2004,Martinez-Valpuesta2006,Saha2013}. 

The paper is organized as follows: the models are described in Sect.~\ref{sec:simulations}. Section\,\ref{sec:fastslow} describes the analysis of the global structure and rotational support of the remnants that were produced in our simulations, and compares the results with the distributions of fast and slow rotators obtained in observational surveys and in previous studies of major merger simulations. In Sect.\,\ref{sec:anisotropy} we also analyse the relation between the anisotropy of velocities and the intrinsic ellipticity in our remnants and compare this again with data and previous simulations. Section\,\ref{sec:bulges} shows how different the intrinsic shape and the rotational support of the central remnant bulge can be from those computed for the galaxy as a whole through $\lambdae$ and $\epse$. Section~\ref{sec:relation} shows the relation between the bulge triaxiality and the global rotational support of the whole remnant, two properties that are usually considered to be strongly related. The limitations of the models are commented on Sect.~\ref{sec:limitations}. Finally, the discussion and the main conclusions are provided in Sects.~\ref{sec:discussion} and \ref{sec:conclusions}. 

\begin{table*}
\begin{minipage}[t]{\textwidth}
\caption{Parameters of the minor and intermediate merger experiments.}
\label{tab:models}
\centering
\begin{tabular}{lccrcc}
\hline\hline
\multicolumn{1}{c}{Model code} &  $M_{\mathrm{sat}}/M_{\mathrm{prim}}$  & $R_{\mathrm{per}}/h_\mathrm{D,prim}$  & \multicolumn{1}{c}{$\theta$} &  \multicolumn{1}{c}{\textrm{$(B/D)_\mathrm{prim}$}} & $\alphaTF$ \\
\multicolumn{1}{c}{(1)}    & \multicolumn{1}{c}{(2)} & \multicolumn{1}{c}{(3)}         & \multicolumn{1}{c}{(4)}    & \multicolumn{1}{c}{(5)}        & \multicolumn{1}{c}{(6)}      \vspace{0.05cm}\\\hline\vspace{-0.3cm}\\
 (a)\,\,\, M6 Ps Db      &  1:6 (M6)   &  0.73 (Ps) &  30 (D) & \textrm{0.5} (b)  & 3.5 \\
 (a2) M6 Ps Db TF3  &  1:6 (M6)   &  0.73 (Ps) &  30 (D) & \textrm{0.5} (b)  & 3.0 \\
 (a3) M6 Ps Db TF4  &  1:6 (M6)   &  0.73 (Ps) &  30 (D) & \textrm{0.5} (b)  & 4.0 \\
 (b)\,\,\, M6 Ps Rb       &  1:6 (M6)   &  0.73 (Ps) & 150 (R) & \textrm{0.5} (b)  & 3.5  \\
 (c)\,\,\, M6 Pl Db       &  1:6 (M6) &  8.25 (Pl) &  30 (D)   & \textrm{0.5} (b)  & 3.5 \\
 (d)\,\,\, M6 Pl Rb       &  1:6 (M6) &  8.25 (Pl) &  150 (R)  & \textrm{0.5} (b)  & 3.5 \\
 (e)\,\,\, M6 Ps Ds       &  1:6 (M6) &  0.87 (Ps) &  30 (D)   & \textrm{0.08} (s) & 3.5 \\
 (f)\,\,\, M6 Ps Rs       &  1:6 (M6) &  0.87 (Ps) & 150 (R)   & \textrm{0.08} (s) & 3.5\\ \vspace{-0.4cm}\\\hline\vspace{-0.3cm}\\
 (g)\,\,\, M9 Ps Db       & 1:9 (M9) &  0.79 (Ps) &  30 (D)   &  \textrm{0.5} (b)  & 3.5 \\
 (g2) M9 Ps Db TF3  & 1:9 (M9) &  0.79 (Ps) &  30 (D)   &  \textrm{0.5} (b)  & 3.0 \\
 (g3) M9 Ps Db TF4  & 1:9 (M9) &  0.79 (Ps) &  30 (D)   &  \textrm{0.5} (b)  & 4.0 \\
 (h)\,\,\, M9 Ps Rb       & 1:9 (M9) &  0.79 (Ps) & 150 (R)  & \textrm{0.5} (b)  & 3.5\\ \vspace{-0.4cm}\\\hline\vspace{-0.3cm}\\
 (i)\,\,\, M18 Ps Db      & 1:18 (M18)&  0.86 (Ps) &  30 (D)   & \textrm{0.5} (b)  & 3.5 \\
 (j)\,\,\, M18 Ps Rb      & 1:18 (M18)&  0.86 (Ps) & 150 (R)   & \textrm{0.5} (b)  & 3.5 \\
 (k)\,\,\,M18 Pl Db      &  1:18 (M18)& 8.19 (Pl) &  30 (D)  & \textrm{0.5} (b)  & 3.5\\
 (l)\,\,\, M18 Pl Rb      &  1:18 (M18)& 8.19 (Pl) &  150 (R) & \textrm{0.5} (b)  & 3.5 \\\hline\\
\end{tabular}
\begin{minipage}[t]{\textwidth}{\small
\emph{Columns}: (1) Model code: M$m$P[l/s][D/R][b/s][TF3/4], see the text. (2) Luminous mass ratio between the satellite and the primary S0 galaxy. (3) Orbital first pericentre distance in units of the primary disc scale-length. (4) Initial angle between the angular momenta of the orbit and the primary disc. (5) Bulge-to-disc ratio of the original primary S0: $B/D=0.5$ (S0b) or $B/D=0.08$ (S0c). (6) Value of $\alphaTF$ assumed for the scaling of the satellite to the primary S0. More details of the models in  EM06 and EM11.}
\end{minipage}
\end{minipage}
\end{table*}

\section{Numerical simulations}
\label{sec:simulations}

Our set of experiments consists of 16 $N$-body collisionless simulations of intermediate and minor mergers on S0 galaxies that are described in detail in \citet[EM06 and EM11 henceforth]{Eliche-Moral2006,Eliche-Moral2011}. Fourteen of these experiments have a primary galaxy matching an S0b galaxy (with 185,000 particles), while the rest are similar to an S0c galaxy (415,000 particles in total). The primary galaxies were built using the \texttt{GalactICS} package \citep{Kuijken1995}, and consist of an exponential disc component, a King bulge, and a dark halo built following an Evans profile. 

When considered as single-component systems, both original S0s (after relaxation) are fast rotators with low rotation (acquired during relaxation) and apparent ellipticities $0.1<\epsilon_e<0.8$ (considering 200 random projections, see Fig.\,1), which could have formed through gas stripping of spirals. These primary progenitors are intrinsically highly flattened oblate systems (EM11; EM12). The progenitor S0c represents an extreme case of a fast-rotator S0 with high intrinsic ellipticity and high rotational support from the ATLAS$^\mathrm{3D}$ sample. The S0b progenitor exhibits an intrinsic ellipticity slightly higher for its rotational support than those observed in real cases, but within the observational range. As observational data in the \lambdae -- \epse\ diagram are not corrected for inclination, our progenitor S0b can also be considered as an extreme, but realistic, case of  a fast-rotator S0 with high intrinsic ellipticity and intermediate rotational support. For more details, see Sect.\,\ref{sec:fastslow}. The radial and vertical surface density profiles of the discs are exponential, with a height-to-length scale ratio of $z_{\rm{D}}/h_{\rm{D}} = 0.1$ at one disc scale-length. The structure of the primary bulges is very stable, remaining nearly spherical during relaxation. Therefore, any deviation from this spherical shape in the remnant bulges must be due to the merger. The numerical disc thickening during relaxation is lower than 10\% and can be neglected in the original galaxies (see EM06 and Secs.\,\ref{sec:rotation} and \ref{sec:limitations}).

All the satellites are scaled replicas of the S0b primary model, to test the effects of the bulge-to-disc ratio of the original primary galaxies. A physically motivated size-mass scaling for the satellites was used to have realistic satellite-to-primary density ratios, based on the Tully-Fisher relation (see \citealt{Gonzalez-Garcia2005} and EM06). Different exponents of this relation in the range of the observational values were considered ($\alphaTF =3.0$, 3.5, and 4.0). Higher $\alphaTF$ values lead to satellites denser than the primary galaxy. The luminous mass ratios under consideration are 6:1, 9:1, and 18:1. The angles between the two discs were set to 30$^\mathrm{o}$ and 150$^\mathrm{o}$ for the direct and prograde orbits, respectively. Orbits with pericentre distances equal to $h_\mathrm{D}$ and $8\,h_\mathrm{D}$ were run. For more information about the initial conditions, see EM06 and EM11. 

The evolution of ten models was computed with the \texttt{TREECODE} \citep{Hernquist1989}, for the rest we used the \texttt{GADGET--2} code \citep{Springel2005}. The energy is preserved to better than 0.1\% and forces were computed within 1\% of those resulting from direct summation in any case (see EM06; EM11). The models were evolved to ensure quasi-equilibrium in the remnants. The main characteristics of each model are summarized in Table\,\ref{tab:models}. All remnants exhibit global morphological, photometric, and kinematic properties typical of S0a-S0b galaxies as in EM06, EM11 and \citet{Eliche-Moral2012,Eliche-Moral2013}; we refer to these studies for more details. 

\section{Results}
\label{sec:results}

\subsection{Rotational support and flattening of the whole remnants}
\label{sec:fastslow}

\subsubsection{Computation of \lambdae\ and \epse}
\label{sec:definitions}

The ratio between the mass-weighted mean of the rotation speed and the random velocity ($V/\sigma$) versus the mean ellipticity of a galaxy has traditionally been used to relate its apparent flattening to its amount of rotation. This relation is known as the  ``anisotropy diagram'' because the location of one galaxy in this plane provides an estimate of the anisotropy of velocities of the galaxy \citep[$\delta$, see][]{Emsellem2007}. During many decades in which only long-slit spectroscopy was available, the $V/\sigma$ ratio was approximated by the ratio between the maximum observed rotational velocity, $V_\mathrm{max}$, and the central dispersion, $\sigma_0$ \citep[see, e.g.,][]{Kormendy1993}. However, \citet{Binney2005} formulated new expressions for a more robust estimation of $\delta$ based on the tensor virial theorem, especially derived for modern 2D spectroscopic data. According to this new formulation, \citet{Emsellem2007} defined a new kinematic parameter, $\lambdae$, as a function of surface-brightness weighted averages of $V$ and $\sigma$ within the half-light radius of the galaxy ($R\leq R_\mathrm{eff,glx}$). This parameter is a proxy for the rotational support of the galaxy, and can be easily derived from 2D spectroscopic data:

\begin{equation}\label{eq:lambda}
 \lambda_\mathrm{e} = \frac{\Sigma _{i=1}^{N_p} F_i R_i |V_i|}{\Sigma_{i=1}^{N_p} F_i R_i \sqrt{V_i^2 + \sigma_i^2}},
\end{equation}

\noindent where $F_i$ , $R_i$ , $V_i$, and $\sigma_i$ are the flux, circular radius, velocity, and velocity dispersion of the $i$--th spatial bin, the sum running on the $N_p$ spatial bins (pixels or Voronoi bins) within an aperture of intrinsic radius $R=R_\mathrm{eff,glx}$ on the galaxy. These authors have demonstrated that $\lambda_\mathrm{e}$ is even more efficient for distinguishing between fast and slow rotators in ETGs than the traditional anisotropy diagram ($V/\sigma$ -- $\epsilon$) if plotted against the light-weighted average ellipticity within $R_\mathrm{eff,glx}$ ($\epsilon _\mathrm{e}$), using SAURON and ATLAS$^{3D}$ data \citep[][E11]{Emsellem2007}.

We analysed the rotational support and the apparent flattening of our merger remnants (see Sect.\,\ref{sec:compare}). To allow a direct comparison with available data from ATLAS$^\mathrm{3D}$, we have mimicked the observational procedure by E11 to estimate $\epsilon _\mathrm{e}$, $(V/\sigma)_\mathrm{e}$, and $\lambda_\mathrm{e}$ directly from our $N$-body data. We considered 200 random projections for each remnant and for the original primary galaxies (following B11).  For each projection, we estimated $\lambda_\mathrm{e}$ in our models through eq.~\ref{eq:lambda}. We also derived $V/\sigma$ as done in E11,

\begin{equation} \label{eq:vsigma}
 \left(\frac{V}{\sigma}\right)_\mathrm{e}^2 \equiv \frac{\langle V^2 \rangle}{\langle \sigma^2 \rangle} =  \frac{\sum_{i=1}^{N_p} F_i\, V_i^2}{\sum_{i=1}^{N_p} F_i\, \sigma_i^2},
\end{equation}

\noindent with $V_i$ and $\sigma_i$ the mean velocity and velocity dispersion in the $i$--th spatial bin, with the sum again running on the $N_p$ spatial bins within an aperture of intrinsic radius $R=R_\mathrm{eff,glx}$ on the galaxy for the view under consideration \citep[see eq.\,10 in][]{Cappellari2007}.

We calculated the moment ellipticity profile of the remnants within radially growing isophotes for each projection ($\epsilon(R)$), again mimicking the procedure by E11. The moment ellipticity $\epsilon$ within one radius $R$ is defined by these authors as a luminosity-weighted average ellipticity in the galaxy, computed via diagonalization of the ellipse of inertia of the galaxy surface brightness inside an isophote enclosing an area $A = \pi\, R^2$,

\begin{equation} \label{eq:epsilon}
 (1 - \epsilon) ^2 = q^2 \equiv \frac{\langle y^2 \rangle}{\langle x^2 \rangle} =  \frac{\sum_{i=1}^{N_p} F_i\, y_i^2}{\sum_{i=1}^{N_p} F_i\, x_i^2}.
\end{equation}

\noindent In the equation above, $i$ runs through all spatial bins in the data within the isophote corresponding to $R$ for a fixed galaxy projection, $F_i$ is the flux contained inside the $i$--th spatial bin, and ($x_i$,\,$y_i$) are the coordinates of this bin considering the galaxy centre as origin and the x-direction parallel to the galaxy photometric major axis \citep[see eq.\,11 in][]{Cappellari2007}. The moment ellipticity at the effective radius of the galaxy (\epse) for each view is obtained in E11 through interpolation of these profiles or curves of growth at $R=R_\mathrm{eff,glx}$.

The resulting values of $\epse$, $(V/\sigma)_\mathrm{e}$, and $\lambda_\mathrm{e}$ for the edge-on view of the remnants and the original S0s are listed in Table\,\ref{tab:kinematics}. We defined the galactic plane as the plane perpendicular to the total stellar angular momentum of the remnant containing its mass centroid. The edge-on view considered here corresponds to the point where this plane intersects with the XY plane of our original coordinates system.

\begin{figure}[t!h]
\centering
   \includegraphics[width=7.5cm]{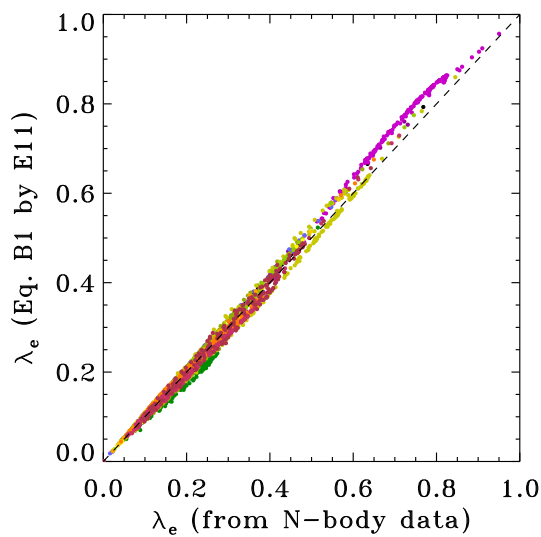}
  \caption{Comparison between the value of the $\lambda_\mathrm{e}$ parameter derived directly from the $N$-body data (using eq.~\ref{eq:lambda}) and the one estimated through eq.~B1 of \citet{Emsellem2007} using $(V/\sigma)_{\mathrm{e}}$ (eq.~\ref{eq:lambdaB1} here), for the 200 projections of our remnants and the original galaxies. The dashed line marks the 1:1 relation. The projections corresponding to the same model are plotted with the same color.}
\label{fig:lambdaelambda}
\end{figure} 

As a test to our computations, we also estimated $\lambda_\mathrm{e}$ from the values $(V/\sigma)_\mathrm{e}$ estimated through eq.\,\ref{eq:vsigma} using the tight relation that \citet{Emsellem2007} found between these two parameters in both real and modelled data,

\begin{equation}\label{eq:lambdaB1}
 \lambda_\mathrm{e} = \frac{\kappa (V/\sigma)_\mathrm{e}}{\sqrt{1 + \kappa^2 (V/\sigma)_e^2}},
\end{equation}

\noindent where $\kappa\sim 1.1$ (eq.~B1 in E11). The values of $\lambda_\mathrm{e}$ derived from this equation for the edge-on projections of our models are also provided in Table\,\ref{tab:kinematics}.

In Fig.~\ref{fig:lambdaelambda}, we plot the $\lambda_\mathrm{e}$ parameter calculated according to eq.\,\ref{eq:lambdaB1} (by means of the $(V/\sigma)_\mathrm{e}$ values directly obtained from our $N$-body data) versus the $\lambda_\mathrm{e}$ directly estimated from the data through eq.\,\ref{eq:lambda}, for the 200 projections of each remnant and the original galaxies. The agreement between the two values is very good, which means that the $\lambda_\mathrm{e}$ -- $(V/\sigma)_\mathrm{e}$ relation derived by \citet{Emsellem2007} is also valid for our merger experiments. 

\begin{figure*}[th]
\centering
   \includegraphics*[width=9.0cm]{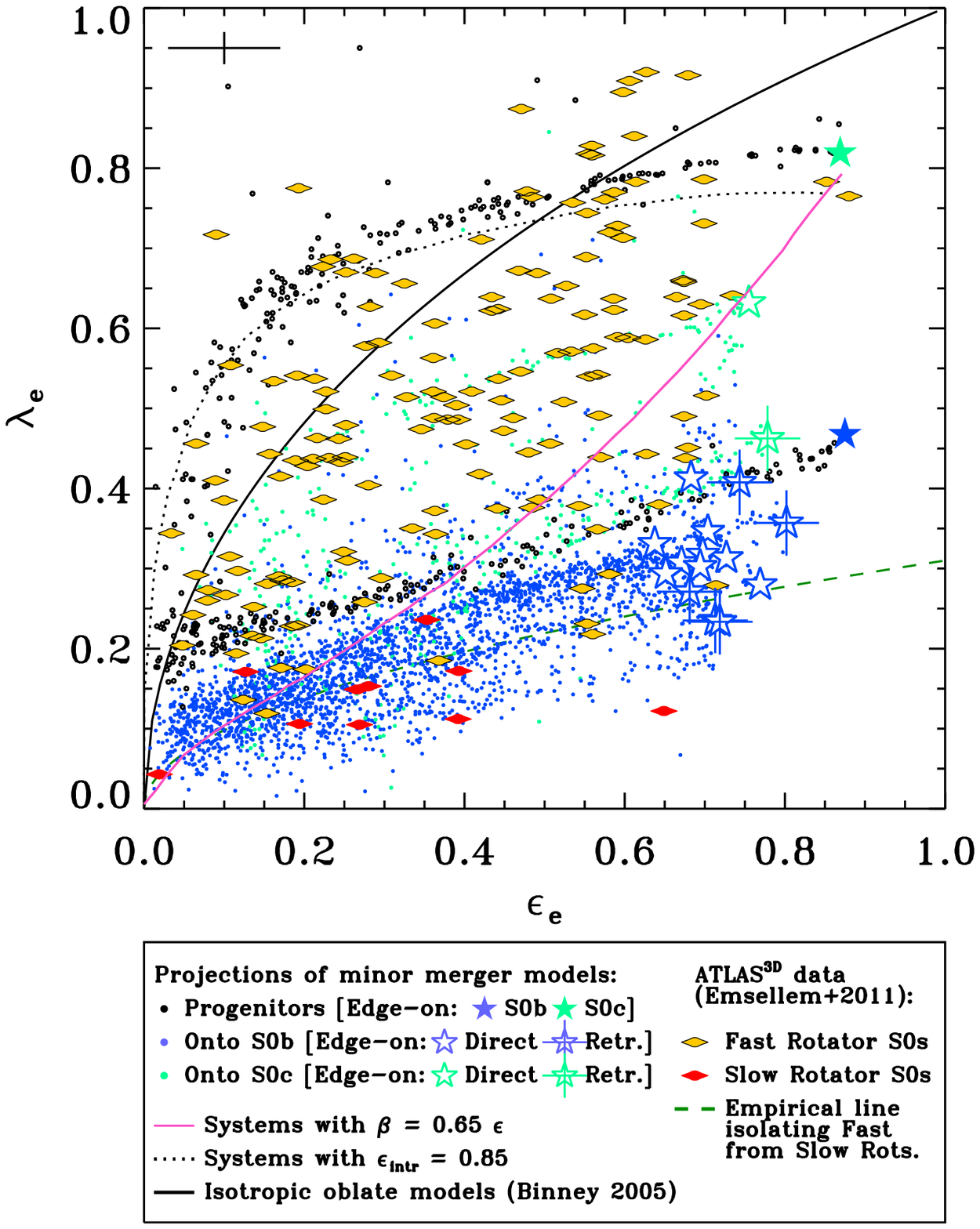}
   \includegraphics*[width=9.0cm]{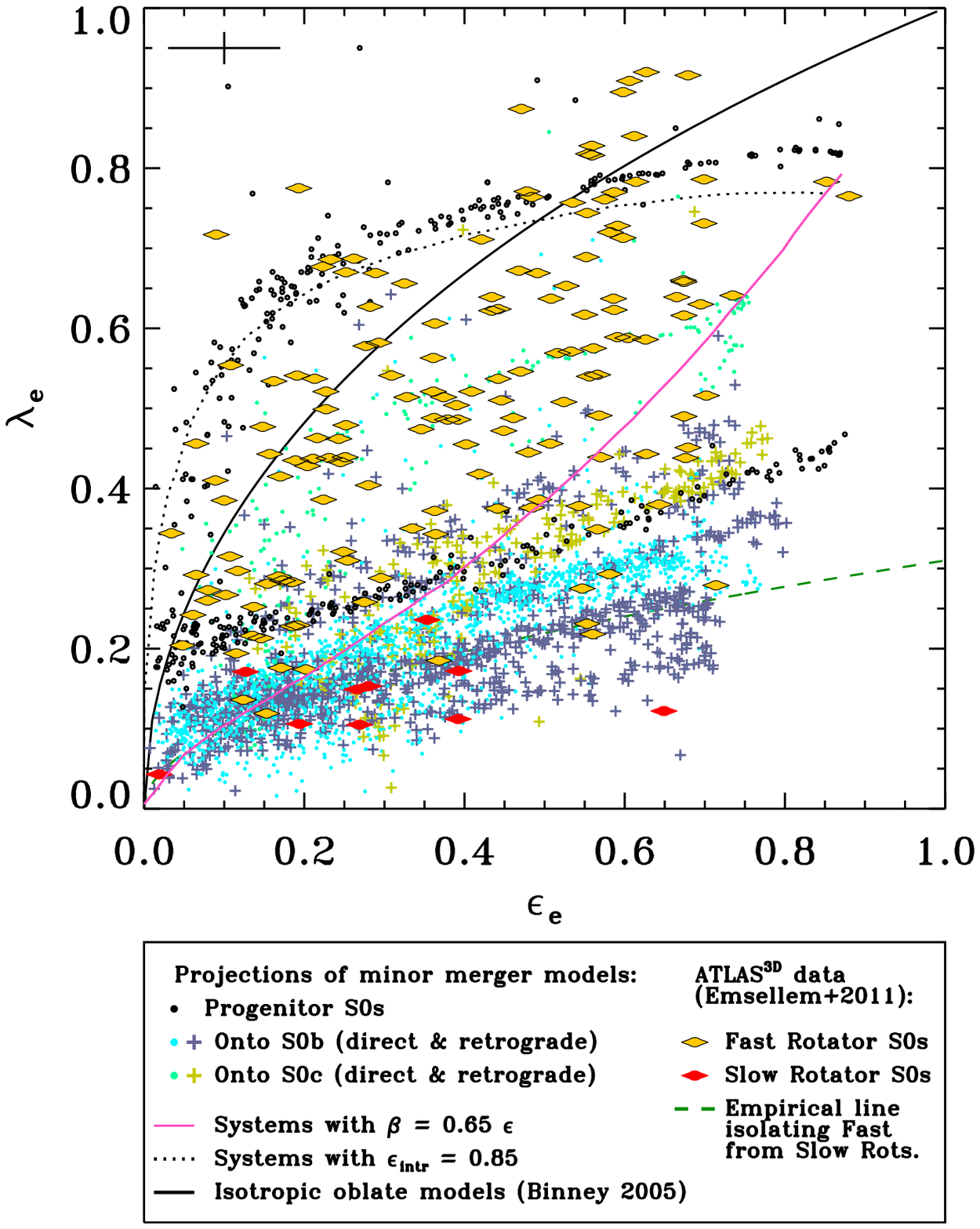}
  \caption{The $\lambdae$ -- $\epse$ diagram considering 200 random projections in each remnant and in the original S0 galaxies, compared with the distributions of fast- and slow-rotator S0s from the ATLAS$^\mathrm{3D}$ sample (yellow and red galaxy-like symbols respectively, see E11).  \emph{Left panel}: for all models, independently of their spin-orbit coupling. The projections of the original S0 galaxies are plotted with black points, blue points correspond to the remnants of the experiments with an original S0b primary galaxy, and green ones to those with an original primary S0c. Stars indicate the location of each remnant in an edge-on view (blue and green filled stars for the original S0b and S0c galaxies; blue and green empty stars for the remnants of the models with an original S0b and S0c galaxy, respectively). The stars marked with a cross correspond to retrograde models.  \emph{Right panel}: the same as in the left panel, but distinguishing between the projections of prograde and retrograde models. The prograde remnants are plotted with points (light blue for those with an original S0b galaxy; light green for the one with an original S0c). The retrogrades are symbolized with crosses (dark blue for original S0b primary galaxies; olive green for the one with an original primary S0c). \emph{Black solid line}: theoretical relation of isotropic oblate systems viewed edge-on from \citet{Binney2005}. \emph{Green dashed line}: empirical limit isolating fast from slow rotators defined by E11 (equivalent to 0.31$\sqrt{\epsilon_\mathrm{e}}$). \emph{Black dotted line}: location of galaxies with an intrinsic ellipticity $\epsilon = 0.82$ when going from a face-on to an edge-on view, to show the effects of inclination. \emph{Magenta solid line}: edge-on view of ellipsoidal galaxies integrated within $r=R_\mathrm{eff,glx}$ for an anisotropy $\beta = 0.65\epsilon$ (for more details, see E11). The error bars on the left upper corner of each frame represent the typical errors in both axes (mean values of all projections).} 
\label{fig:lambdaobs}
\end{figure*}

\subsubsection{Location of the models in the \lambdae\ --\epse\ diagram}
\label{sec:compare}

In Fig.~\ref{fig:lambdaobs}, we plot the locations in the $\lambdae$ -- $\epse$ plane for the 200 projections of each remnant and the primary S0 galaxies. We remark that the $\lambdae$ values plotted here are derived directly from the N-body data, using the definition of this parameter as provided by E11 (eq.~\ref{eq:lambda}, see Sect.\,\ref{sec:definitions}). In the two panels, we compare the location of our remnants with the distributions of fast- and slow-rotator S0s reported by the ATLAS$^\mathrm{3D}$ project (E11). The information plotted in the two panels is similar, but the right one distinguishes between the retrograde and prograde models using different symbols (dots for the projections of the prograde models and crosses for the retrograde ones) and does not show the location of the edge-on views of each model, which is marked with stars in the left panel. The green dashed line represents the empirical threshold of $\lambdae$ for each $\epse$ that distinguishes between fast and slow rotators, as defined by E11. Galaxies above this line are fast rotators, whereas the ones below it are slow rotators with $\sim 90$\% of likelihood.

The left panel of Fig.~\ref{fig:lambdaobs} shows that our two original primary S0 galaxies would be classified as highly flattened fast rotators (black dots). The remnants of the models with S0c progenitors and some with S0b ones would keep this classification, but many remnants with S0b primary galaxies move towards the transition zone between fast and slow rotators (they accumulate around the green line defined by E11), exhibiting \epse\ values up to $\sim 0.8$.  Thus, minor mergers are a feasible process to produce S0 systems with hybrid kinematic properties in the $\lambdae$ --  $\epse$ diagram.

As expected, the merger events tend to decrease the intrinsic ellipticity of the original galaxies, making the remnants less flattened \emph{intrinsically} than the original primary galaxies. This can be deduced from the location of the edge-on views of the remnants. The remnants of the models with an original S0b galaxy (blue empty stars) have lower intrinsic ellipticities than the primary S0b (blue filled star). The same is observed in the models with an original S0c (green empty stars) compared with the edge-on location of the original S0c galaxy (filled green star). Note that these intrinsic ellipticities continue to be high ($0.6 <\epse < 0.8$). 

We also confirm that models with retrograde orbits tend to produce remnants with lower values of $\lambdae$ (and thus, with lower global rotational support) than their prograde analogs, as already observed in major and intermediate merger simulations (see, e.g., B11). In the right panel of Fig.\,\ref{fig:lambdaobs}, the projections of the retrograde models (crosses) show lower values of $\lambdae$ than those of prograde models (dots) in general, accumulating in the limiting region between fast and slow rotators. In fact, most projections with $\epse>0.2$ that would be observationally classified as slow rotators (because they lie below the limiting line) correspond to a retrograde model (M6PlRb).

\begin{figure*}[th!]
\centering
 \includegraphics[height=6.5cm]{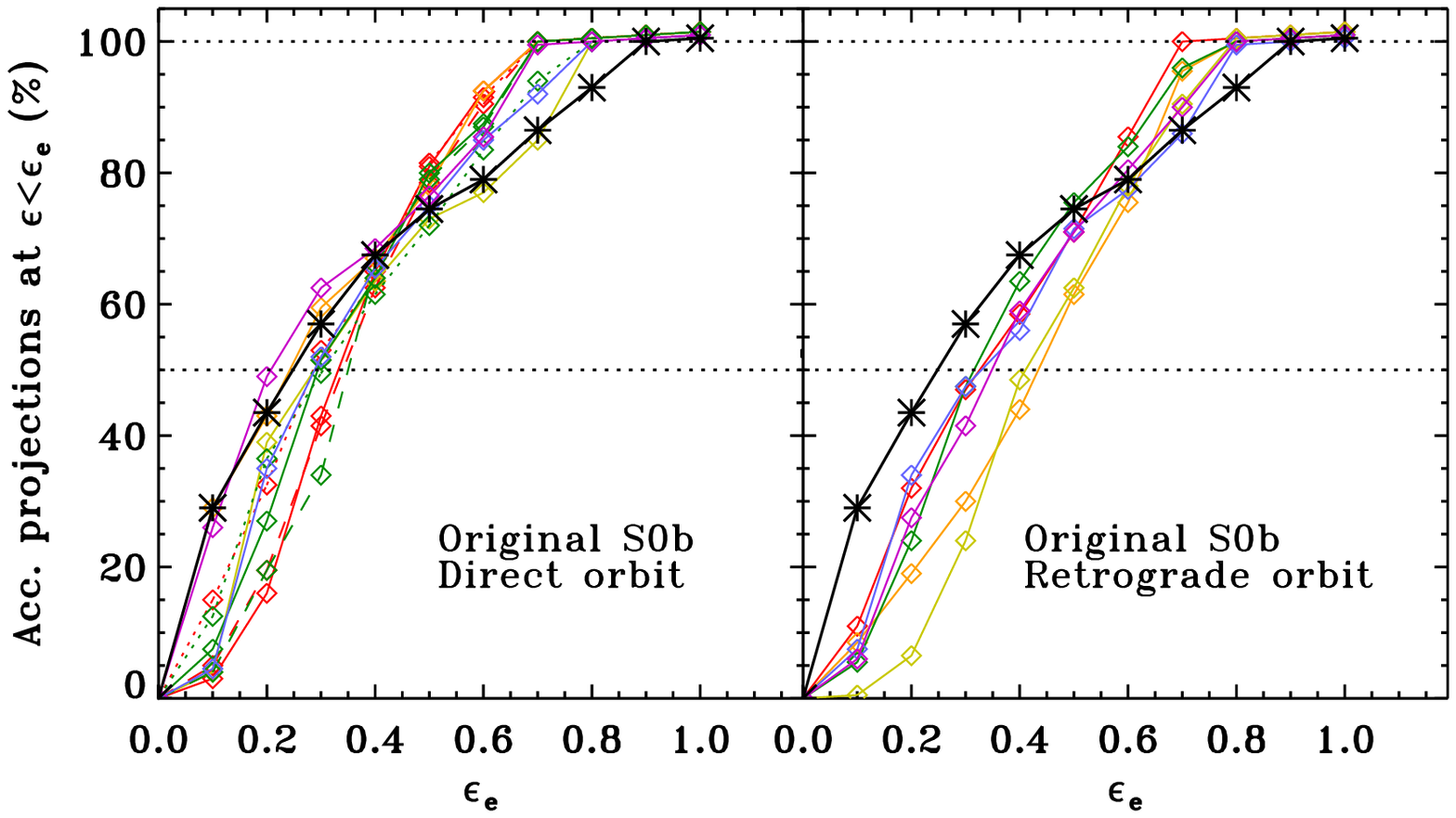}
 \includegraphics[height=6.5cm]{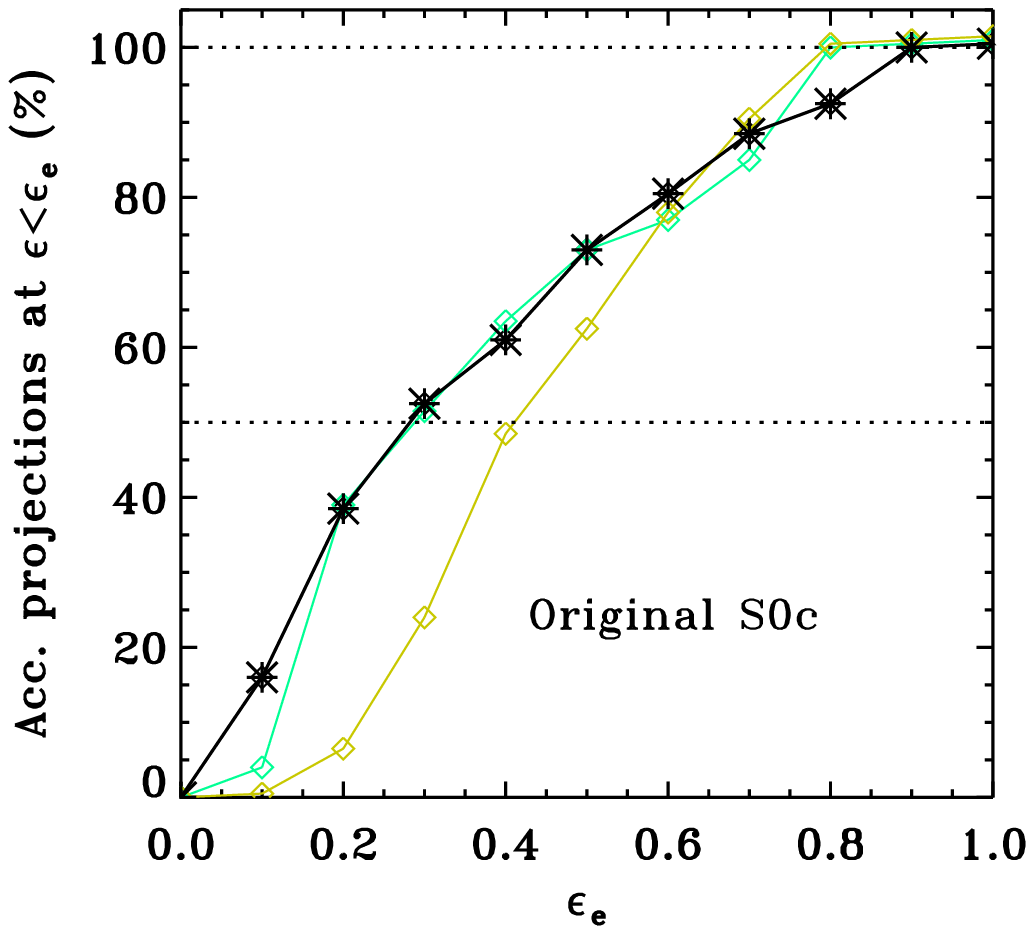}
  \caption{Cumulative percentage of random projections with apparent ellipticity lower than or equal to a given value $\epse$ for the models with an S0b primary galaxy (left panels) and with an S0c primary galaxy (right panel). The dotted lines mark the values corresponding to 50\% and 100\% of all projections in each model, as a reference. \emph{Color-coding for the left panels}: \emph{black}: S0b primary galaxy. \emph{Red}: mass ratio 6:1, short pericentre, primary S0b. \emph{Orange}: mass ratio 6:1, long pericentre, primary S0b. \emph{Olive green}: mass ratio 6:1, short pericentre, primary S0b. \emph{Dark green}:  mass ratio 9:1, short pericentre, primary S0b. \emph{Blue}: mass ratio 18:1, short pericentre, primary S0b. \emph{Purple}: mass ratio 18:1, long pericentre, primary S0b.  \emph{Color-coding for the right panel}: \emph{black}: S0c primary galaxy. \emph{Olive green}: mass ratio 6:1, short pericentre, primary S0c, direct orbit.  \emph{Light green}: mass ratio 6:1, short pericentre, primary S0c, retrograde orbit.}
\label{fig:percentages}
\end{figure*}

In Fig.\,\ref{fig:percentages}, we represent the cumulative number of projections with $\epsilon < \epse$ for each model compared with the original distributions in the corresponding primary galaxies. It might be expected that, because the intrinsic ellipticity of all remnants decreases with respect to the original values, the number of random projections that exhibit very low \epse\ values ($\epse < 0.4$) increases in the remnants. But the behaviour is the opposite: almost all the remnants exhibit a lower fraction of projections with $\epse < 0.4$ than the original galaxies (black line in each panel), keeping this tendency up to a threshold \epse\ value. At the same time, the intrinsic ellipticity in the remnants is always lower than the original primary galaxy. The only exceptions are the models that correspond to accretions in direct orbits with long pericentre distances onto S0b progenitors (purple and orange lines in the first panel of the figure). In the models with an S0b primary galaxy, this threshold \epse\ value is $\sim 0.45$ for the prograde encounters and $\sim 0.55$-0.6 for the retrograde ones, while in those with an S0c progenitor the threshold \epse\ is $\sim 0.7$.

In Table\,\ref{tab:projections}, we list the percentage of the 200 random projections considered for each model in five bins of apparent ellipticities. While the original galaxies have $\sim 40$\% of projections at $\epse \leq 0.2$, the resulting remnants typically exhibit a lower percentage of projections at this ellipticity bin. However, the fractions of random projections increases in most remnants with respect to the original progenitors at $0.2<\epse\leq 0.4$ and $0.4<\epse\leq 0.6$ (see Cols.\,2 -- 4 in the Table). In the $0.6<\epse\leq 0.8$ bin, the behaviour depends on the model. The experiments with an S0b progenitor and direct orbits tend to decrease or keep the percentage, whereas those with retrograde orbits increase it. In the models with an S0c, this percentage increases. In the final bin ($\epse > 0.8$), the remnants do not exhibit projections, whereas the original progenitor S0s had $\sim 7$\%.

This means that our remnants tend to exhibit more intermediate \epse\ values at random projections than their corresponding progenitor galaxies. Although the intrinsic (i.e., maximum) ellipticities decrease in all remnants, the percentage of random projections with low apparent ellipticities does not increase (contrary to expectations). This is due to the formation of triaxial structures in the center of the remnants (ovals or weak bar-like distortions, see EM12) and to the intrinsic thickening of the whole galaxy structure induced by the encounter (see EM06). In inclined views, these non-axisymmetric distortions cause the isophotes to be more elliptical than originally. The two exceptions to the trend confirm this, as these accretions (in long pericentre, direct orbits) have a higher spin-orbit coupling than the other cases and  take longer to be complete. Consequently, the satellite undergoes higher orbital circularization while it is disrupted, the accretion is smoother, and hence the oval distortion is weaker (EM11). Therefore, Figs.~\ref{fig:lambdaobs} and \ref{fig:percentages} show that minor and intermediate dry mergers can evolve highly flattened fast-rotator S0s into systems that are intrinsically less flattened and less axisymmetric, that is, it converts them into more triaxial systems.

\begin{figure*}[th]
\centering
   \includegraphics*[width=9.0cm]{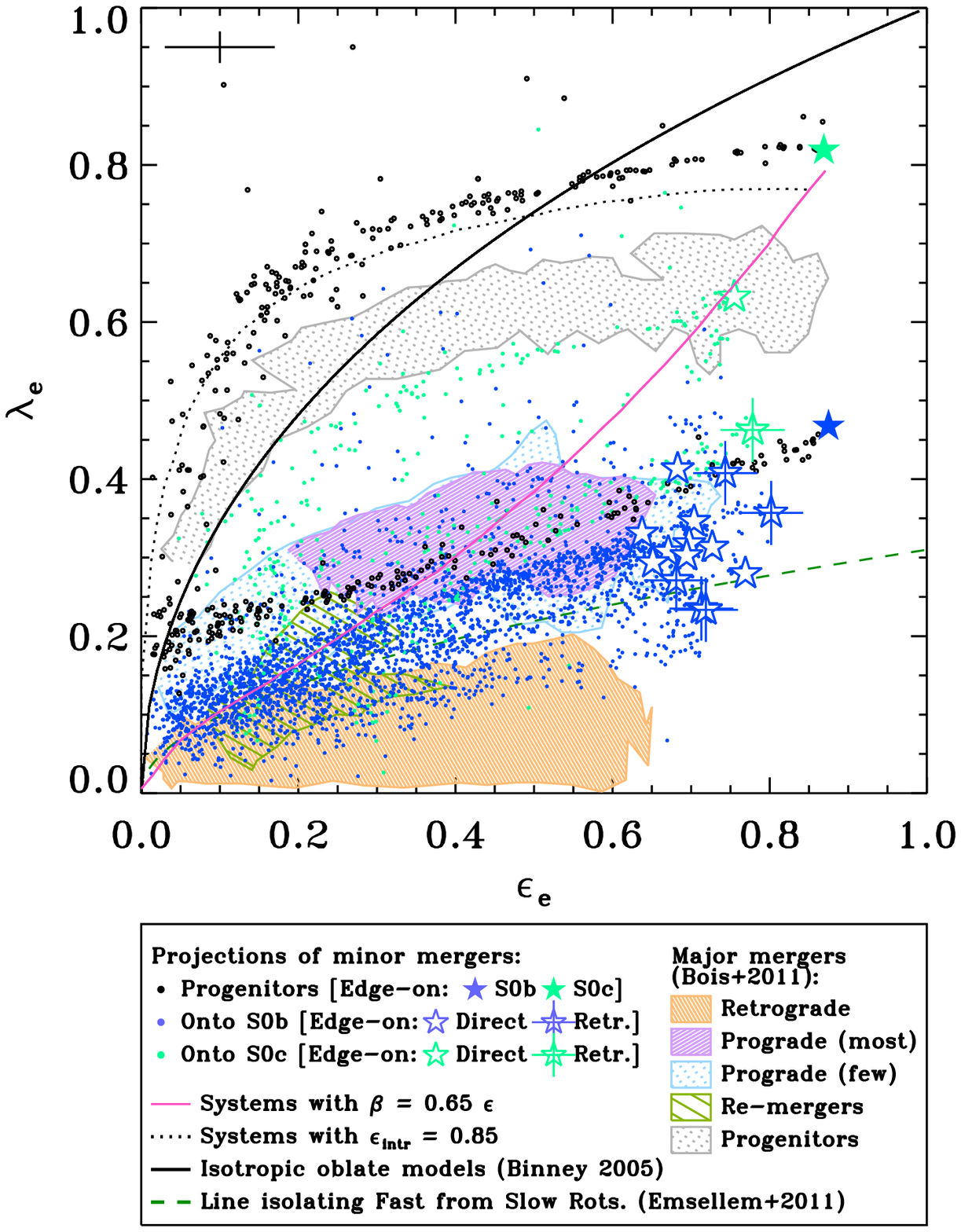}
   \includegraphics*[width=9.0cm]{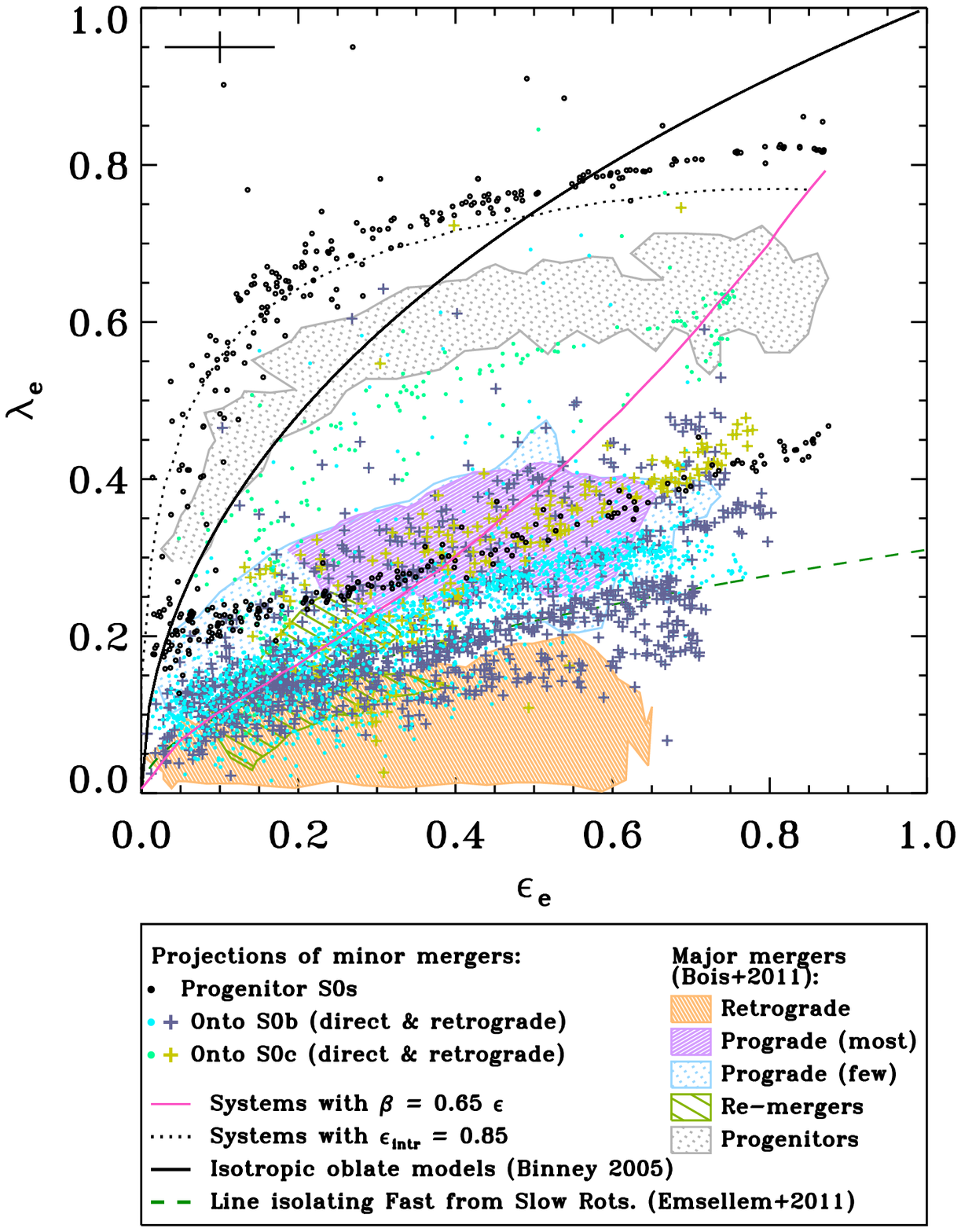}
  \caption{The $\lambda_{\mathrm{e}}$ -- $\epsilon_\mathrm{e}$ diagram for 200 random projections of our remnants and the original S0 galaxies, now compared with the distribution of the remnants resulting in major and intermediate merger simulations by B11. The symbols represent the projections of the remnants of our dry intermediate and minor mergers, following the same legend as in the panels of Fig.\,\ref{fig:lambdaobs}. The locations of the remnants by B11 are plotted with contours for their fast-rotator remnants resulting in prograde encounters (purple and light blue, see Fig.~7 in B11 and Fig.~16 in E11), slow-rotator remnants from their retrograde mergers (orange, from Fig.~2 in B11), re-merger remnants (green, see their Fig.~8), and their spiral progenitors (grey, from their Fig.~1). The region in purple denotes where most fast-rotator remnants from major merger simulations by B11 are located, while the one in light blue is only reached under very specific projections of these remnants. \emph{Left panel}: for all our models, independently of their spin-orbit coupling. \emph{Right panel}: the same as in the left panel, but distinguishing between prograde and retrograde encounters. The lines and symbols represent the same as in Fig.\,\ref{fig:lambdaobs}. }
\label{fig:lambdamodel}
\end{figure*}

In Fig.~\ref{fig:lambdamodel}, we compare the location of the 200 projections in the $\lambdae$ -- $\epse$ diagram of our models with the distribution of the remnants obtained in the simulations of major-to-intermediate mergers by B11, who also considered 200 random projections in their remnants (contours in the figure). The blue contour delimits the region that can be achieved by their prograde mergers considering all projections (see their Fig.~7), while the purple one indicates where most of these projections accumulate for major mergers (see Fig.~16 by E11). Orange and green contours mark where the projections of their retrograde and re-merger experiments are located (Figs.~2 and 8 in B11). As in Fig.\,\ref{fig:lambdaobs}, the left panel shows the location of the edge-on projections for our remnants and progenitors, while the right panel distinguishes between the projections of our prograde and retrograde ones (dots and crosses, respectively).

The left panel of Fig.\,\ref{fig:lambdamodel} shows that our simulations with S0b progenitors can reproduce the region in the diagram near the empirical limit defined by E11 for isolating fast from slow rotators in the plane, a region that the major merger experiments by B11 can achieve only through re-mergers (green region) or in very few projections of their prograde encounters (region filled in light blue). The remnants of the S0c progenitors show $\lambdae$ values higher than or similar to the prograde major merger models by B11 (purple region). These models suggest that any location in the diagram occupied by an ETG can be achieved through major and minor mergers, or a combination of both processes \citep[see also][]{Naab2013}. 

The right panel of the figure highlights a key difference between major and minor merger events. Major mergers are such violent processes that retrograde encounters always result in slow rotators \citep[][B11]{Gonzalez-Garcia2005,Jesseit2009}. Nevertheless, in minor mergers a retrograde encounter onto a fast-rotator S0 can result in another fast rotator, because the decrement in the original rotational support of the galaxy induced by the encounter may be quite low. In fact, many of the projections of the retrograde models (crosses in the right panel of Fig.\,\ref{fig:lambdaobs}) are located in the region of fast rotators, and in some cases the projections can even show apparent \lambdae\ values higher than the progenitor (see Sect.\,\ref{sec:modelparameters}). This again supports the idea that any location of the diagram can be achieved through mergers. 

In conclusion, dry minor mergers can evolve highly flattened fast-rotator S0s (which in turn might derive from prograde major mergers occurred at earlier epochs or from gas-stripped spirals) into systems with lower intrinsic ellipticities, inducing weaker changes in their rotational support than major encounters. The remnants exhibit a higher fraction of random projections with intermediate apparent ellipticities than the progenitors and a lower one at low \epse\ values due to the formation of oval distortions in the centres and to the disc thickening. The present models show that minor mergers can provide a plausible explanation to the existence of S0s with hybrid kinematics in the \lambdae\ -- \epse\ diagram that major mergers and cosmological simulations find difficult to reproduce.

\subsubsection{Dependence on initial conditions}
\label{sec:modelparameters}

\begin{figure*}[th]
\centering
  \includegraphics*[width=4.5cm]{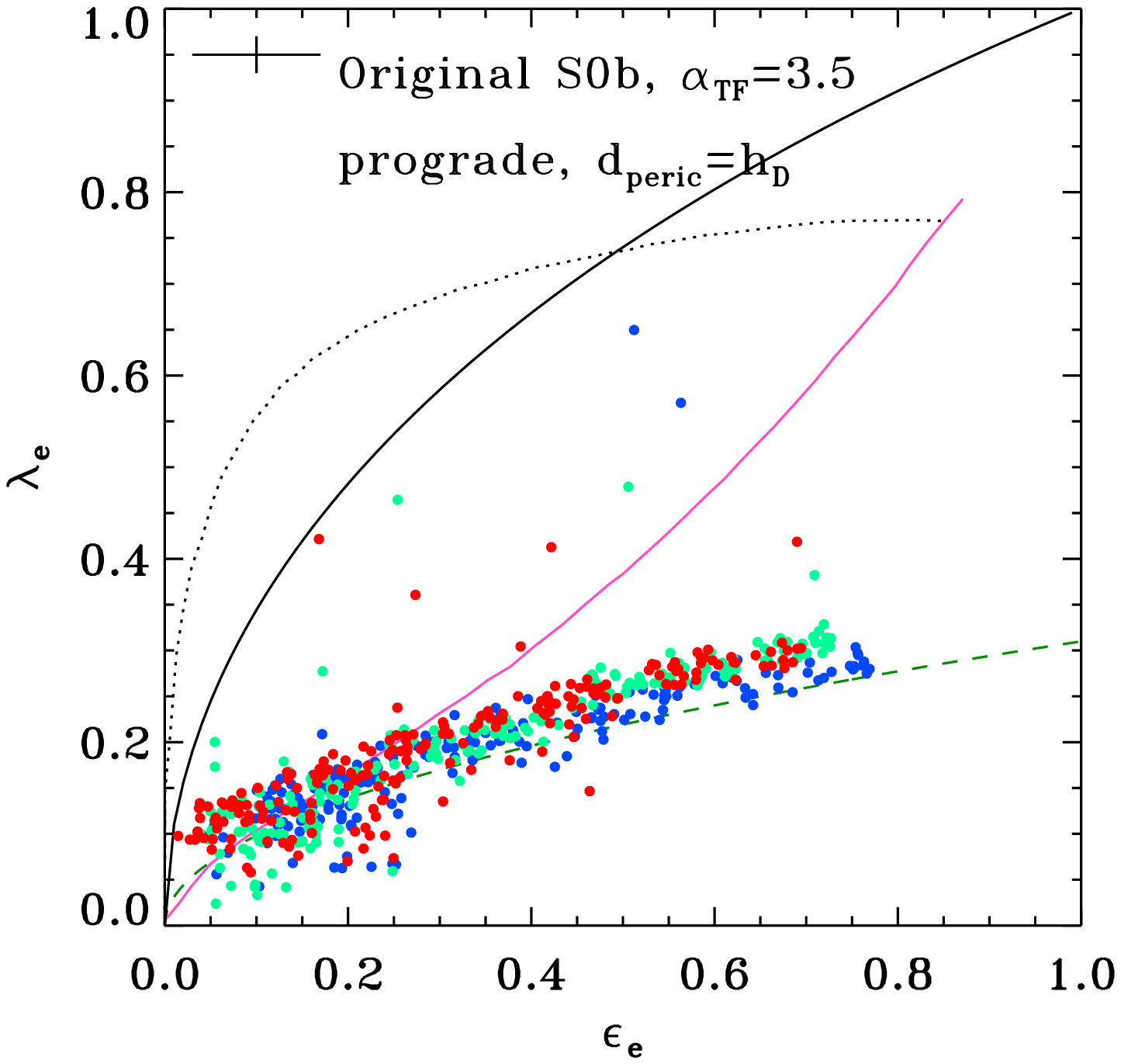}
  \includegraphics*[width=4.5cm]{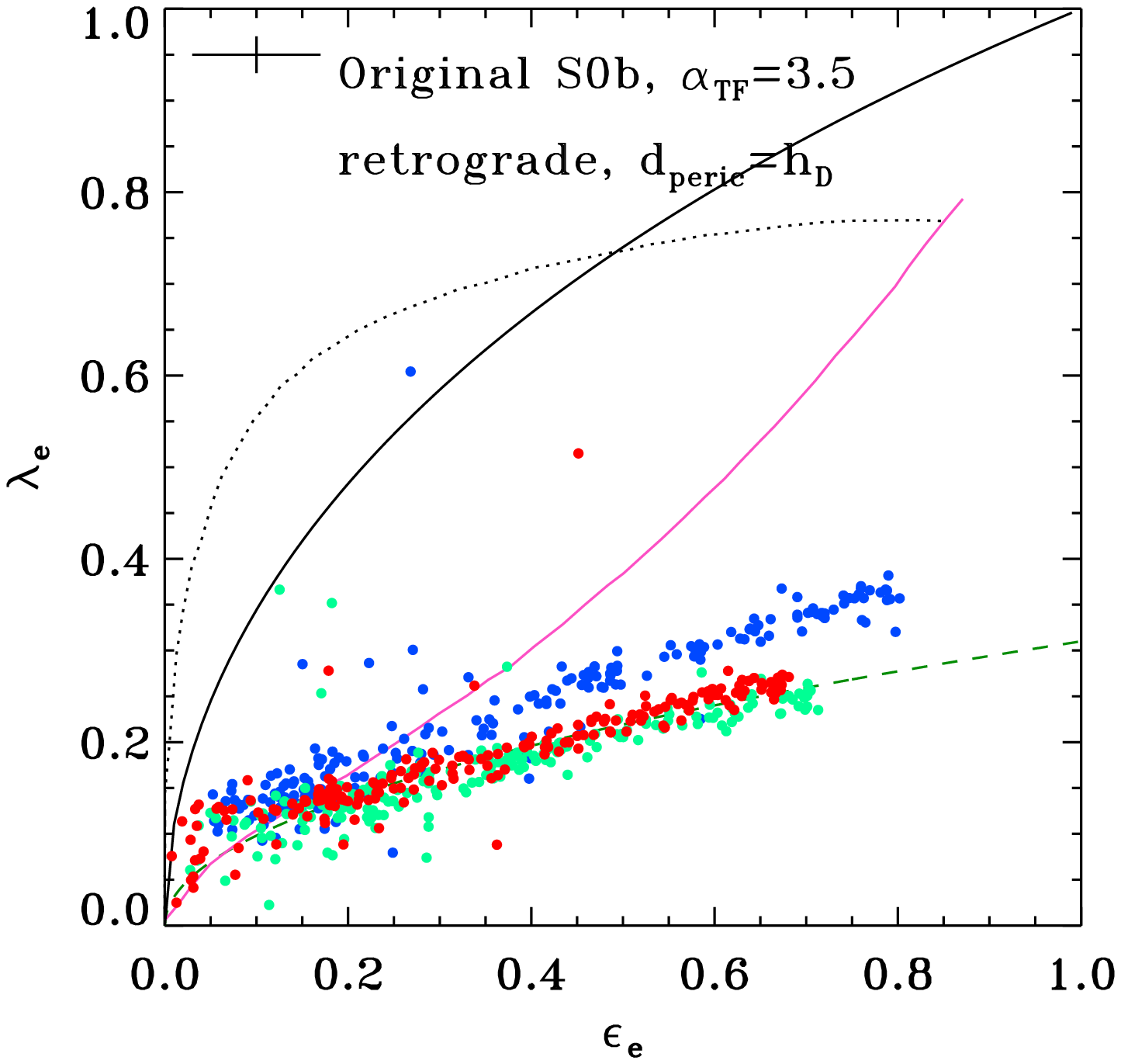}
  \includegraphics*[width=4.5cm]{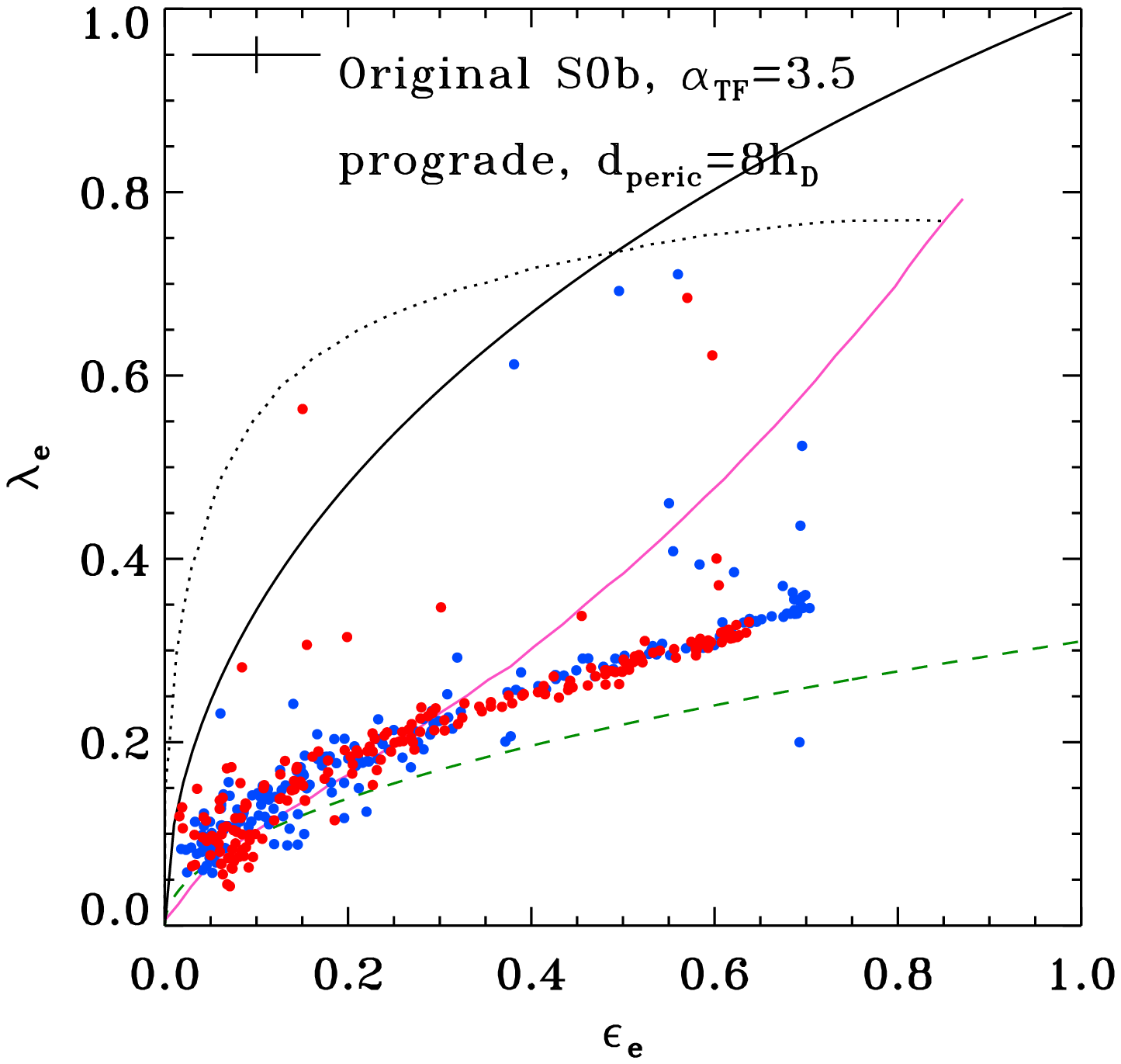}
  \includegraphics*[width=4.5cm]{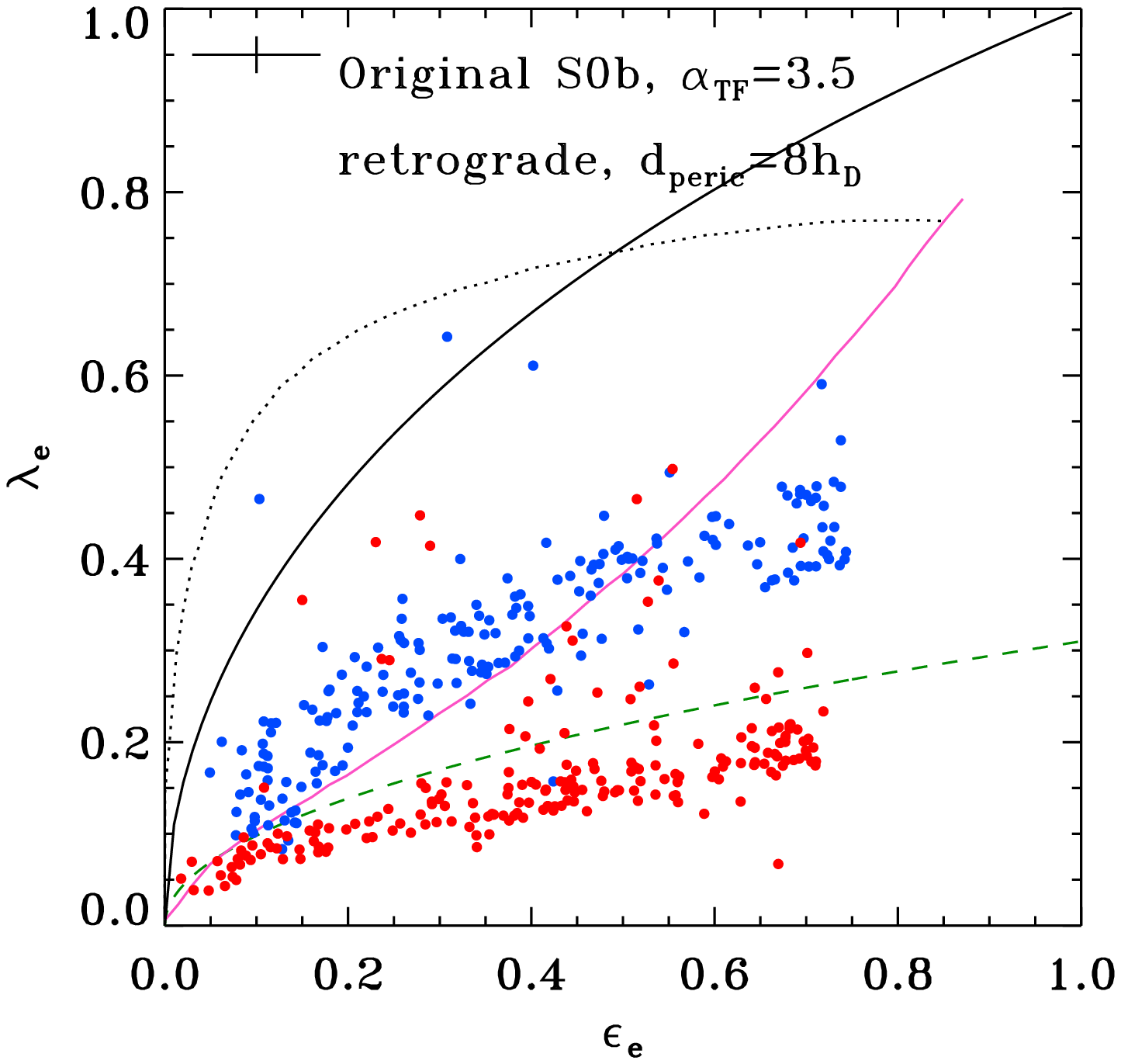}
  \caption{Dependence on the mass ratio of the location in the $\lambda_\mathrm{e}$ -- $\epse$ diagram of the 200 random projections of our models, for an identical set of initial conditions (indicated at each frame). \emph{Red dots}: model with 6:1. \emph{Green dots}: model with 9:1. \emph{Blue dots}: model with 18:1. The legend for the lines is the same as in Fig.\,\ref{fig:lambdaobs}. The error bars on the left upper corner of each frame represent the typical errors in both axes.}
\label{fig:massratio}
\end{figure*}

We studied the dependence of the distribution in the $\lambda_\mathrm{e}$ -- $\epse$ diagram on the initial conditions in our models. Figure\,\ref{fig:massratio} shows the distribution of models that only differ in the mass ratio of the encounter in each frame. In the models with prograde encounters (first and third panels), there is no difference in the distribution of projections of models with different mass ratios (the different models lie within 1$\sigma$ distribution in both panels). However, in the case of retrograde orbits (second and fourth panels), encounters with lower mass ratios (i.e., with more massive satellites) seem to decrease the intrinsic rotational support of the galaxy more. 

\begin{figure}[!th]
\centering
  \includegraphics[width=7.5cm]{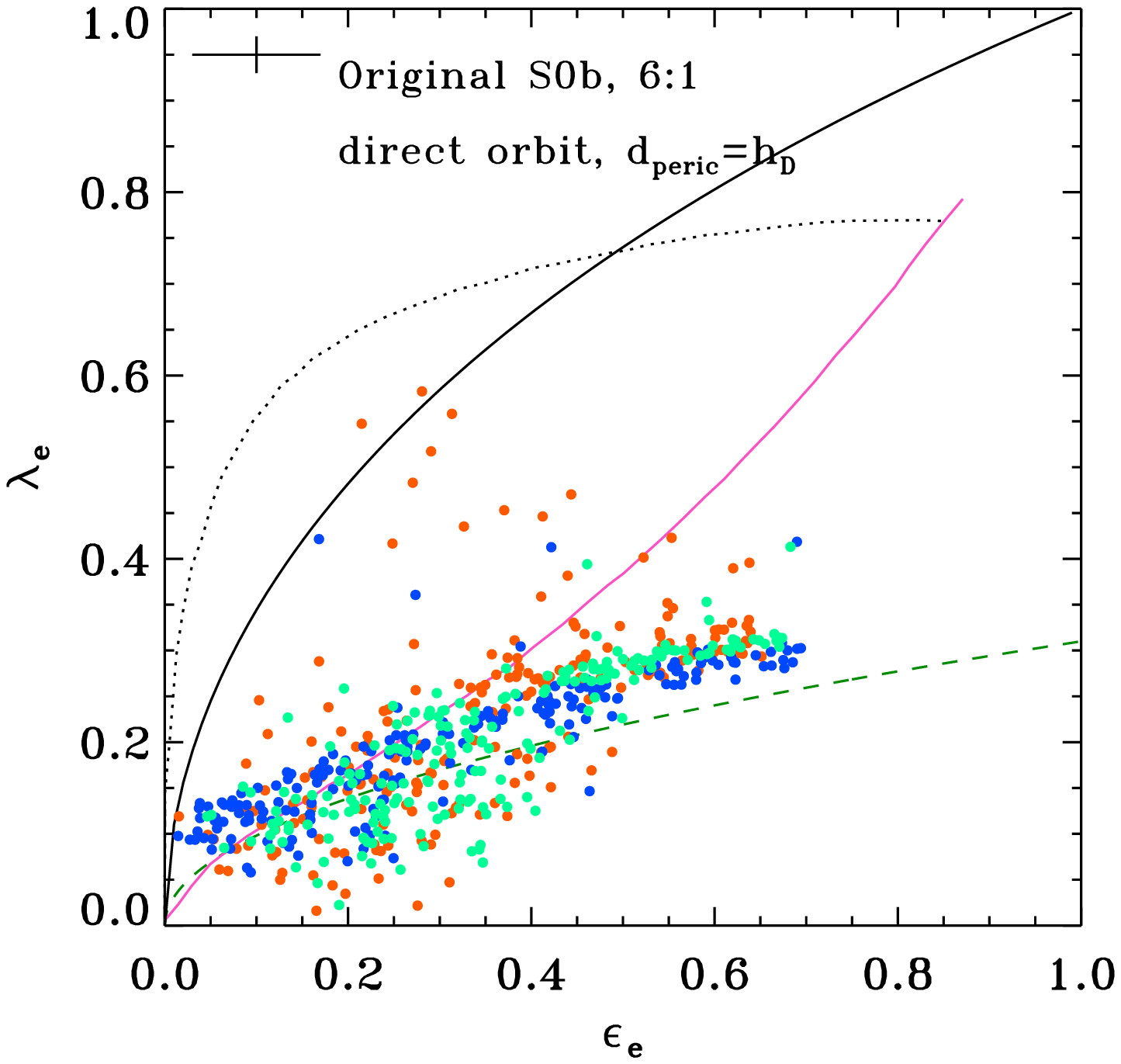}
  \includegraphics[width=7.5cm]{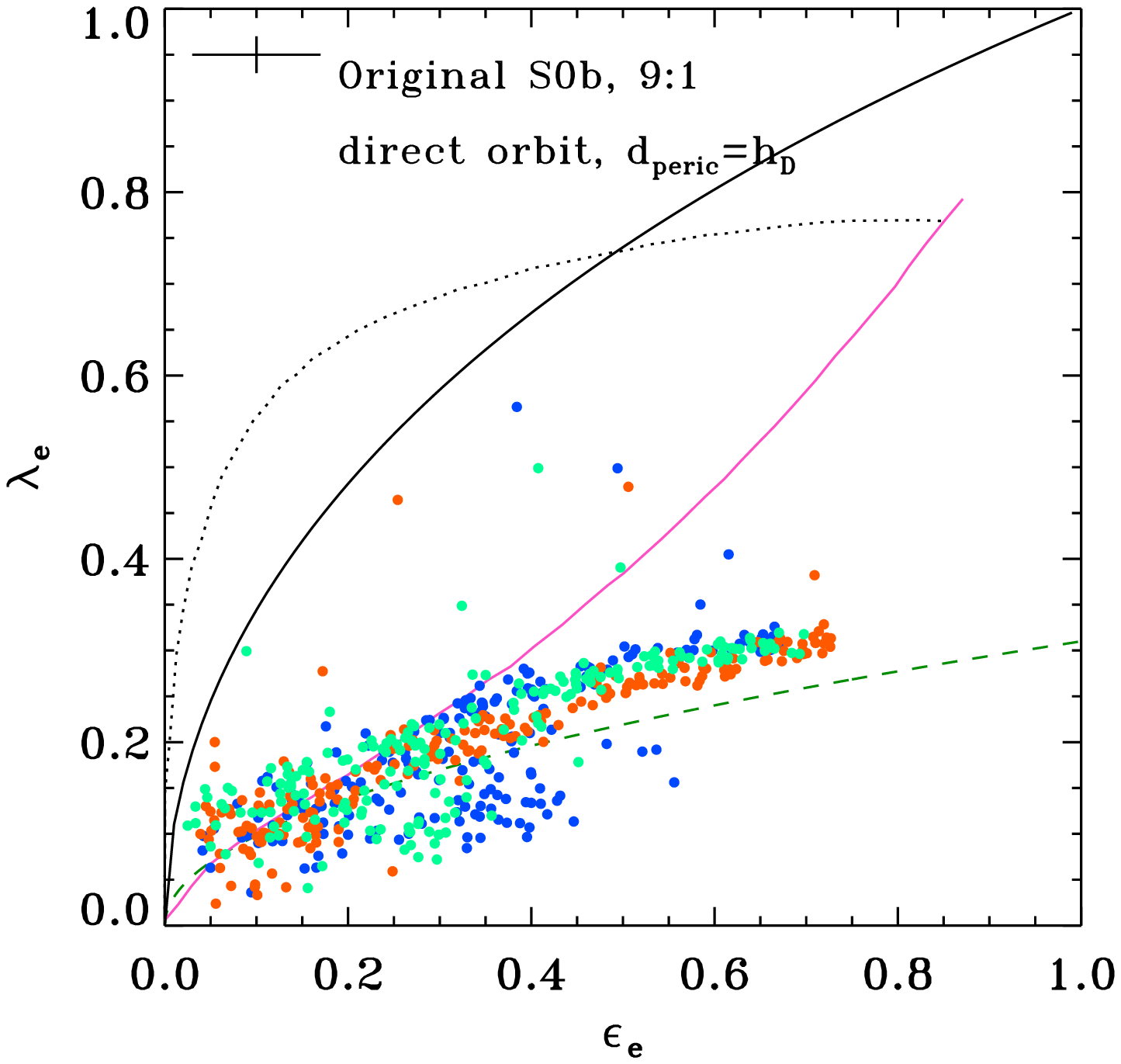}
  \caption{Dependence on the satellite-to-primary density ratio (as provided for the $\alphaTF$ parameter) of the location in the $\lambda_\mathrm{e}$ -- $\epse$ diagram of the 200 random projections of our models for an identical set of initial conditions (indicated in each frame). \emph{Red dots}: model with $\alphaTF = 4.0$. \emph{Green dots}: model with $\alphaTF = 3.5$. \emph{Blue dots}: model with $\alphaTF = 3.0$. Higher values of $\alphaTF$ imply satellites denser than the primary galaxy. The legend for the lines is the same as in Fig.\,\ref{fig:lambdaobs}. The error bars in the left upper corner of each frame represent the typical errors in both axes.
  }
\label{fig:tullyfisher}
\end{figure}

\begin{figure*}[th]
\centering
  \includegraphics*[width=4.5cm]{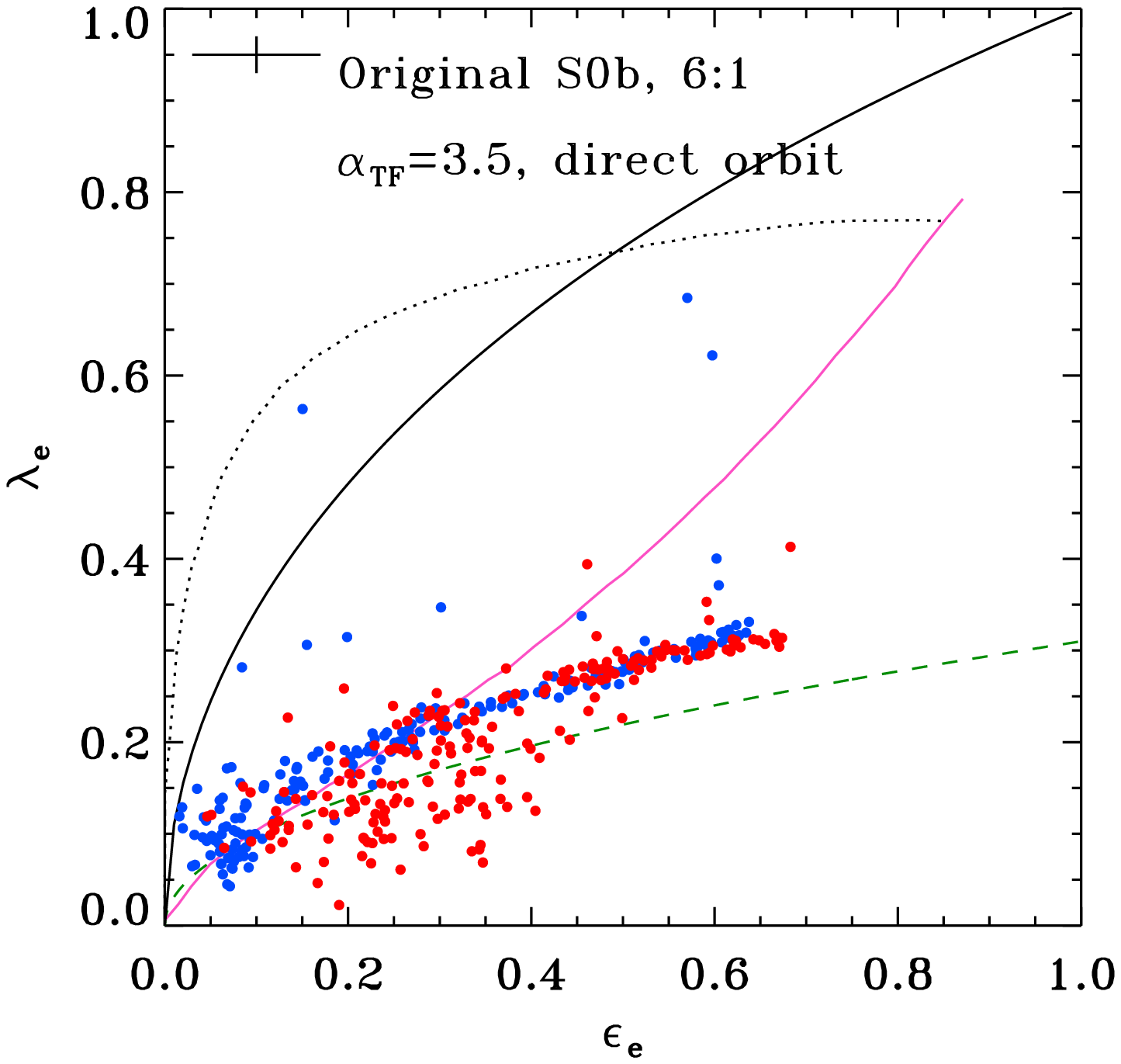}
  \includegraphics*[width=4.5cm]{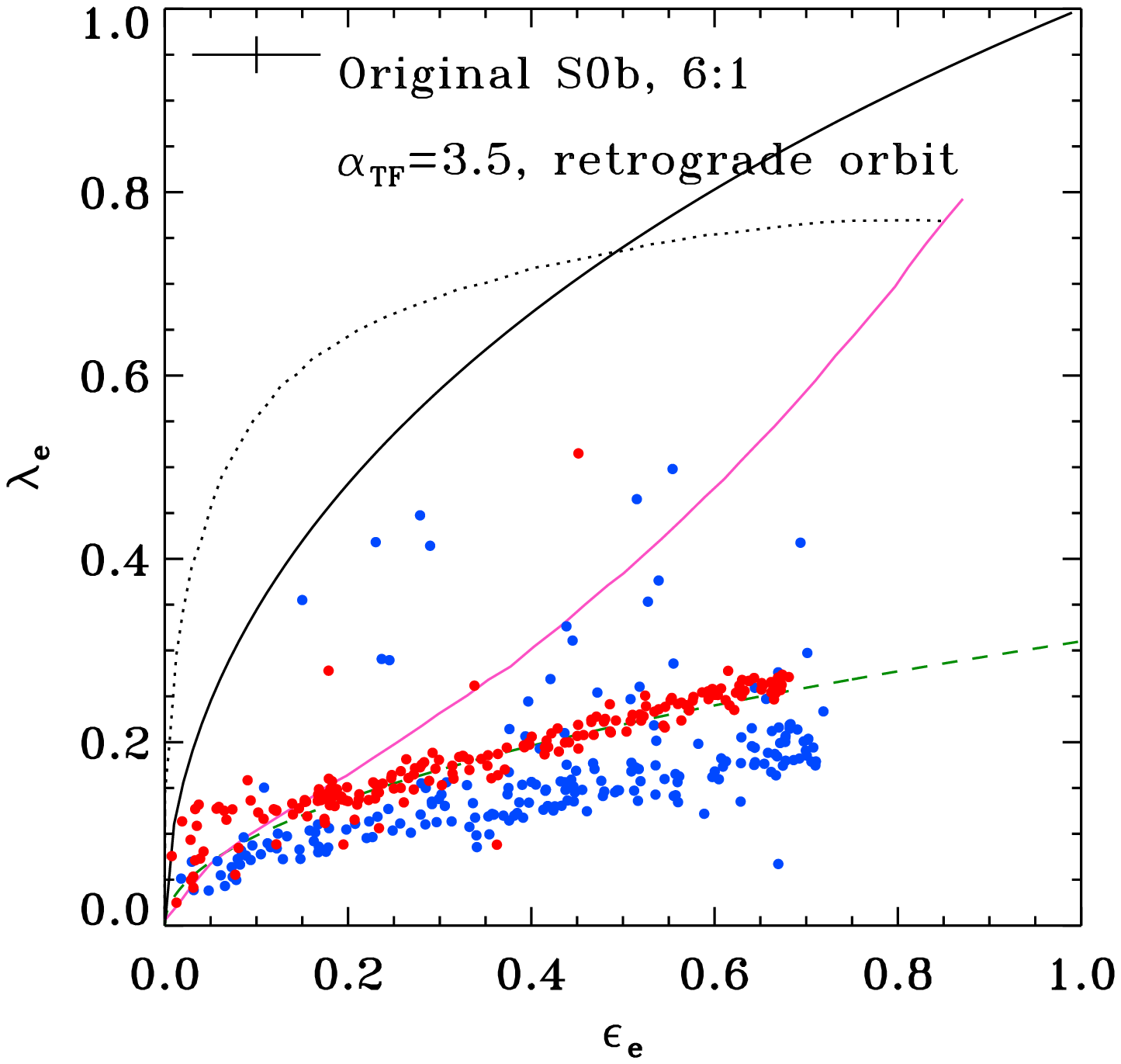}
  \includegraphics*[width=4.5cm]{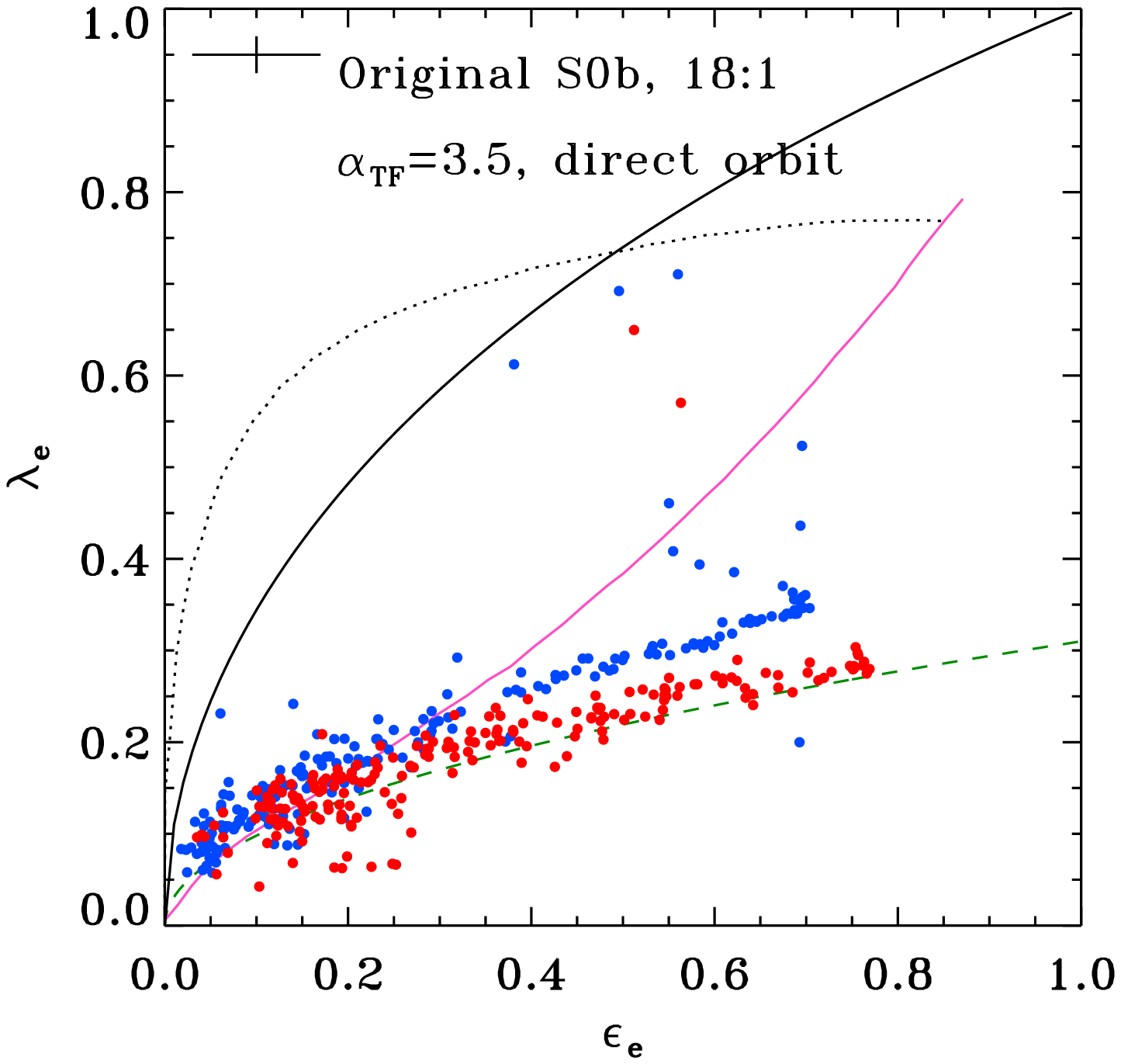}
  \includegraphics*[width=4.5cm]{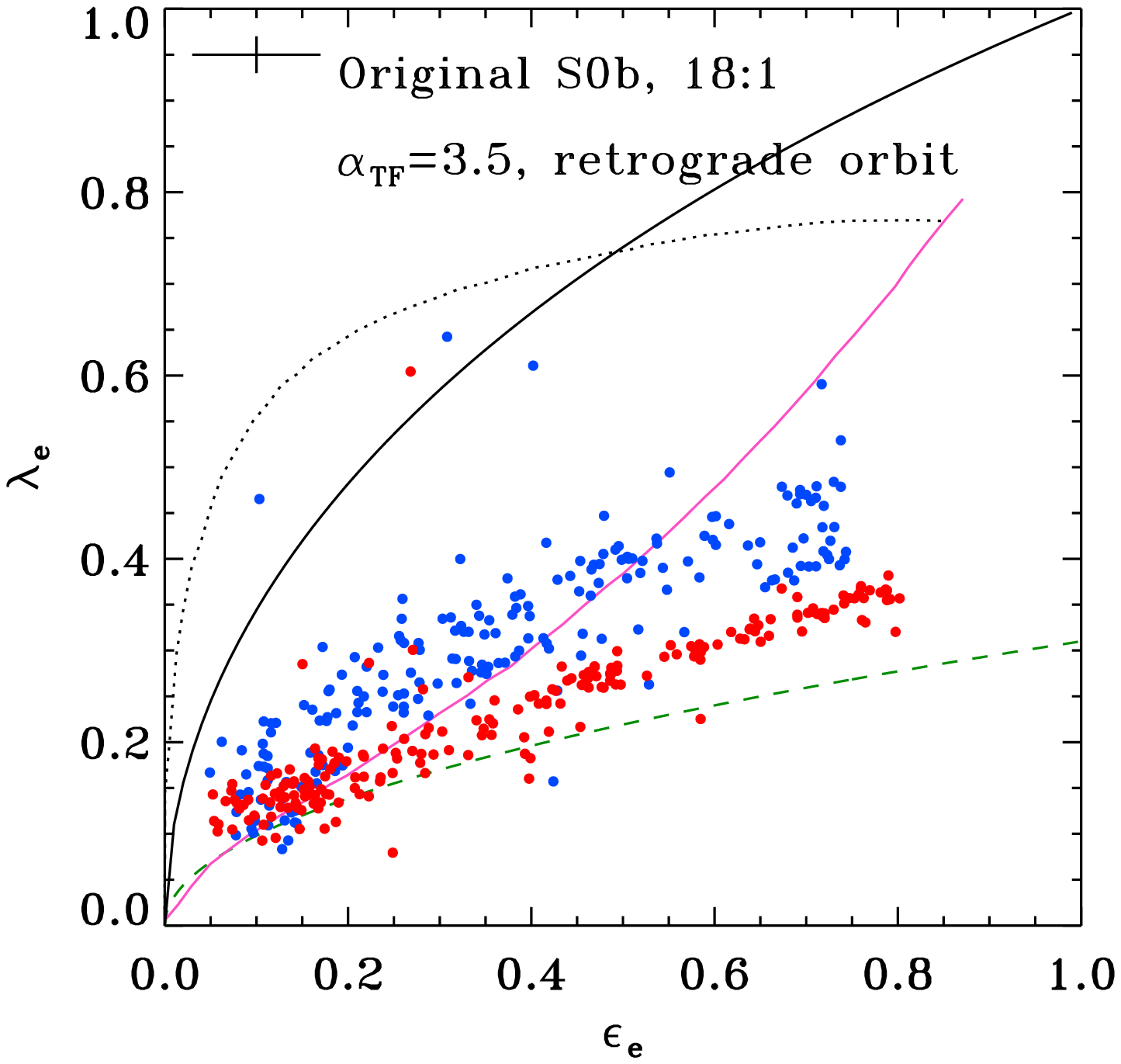}
  \caption{Dependence on the pericentre distance of the location in the $\lambda_\mathrm{e}$ -- $\epse$ diagram of the 200 random projections of our models for an identical set of initial conditions (indicated at each frame). \emph{Red dots}: model with $d_\mathrm{per} = h_\mathrm{D}$ (short pericentre distance). \emph{Blue dots}: model with $d_\mathrm{per} = 8\,h_\mathrm{D}$ (long pericentre distance). The legend for the lines is the same as in Fig.\,\ref{fig:lambdaobs}. The error bars on the left upper corner of each frame represent the typical errors in both axes.}
\label{fig:pericentre}
\end{figure*}

In prograde orbits, the satellite material typically co-rotates with the primary galaxy stars (see EM06; EM11). The velocity field is thus controlled by the primary stellar material and the contribution of the accreted mass to the mass-weighted average velocities is low, independently of the mass of the accreted satellite. Therefore, the decrease in the rotational support induced by the merger in prograde encounters is quite similar.  However, the maximum ellipticity in the models with lower mass ratios (6:1) tends to be lower than in those with higher mass ratios (18:1). In retrograde cases, the satellite material typically counter-rotates with the primary galaxy stars, and hence, the rotational support decreases more if more stars (from the satellite) counter-rotate (i.e., in the cases with lower mass ratios). 
We plot the dependence of the location of the models in the $\lambda_\mathrm{e}$ -- $\epse$ plane with the satellite-to-primary galaxy density ratio in Fig.\,\ref{fig:tullyfisher}. As shown there, the central satellite density does not significantly affect the results for the range of values studied here. However, the dependence on the pericentre distance is much more complex. In Fig.\,\ref{fig:pericentre}, we plot the distribution of projections of models that only differ in the pericentre distance of the initial orbit in each panel. The trends vary depending on the parameters of the simulations: while there is no trend with the pericentre in the prograde encounters with mass ratio 6:1 (first panel of the figure), encounters with long pericentre distances in mergers with mass ratio 18:1 (third panel) induce a lower decrement of the original rotational support of the progenitor than those with short pericenters. In the cases of retrograde orbits, the trend with the pericentre distance is exactly the opposite depending on the mass ratio (compare the second and fourth panels in Fig.\,\ref{fig:pericentre}). The effects of the mass ratio and the pericentre distance on the resulting \lambdae\ thus seem to be of similar order. Nevertheless, the maximum apparent ellipticities achieved in each remnant seem to be quite similar.

\begin{figure}[th]
\centering
  \includegraphics[width=7.5cm]{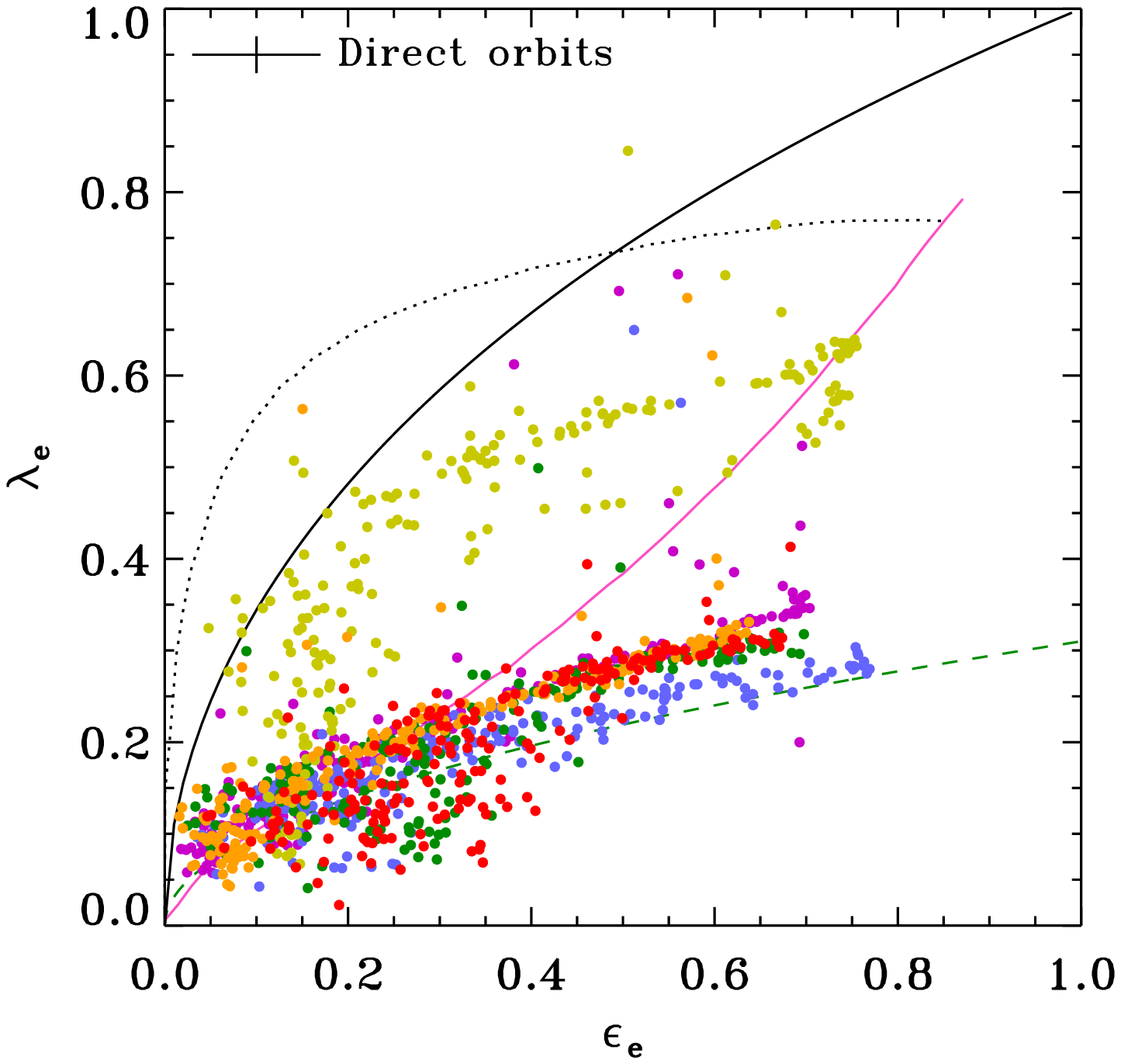}
  \includegraphics[width=7.5cm]{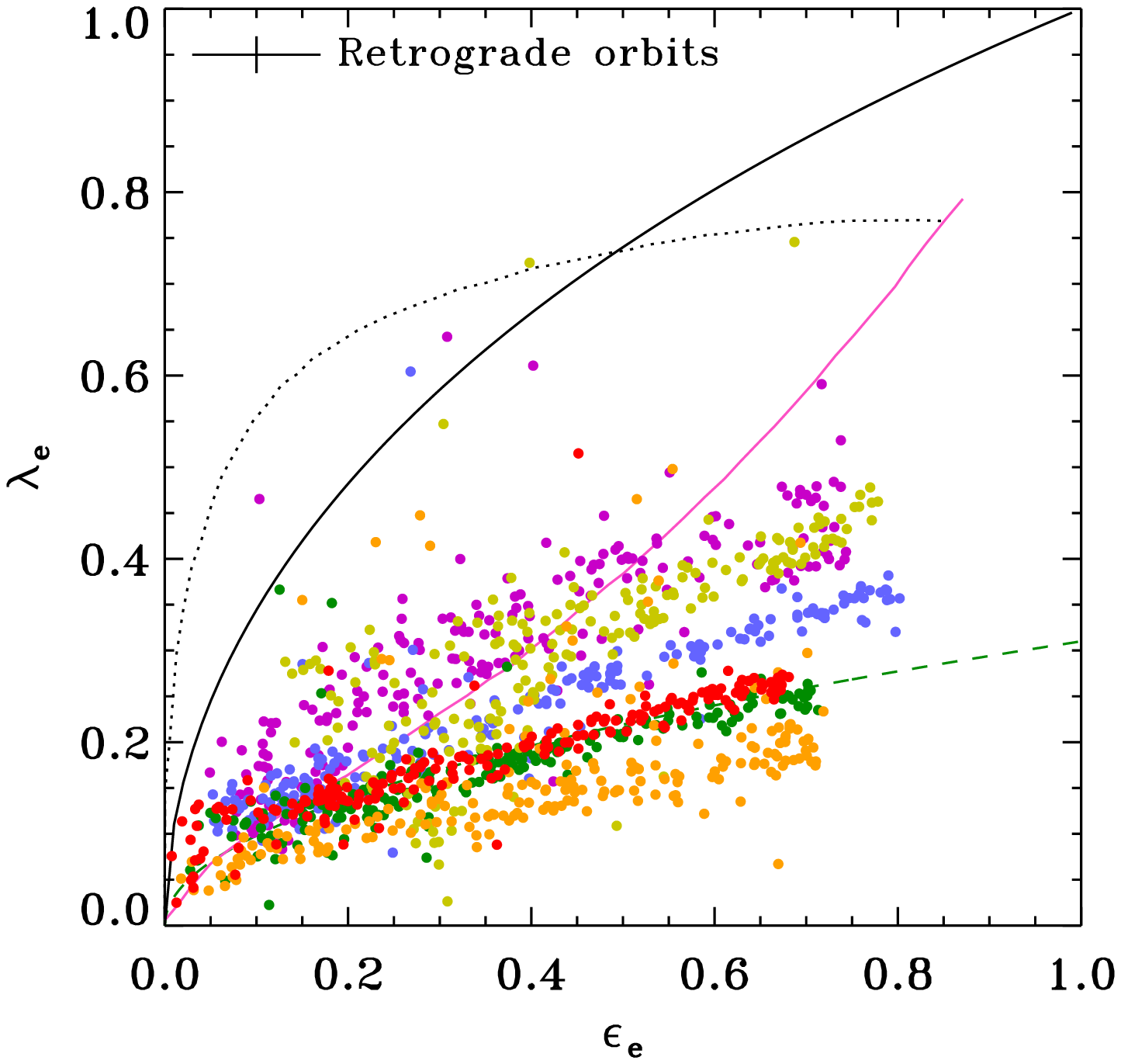}
  \caption{Dependence on the spin-orbit coupling of the location in the $\lambda_\mathrm{e}$ -- $\epse$ diagram of the 200 random projections of our models. Prograde models are represented in the \emph{top panel}, while retrograde ones are plotted in the \emph{bottom panel}. The same color in each panel signifies the same set of initial parameters ($\alphaTF = 3.5$ in all cases). The color-coding is the same as the one used in the left panels of Fig.\,\ref{fig:percentages} (consult the caption there). The legend for the lines is the same as in Fig.\,\ref{fig:lambdaobs}. The error bars in the left upper corner of each frame represent the typical errors in both axes.}
\label{fig:spinorbit}
\end{figure}

\begin{figure}[th]
\centering
  \includegraphics[width=7.5cm]{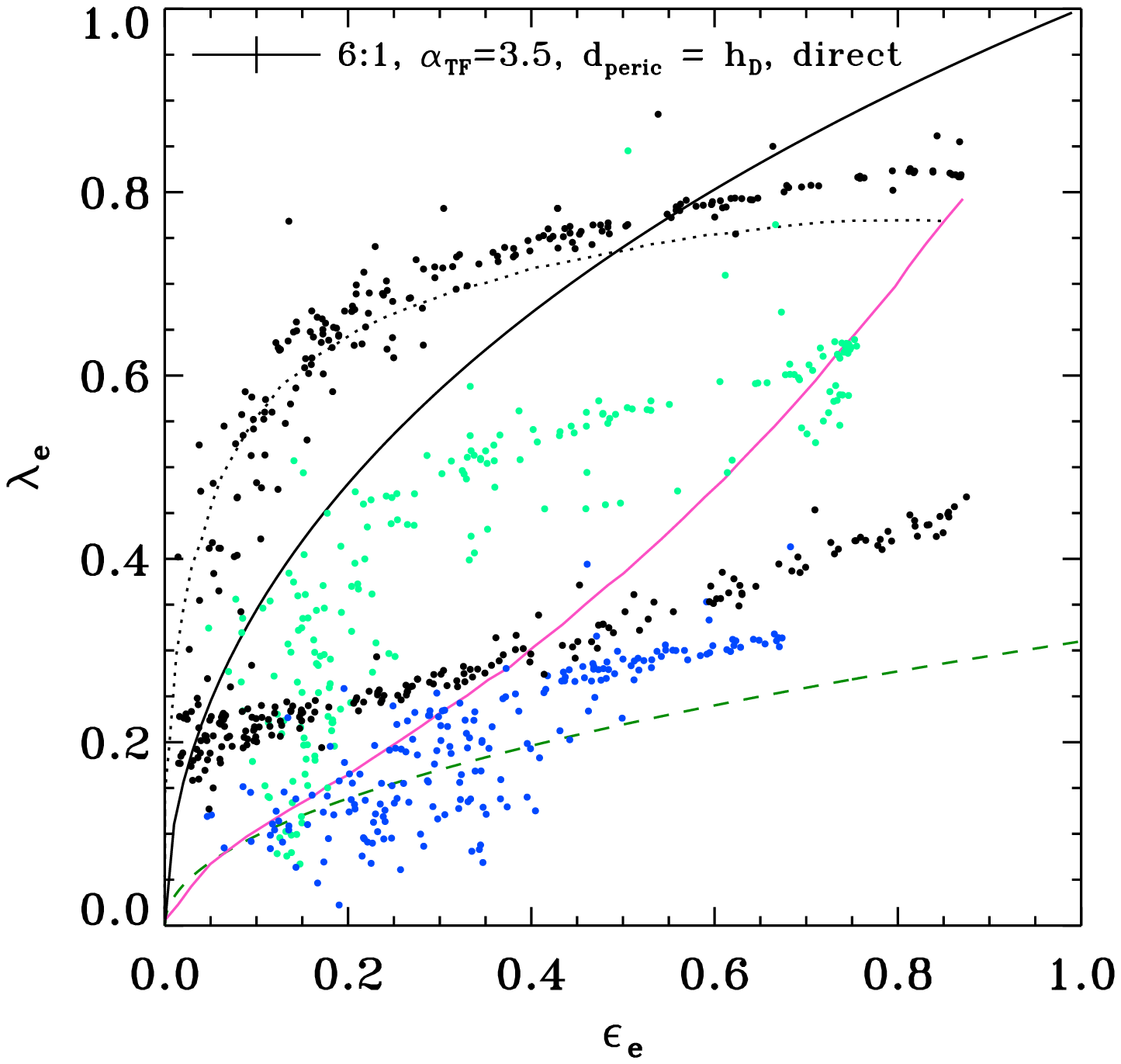}
  \includegraphics[width=7.5cm]{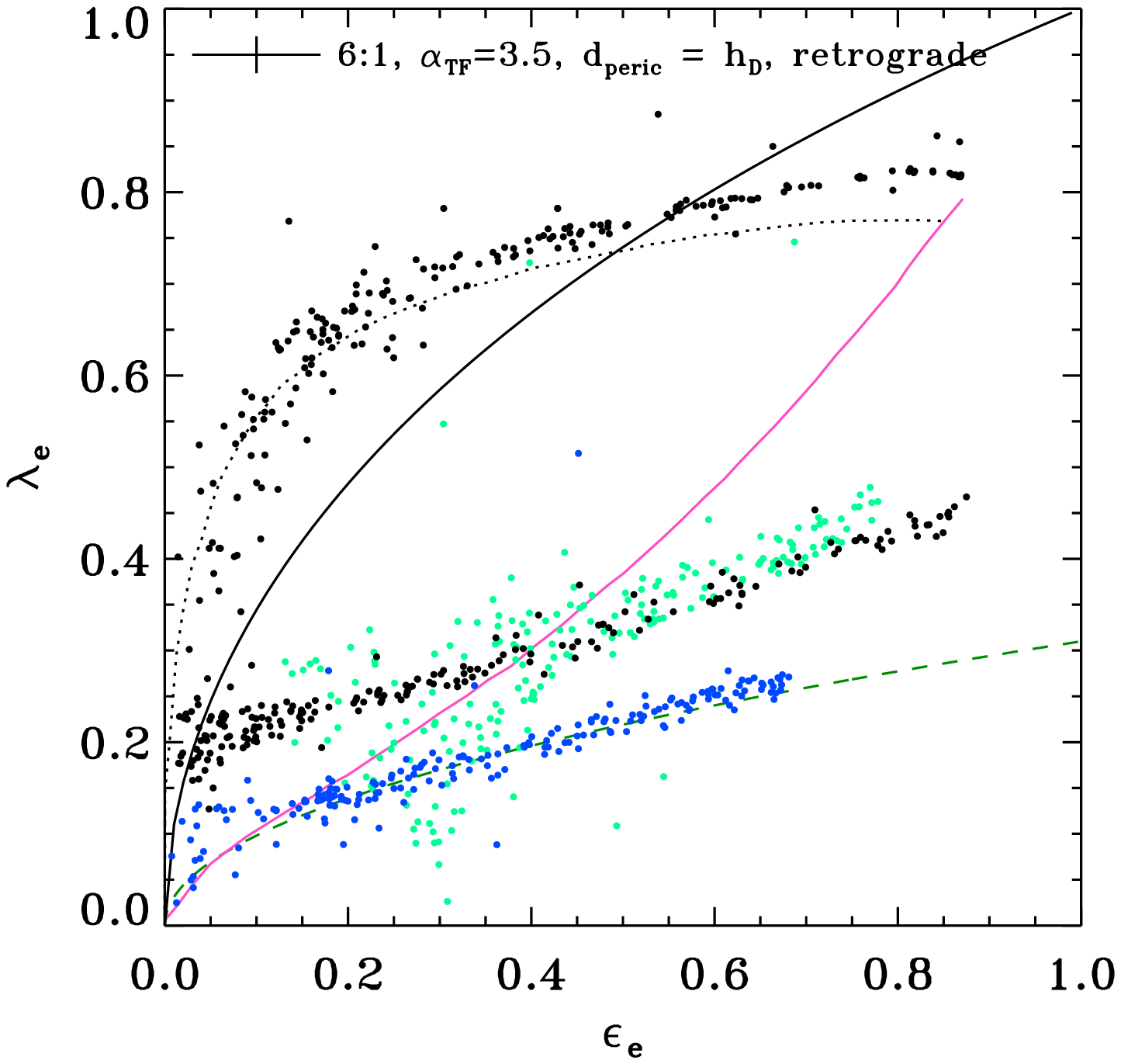}
  \caption{Dependence on the original S0 model of the location in the $\lambda_\mathrm{e}$ -- $\epse$ diagram of the 200 random projections of our models. In each panel, we plot the progenitors and remnant resulting for models with identical initial conditions and accreted satellites, but with different initial S0 galaxies (S0b or S0c).  The projections for the original S0s are represented in both panels for comparison (black dots). Blue and green dots correspond to the remnants of the models using initial S0b and S0c galaxies, respectively. \emph{Top panel}: projections of the models with mass ratio 6:1, $\alphaTF = 3.5$, small pericentre distance, and prograde orbits. \emph{Bottom panel}: projections of the models with mass ratio 6:1, $\alphaTF = 3.5$, small pericentre distance, and retrograde orbits. The legend for the lines is the same as in Fig.\,\ref{fig:lambdaobs}. The error bars in the left upper corner of each frame represent the typical errors in both axes.}
\label{fig:massbulge}
\end{figure}

As commented before, the dynamical state of the remnants strongly depends on the spin-orbit coupling of the encounter. We plot the $\lambda_\mathrm{e}$ -- $\epse$  diagram for the 200 projections considered for the models with prograde orbits and with retrograde orbits separated in the two panels of Fig.\,\ref{fig:spinorbit}. The same color has been used in both panels for a given set of initial conditions, so the distributions resulting for direct and retrograde orbits for the same set of initial conditions can be easily compared. Although the intrinsic ellipticities (i.e., the maximum of all the projections for a given model) seem to be similar for both prograde and retrograde experiments for a given set of initial parameters, the intrinsic rotational support of the remnant depends noticeably on the coupling of the angular momenta of the galaxies and the orbit. In general, retrograde cases exhibit lower values of $\lambda_\mathrm{e}$ than direct cases for similar projections, as expected. The only two exceptions are the retrograde models with the lowest mass ratio (18:1), which surprisingly exhibit higher values of \lambdae\ than their direct analogs. The reason might be that these two remnants present the least disrupted distribution of satellite mass in the centres of all models (see Figs.\,4 and 5 in EM11), although they were the most evolved in time of all models. Therefore, the remnants might not be fully relaxed.
In Fig.\,\ref{fig:massbulge}, we compare the projections in the $\lambda_\mathrm{e}$ -- $\epse$ diagram of models that only differ in the original primary galaxy in each frame. The model with an original S0b galaxy is more stable against the dynamical changes induced by the accreted satellite than the model with an original S0c galaxy for any set of initial conditions, as central mass concentrations tend to stabilize the galaxy disc (see \citealt{Gonzalez-Garcia2005} and references in \citealt{Eliche-Moral2012}).

To summarize, for the space of parameters covered here, we find a significant dependence of the results with the mass ratio of the encounter, the spin-orbit coupling of the model, the pericentre distance, and the bulge-to-disc ratio of the primary galaxy. The satellite density only moderately affects the location of the projections of each model (for identical initial conditions). Some systematic trends are found in some cases, but the dynamical state of the final remnant is a complex combination of all these initial conditions.

\subsection{Anisotropy of velocities}
\label{sec:anisotropy}

\subsubsection{General definitions and considerations}
\label{sec:anisotropydef}
The anisotropy of velocities ($\delta$) is a measurement of how relevant rotation is for the stellar dynamical state of a galaxy, that is, whether its spatial structure is caused by a flattening of the velocity ellipsoid in any spatial direction or not. According to \citet{Binney1978}, $\delta$ is defined as,

\begin{equation}\label{eq:delta}
 \delta = 1 - \frac{\Pi_\mathrm{zz}}{\Pi_\mathrm{xx}},
\end{equation}

\noindent where $\Pi_\mathrm{ii}$ are the diagonal elements of the velocity dispersion tensor in the i$^\mathrm{th}$ direction (in axisymmetric systems, $\Pi_\mathrm{xx}=\Pi_\mathrm{yy}$). By definition, $z$ is in the direction of the symmetry axis of the galaxy (i.e., the line of sight in which the galaxy is seen face-on). The elements of the velocity dispersion tensor are estimated through integrals of the distribution function of the system $f(\mathbf{w})$, as

\begin{equation}\label{eq:pi}
\Pi _{i,j} \equiv \int (v_i - \overline{v_i})\,(v_j - \overline{v_j})\,f(\mathbf{w})\,d^6\mathbf{w},
\end{equation}

\noindent where $\mathbf{w}$ represents a vector ($\mathbf{r}$,$\mathbf{v}$) in the phase space. The bar onto any quantity represents an average over the velocity space \citep[see also][]{Binney2005},

\begin{equation}\label{eq:meanv}
\overline{v_i}(\mathbf{x}) \equiv \frac{1}{\rho(\mathbf{x})} \int v_i \,f(\mathbf{w})\,d^3\mathbf{v},
\end{equation}

\noindent and $\rho (\mathbf{x})$ is the density distribution of the galaxy, defined as 

\begin{equation}\label{eq:density}
\rho(\mathbf{x}) \equiv \int f(\mathbf{w})\,d^3\mathbf{v}.
\end{equation}

We estimated $\delta$ directly from our N-body data for each model using eqs.\,\ref{eq:delta} -- \ref{eq:density}, restricted to the stellar particles within $R_\mathrm{eff,glx}$ to allow the comparison with observational data. The resulting $\delta$ values are provided in Table\,\ref{tab:kinematics}. The errors in $\delta$ correspond to the propagation of the estimated errors of numerical integration and the limited resolution in the phase space. We remark that these values were obtained considering the galaxies as single-component systems.
We compared the distribution of our models in the $\delta$ -- $\epsilon_\mathrm{e,intr}$ (considering the intrinsic ellipticity $\epsilon_\mathrm{e,intr}$) plane with the one shown by real ETGs from the SAURON and ATLAS$^\mathrm{3D}$ samples in Sect.\,\ref{sec:anisotropycomp}. The 3D spectroscopic data of these galaxies were used to construct complex axisymmetric JAM dynamical models that reproduce in detail both the galaxy morphology and its stellar kinematics out to $R_\mathrm{eff,glx}$ \citep[JAM stands for Jeans anisotropic multi-gaussian expansion models, see][]{Cappellari2013}. These JAM models provide estimates of $\delta$ and $\epsilon_\mathrm{e,intr}$ for these galaxies under the assumption of axisymmetry. 
The $\delta$ values from the SAURON sample that we have used in Sect.\,\ref{sec:anisotropycomp} were directly taken from \citet{Burkert2008} and  mostly refer to elliptical galaxies. These authors used the JAM models published by \citet{Cappellari2007} for 22 ETGs to derive $\delta$ and $\epsilon_\mathrm{e,intr}$ for them (16 ellipticals and 6 S0s). We increased this observational sample in Sect.\,\ref{sec:anisotropycomp} by adding the JAM models recently published for 15 galaxies (12 S0s and 3 ellipticals) of the ATLAS$^\mathrm{3D}$ sample \citep{Cappellari2013}. Because the values of ($\beta_z$,$\gamma$) provide a higher likelihood for the modelling of each galaxy reported by these authors, we can derive $\delta$ from eq.\,7 in \citet{Cappellari2007},
\begin{equation}\label{eq:deltaforobservationaldata}
 \delta = \frac{(2\beta_z) - \gamma)}{2 - \gamma}.
\end{equation}

\noindent In the general equation above, $\beta_z$ and $\gamma$ are two additional anisotropy parameters defined by \citet{Cappellari2007} to describe the shape of the velocity-dispersion tensor in different planes (see there for specific definitions). The $\epsilon_\mathrm{e,intr}$ of these JAM-modelled galaxies have been derived using the apparent \epse\ values and galaxy inclinations provided in Table\,1 by \citet{Cappellari2013} with the following relation \citep[eq.\,13 in][]{Cappellari2007}:
\begin{equation}\label{eq:epseintr}
 \epsilon_\mathrm{e,intr} = 1 - \sqrt{ 1 + \epse \, \frac{\epse - 2}{\sin^2 (i)}}. 
\end{equation}

We also compared our results with the theoretical trends expected for oblate axisymmetric models in the  $\delta$ -- \epse\ diagram \citep{Binney2005}. In these systems, $\delta$ and the intrinsic values of $V/\sigma$ and $\epsilon$ are related according to
 
\begin{equation}\label{eq:deltaBinney}
 \delta = 1 - \frac{1 + (V/\sigma)^2}{[1 - \alpha (V/\sigma)^2 ]\Omega(e)},
\end{equation}

\noindent where 
\begin{equation}\label{eq:e}
e = \sqrt{1 - (1 - \epsilon)^2}, 
\end{equation}

\noindent and 

\begin{equation}\label{eq:omega}
\Omega(e) = \frac{0.5 (\arcsin e - e \sqrt{1 - e^2})}{e\sqrt{1-e^2} - (1-e^2) \arcsin e}.
\end{equation}
In the equations above, $\alpha$ is a dimensionless parameter that quantifies the contribution of streaming motion to the line-of-sight velocity dispersion.  \citet{Cappellari2007} found for isotropic galaxy models  $\alpha \sim 0.15$, and that it varies very little in different galaxies, despite depending on the stellar density distribution and the shape of the rotation curve.  Therefore, the anisotropy of velocities can be estimated from a set of edge-on (i.e., intrinsic) parameters for oblate axisymmetric models (see Sect.\,\ref{sec:anisotropycomp}).

\subsubsection{Comparison with data and major merger simulations}
\label{sec:anisotropycomp}

In Fig.\,\ref{fig:anisotropy}, we compare the distribution of our models in the $\delta$ -- $\epsilon_\mathrm{e,intr}$ diagram with those of real elliptical and S0 galaxies (left panel) and with elliptical remnants resulting from major merger and cosmological simulations (middle and right panels). In the first panel, we can see that the distributions of real ellipticals and S0s  in the $\delta$ -- $\epsilon_\mathrm{e,intr}$ diagram overlap widely. However, while ellipticals (purple circles) follow a tight linear correlation in the $\delta$ -- $\epsilon_\mathrm{e,intr}$ plane (except for one outlier), S0s spread at intermediate-to-high intrinsic ellipticities (magenta squares). The trend of the ellipticals was already reported by \citet{Burkert2008}, but the wide spread of the S0s in the diagram was not deduced there, because the SAURON sample only contained six S0s. Nevertheless, when additional data of S0s from ATLAS$^\mathrm{3D}$ are considered, it is clear that $\delta$ and $\epsilon_\mathrm{e,intr}$ are linearly correlated in elliptical galaxies, whereas they do not follow any trend for S0s, which are widely dispersed in the region of $0.4<\epsilon_\mathrm{e,intr}<0.9$. However, note that some S0s follow the linear trend drawn by the ellipticals.

The remnants of our minor mergers (which are S0s) are also spread in the region of high \epse\ and $\delta$ values (see the stars in the first panel of the figure). The remnants from experiments with an S0b progenitor lie nearby the trend of the ellipticals (maybe because the S0b progenitor already does), but those resulting from S0c progenitors lie high above this trend. Considering all models globally, this means that the S0s resulting from a minor merger can be quite dispersed in the $\delta$ -- $\epsilon_\mathrm{e,intr}$ diagram, depending on the progenitor.

We have overplotted a linear fit performed to the data of real ellipticals in the first panel of Fig.\,\ref{fig:anisotropy}. The resulting slope has been $d\delta / d\epsilon \sim$ 0.52 $\pm$ 0.08, with a correlation coefficient of $\rho = $ 0.84 (purple dashed line). The slope is very similar to the one obtained by \citet{Burkert2008} from fitting all SAURON galaxies in this panel ($d\delta / d\epsilon \sim$ 0.55 $\pm$ 0.10), even though they included their six S0s (empty magenta squares).  

In the second and third panels of Fig.\,\ref{fig:anisotropy}, we compare the location in the $\delta$ -- $\epsilon_\mathrm{e,intr}$ plane of our minor merger experiments with the ellipticals resulting from the major mergers and cosmological simulations by \citet{Burkert2008}. We have overplotted the linear fit performed to the real ellipticals in the first panel of the figure (purple dashed line), as well as the fits performed by \citeauthor{Burkert2008} to the major merger simulations in each panel (see the legends). As reported by these authors, the elliptical remnants of major merger simulations reproduce the linear trend found for real ellipticals pretty well, independently of the initial conditions of the encounter, the recipes used to simulate dissipative processes, the insertion of black-hole physics, and the consideration of a cosmological framework.

Some of our minor merger remnants (those with an S0b primary progenitor, blue stars) overlap with the distribution of major merger simulations in these two panels of Fig.\,\ref{fig:anisotropy}, extending the linear trend followed by real and simulated ellipticals towards higher \epse\ values. However, the remnants of our minor mergers with S0c progenitors (green stars) lie above this trend, contributing to increase the dispersion of the simulated S0s in the diagram. This means that a larger sample of minor merger experiments (considering different initial conditions) will probably disperse the resulting S0 remnants in the $\delta$ -- $\epsilon_\mathrm{e,intr}$ diagram, reproducing the spread distribution of real S0s observed in the first panel of  Fig.\,\ref{fig:anisotropy}.

To summarize, we find that if a representative sample of S0s is considered, ellipticals and S0 galaxies behave differently in the $\delta$ -- $\epsilon_\mathrm{e,intr}$ diagram. Ellipticals tend to follow a well-defined linear relation in this plane (which is accurately reproduced by major merger simulations), whereas S0s spread into a wide region at $0.4<\epsilon_\mathrm{e,intr}<0.9$. Although the remnants of our minor merger models with S0b progenitors extend this linear trend towards higher $\epsilon_\mathrm{e,intr}$ values, those with S0c progenitors increase the dispersion in the diagram. This fact makes minor mergers good candidates to generate the sparse distribution exhibited by real S0s in this diagram. 

\subsection{Rotational support and flattening of the final bulges}
\label{sec:bulges}

The $\lambda_\mathrm{e}$ parameter is derived considering galaxies as single-component systems. Studies conducted before \citet{Emsellem2007} used the classical anisotropy diagram to analyse the dynamical state of the spheroidal components of galaxies \citep[i.e., the bulges of disc galaxies and elliptical galaxies, see, e.g.,][]{Kormendy1982,Kormendy1983,Kormendy1993}. The $V/\sigma$ measurements there were thus estimated from the data within the effective radius of the spheroidal component, meaning that their results refer to a galaxy sub-component in the cases of disc galaxies (the bulges), and not to the galaxy as a whole (as in the procedure followed in the studies by \citeauthor{Emsellem2007}).  Therefore, the results and conclusions derived for the bulges do not apply to their host galaxies as single-component systems in general. Obviously, the case of elliptical galaxies is an exception, because the whole galaxy is basically the spheroidal component \citep[see the discussion in Sect.\,2 of][]{Kormendy2005}. 

Column 2 of Table\,\ref{tab:kinematics} emphasizes the difference between the half-light radius of our S0 remnants and the effective radius of their central bulges. The galaxy region considered for computing the values of $\epse$, $(V/\sigma)_\mathrm{e}$, and $\lambda_\mathrm{e}$ in this Table following the procedure by \citeauthor{Emsellem2007} (i.e., considering the galaxy as a single component) is $\sim 4$-10 times larger than the characteristic scale-length of the bulge component, which is the region analysed in studies of Kormendy~and~collaborators for disc galaxies. 

Consequently, we also studied the intrinsic structural and kinematic properties of the bulges in our remnants using the traditional anisotropy diagram, mimicking the observational techniques used in classical studies (based on 1D spectroscopy). This has allowed us to compare the shape and rotational support of the bulges that result from our minor merger remnants with published data on real bulges and elliptical galaxies, as well as on elliptical remnants that result from major merger simulations. The results are shown in this section.

\subsubsection{Shape of the remnant bulges}
\label{sec:morphology}

We derived the bulge effective radius $r_\mathrm{eff,bulge}$ from Sersic$+$exponential disc decomposition of the azimuthally averaged surface density profiles of all the stars in the remnants, using face-on views \citep[][]{Eliche-Moral2012}. We determined the 3D shape of each bulge remnant following the procedure described by \citet[C06 hereafter]{Cox2006}, where the axis ratios of the bulge are computed by diagonalization of the inertia tensor of all stars within $r_\mathrm{eff,bulge}$ in the galaxy. The semi-axes then correspond to the square root of the obtained eigenvalues. The major and minor axial ratios are defined as $b =  B/A$ and $c=C/A$ respectively, where $A$, $B$, and $C$ are the semi-axes in the directions of the eigenvectors in descending order ($A>B>C$). The resulting axial ratios for each model are listed in Table\,\ref{tab:bulges}. 

In the first panel of Fig.~\ref{fig:axialratios}, we show the minor versus major axial ratios of the remnant bulges compared with the elliptical remnants obtained by C06 from 1:1 major merger simulations. All our remnants have $0.9 \leq b \sim c \leq 1.0$, indicating that our remnant bulges continued to be nearly spherical (as initially), in contrast with the wide range of shapes exhibited by the ellipticals resulting in the major merger simulations of C06 \citep[see also Fig. 12 in][]{Gonzalez-Garcia2005}.

In the second panel of Fig.~\ref{fig:axialratios}, we compare the structure of our remnant bulges with those of real bulges. In this panel, we show the major and minor axial ratios of our remnant bulges over-plotted on a sample of 115 bulges of S0-Sb galaxies from \citet{Mendez-Abreu2010}. These authors used a different criterion from that used by \citet{Cox2006} to determine the axial ratios, with semi-axes $A$ and $B$ contained within the equatorial plane of the galaxy ($A>B$) and the $C$ semi-axis perpendicular to it. The axial ratios according to this new criterion were also estimated in our remnants (noted as $\hat{b}=B/A$ and $\hat{c}=C/A$). The results are also indicated in Table\,\ref{tab:bulges}. The figure shows that our remnant bulges exhibit $\hat{b}$ and $\hat{c}$ $\sim$ 1.0, meaning that the semi-axis in the polar direction of the galaxy ($C$) barely changes with respect to the highest semi-axis in the galaxy equatorial plane ($A$) after the merger. However, the lowest semi-axis in the equatorial plane ($B$) always decreases. This evolution of the 3D structural semi-axes of the bulges traces the formation of weak central ovals in all cases. In EM11, we reported that all remnants developed them, as relics of transient bars induced by the encounters. Because the models lack gas (and thus star formation), the ovals are weak. But nevertheless they are very frequent sub-components of real S0 galaxies \citep{Laurikainen2005,Laurikainen2009,Laurikainen2013}. 

The most noticeable changes in the $B$ semi-axis correspond to the models with small original primary bulges (S0c) and to the models with original S0b galaxies that have long-pericentre retrograde orbits and the lowest mass ratios. The first result can be understood by considering that high central mass concentrations tend to stabilize discs, so we can expect stronger oval instabilities in the models with S0c primary galaxies than in those with S0b ones \citep[see references in][]{Eliche-Moral2012}. Concerning the other cases of noticeable ovals, they might be related to the fact that the bar formation time-scale decreases because the tidal forces in a merger are stronger \citep{Gerin1990}. This means that the bar formation is delayed in encounters with lower tidal forces, but not inhibited (see the previous reference). Therefore, retrograde orbits, lower mass ratios, and longer orbital pericentres tend to destabilize the discs later during the interaction than other cases  (because the tidal response is lower), which makes the oval instabilities more noticeable at advanced stages of the simulations. This might explain why the original S0b galaxies that accrete a satellite in retrograde, long-pericentre orbits develop more elongated bulges than in other cases at the final stage.
 
We estimated the triaxialities of the remnant bulges, defined as \citep{Franx1991,Mendez-Abreu2010}

\begin{equation}\label{eq:T}
 T = \frac{1 - b^2}{1 - c^2},
\end{equation}

\noindent where $b$ and $c$ are the major and minor axial ratios defined according to the \citet{Cox2006} criterion ($A>B>C$). The resulting triaxialities are listed in Table\,\ref{tab:bulges}.

The triaxialities for our remnants are plotted versus the major and minor axial ratios of the bulges ($b$ and $c$) in the right panels of Fig.~\ref{fig:axialratios}. It is remarkable that even though our remnant bulges are almost spherical, they exhibit a wide range of triaxialities ($0.20<T<1.00$). This is due to the definition of this shape parameter, which makes it too sensitive in nearly spherical systems, where weak variations in one of the semi-axes can result in very different $T$ values. This should be taken into account when this parameter is used as an intrinsic shape indicator in nearly spherical bulges. 

Attending to the values of $T$, our remnant bulges are triaxial, mostly prolate systems ($T\sim 1$). Four models are more oblate ($T\sim0$), corresponding to the ones with the most noticeable oval distortions at the centre (see above). This strong triaxiality agrees with the structure of elliptical remnants resulting from major-to-intermediate merger experiments in the two right panels of Fig.~\ref{fig:axialratios}, in the sense that they tend to be quite triaxial \citep{Cox2006,Jesseit2009}.

\subsubsection{Rotational support of the remnant bulges}
\label{sec:rotation}

Traditionally, studies based on 1D spectroscopic data used $V_\mathrm{max}$ and $\langle\sigma\rangle$ as surrogates of the $V$ and $\sigma$ defined by \citet{Binney1978}. Moreover, these studies centered on edge-on disc galaxies, to be capable of locating the slits in a bulge region free of disc contamination \citep[see, e.g.,][]{Kormendy1982,Kormendy1983,Kormendy1993}. \citet{Binney2005} demonstrated that these observational approximations worked well, and that the location of bulges and ellipticals in the observational anisotropy diagrams provided a realistic estimate of their $\delta$ values, and thus, of their dynamical state. 

Therefore, we computed $\epsilon$ and $V/\sigma$ for our remnant bulges by reproducing the observational procedure used by \citet{Kormendy1983}, considering all stars within $r= r_\mathrm{eff,bulge}$ in an edge-on view of each galaxy. We estimated the intrinsic rotational support of the original and remnant bulges in our experiments by estimating $V_\mathrm{rot,max}/\langle \sigma\rangle$ as a function of their intrinsic average ellipticities ($\langle \epsilon\rangle$), where $V_\mathrm{rot,max}$ and $\langle \sigma\rangle$ represent the maximum rotational velocity and the mean central velocity dispersion within one effective radius of the bulge, and $\langle \epsilon\rangle$ corresponds to the average ellipticity of the isophotes within this region. The values of $V_\mathrm{rot,max}$ and $\langle\sigma\rangle$ were derived from a set of slits placed parallel to the galaxy plane in an edge-on view, located at different heights over this plane since twice the median vertical scale-length of the disc up to $z=r_\mathrm{eff,bulge}$ to avoid disc contamination. The different values of $V_\mathrm{rot,max}$ obtained for different heights over the disc are extrapolated towards $z=0$, assuming that they linearly increase towards the centre, as was also done by \citet{Kormendy1982}.

The effects of the numerical thickening in the estimated values of $\langle \epsilon\rangle$ and $V_\mathrm{rot,max}/\langle \sigma\rangle$ were quantified using new $N$-body models that reproduce some of our experiments with a factor $\times 3$ and $\times 10$ more particles \citep[][Tapia et al., in preparation]{Tapia2010a,Tapia2010b}. We find that models with 10 times more particles have lower $V_\mathrm{rot,max}/\langle \sigma\rangle$ values by up to $\sim 20$\% on average. The ellipticities can be affected by up to $\sim 25$\%, although no trend is observed with the number of particles of the simulation. Therefore, we applied an average correction to our measurements, decreasing all $V_\mathrm{rot,max}/\langle \sigma\rangle$ values by 20\% and quadratically adding the previous uncertainties to the errors of $\epsilon$ and $V/\sigma$. Final values of $\langle \epsilon\rangle$ and $V_\mathrm{rot,max}/\langle \sigma\rangle$ for our remnant bulges corrected for numerical thickening are listed in Table\,\ref{tab:bulges}. 

We plot the location of our remnant bulges in the $V_\mathrm{rot,max}/\langle \sigma\rangle$ -- $\langle \epsilon\rangle$ plane in Fig.~\ref{fig:vsigmak}. Data of real bulges and ellipticals obtained according to the observational technique described above are plotted for comparison. The bulges of the original S0 galaxies have low intrinsic rotational support by construction. The original S0b bulge has low intrinsic ellipticity, whereas the initial S0c bulge is intrinsically flattened because it is too deeply embedded in the primary disc. 

The figure shows that our minor and intermediate dry mergers induce an increment in the rotational support of the remnant bulges by up to a factor of $\sim 4$ in the models with primary S0b galaxies, with final $V_\mathrm{rot,max}/\langle \sigma\rangle$ values ranging from 0.3 to 0.9. This increase of rotational support of the bulge occurs because, although $V_\mathrm{rot,max}$ rises or falls depending on the model (by up to a factor of $\sim 2$), $\sigma$ in the bulge always decreases after the merger in an even more significant amount. Nevertheless, $V_\mathrm{rot,max}/\langle \sigma\rangle$ remains nearly constant after the merger in the models with initial S0c galaxies. The reason is that both quantities increase by a similar fraction in these models. 

We studied the contribution to the final stellar rotation maps of the stars that originally belonged to different galaxy components in EM11. There, we showed that the satellite stars (from its bulge and disc) formed central dynamically cold inner components (discs and rings) in the S0b models, with scale-heights quite similar to those exhibited by the primary disc (see Fig.~3 in EM11). Moreover, some primary disc stars are injected towards the centre. As this component is heated during the encounter (and thus, thickens), this rotating material contributes to the slits placed above the main disc, decreasing the mean velocity dispersion of the bulge in these models (and thus increasing its rotational support, as provided by $V/\sigma$).

In the models with original S0c galaxies, the bulge of the accreted satellite sinks towards the centre of the remnant without being disrupted (see again Fig.~3 in EM11). Even though there is also an inflow of rotating material from the primary disc and satellite disc stars in these models (in fact, $V_\mathrm{rot,max}$ increases in both remnants compared to its initial value), the undisrupted satellite core increases the velocity dispersion in the bulge as to keeps $V_\mathrm{rot,max}/\langle \sigma\rangle$ nearly constant.   

Fig.~\ref{fig:vsigmak} also shows that the ellipticity in the models with primary S0b galaxies remains low after the merger ($\langle\epsilon\rangle\lesssim 0.3$), while it decreases noticeably in the mergers with original S0c galaxies. The fact that the spheroidal bulge of an (initially) S0b galaxy does not experience a significant change of shape after accreting a satellite might be expected considering the lack of dissipative effects in our simulations and the massive original bulge \citep[to understand the role of gas in the structure of major merger remnants, see][]{Jesseit2007}. In contrast, the bulge of the accreted satellite sinks towards the centre of the remnant without being disrupted in the S0c models (see EM11). This additional spheroidal sub-component deposited in the galaxy centre decreases the flattening of the bulge region in the experiments with an original S0c. 

The simulated mergers thus tend to move the original bulges in Fig.~\ref{fig:vsigmak} towards the theoretical line of the rotationally flattened oblate spheroids with isotropic velocity dispersions ($\delta =0$), around which most real bulges accumulate. Some remnant bulges are even above this line. This means that dry minor and intermediate mergers onto S0s tend to decrease the anisotropy of the bulges, which causes their flattening agree better with their rotational support. We refer here to the dynamical state of the bulge subcomponent, whereas in Sect.\,\ref{sec:fastslow} we studied the whole galaxy considering it as a single component.

Summarizing, dry minor mergers onto S0b galaxies can increase the intrinsic rotational support of the bulges noticeably, through the transport of orbital angular momentum towards the inner parts, which affects their original shape only negligibly. In contrast, minor mergers onto initial S0c galaxies do not modify the rotational support of the bulges significantly, but may decrease their intrinsic ellipticities. These two evolutionary mechanisms contribute to decrease the velocities anisotropy of the central bulge, which means that the rotational support of the bulge is enough to explain its flattening. These results apply to primary galaxies that originally have spherical bulges. The effects of using non-axisymmetric initial bulges will be analysed in a forthcoming paper (Tapia et al., in preparation).

\begin{figure*}[th]
\centering
   \includegraphics*[width=6.0cm]{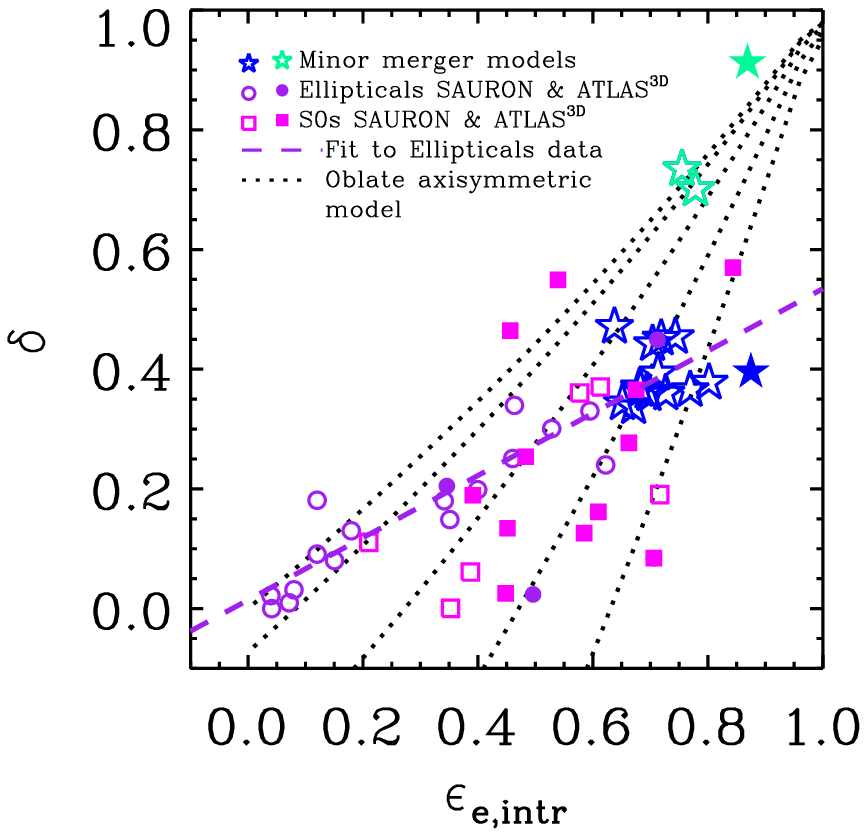}
   \includegraphics*[width=6.0cm]{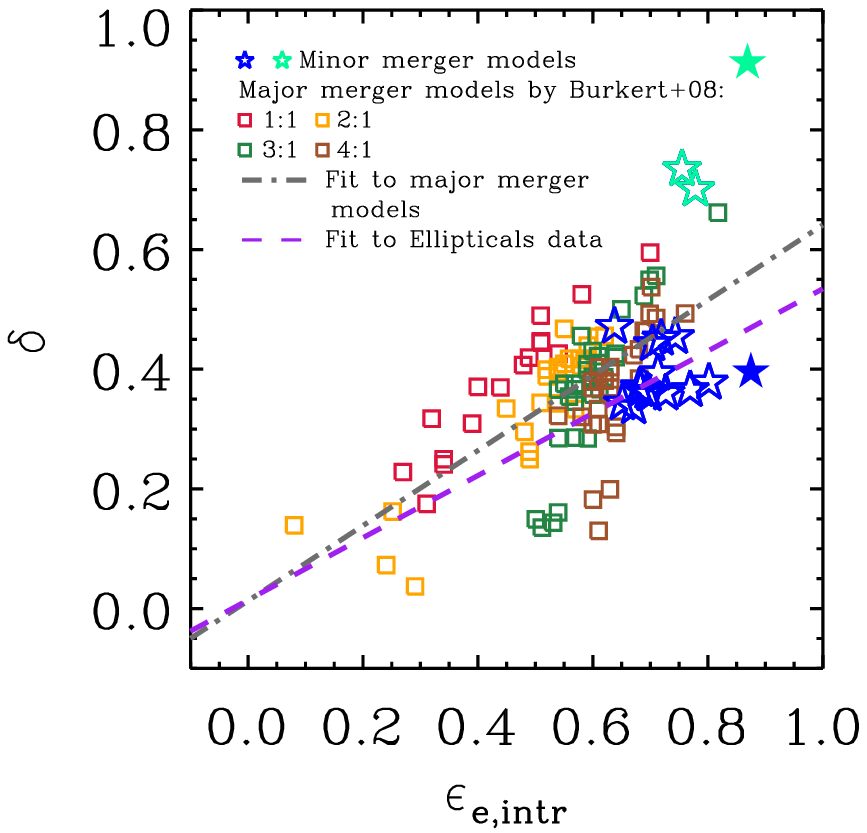}
   \includegraphics*[width=6.0cm]{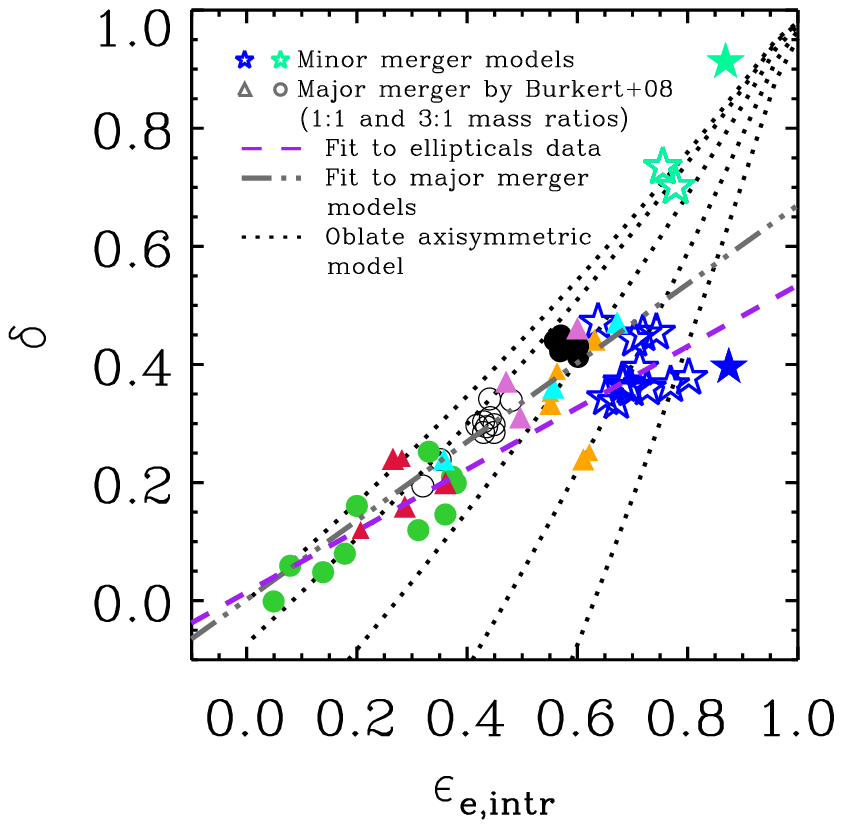}
  \caption{Anisotropy of velocities of our minor merger remnants ($\delta$) as a function of the intrinsic ellipticity at $R=R_\mathrm{eff,glx}$ ($\epsilon_\mathrm{intr}$). Stars represent the location of our models (the original primary S0b and S0c galaxies are the filled blue and green, and their remnants are the empty blue and green stars, respectively). \emph{Left panel}: comparison with the distribution of real ellipticals (purple empty and filled circles) and S0s (magenta empty and filled squares) from the SAURON and ATLAS$^\mathrm{3D}$ samples \citep{Burkert2008,Cappellari2013}. \emph{Middle panel}: comparison with to the elliptical remnants that result in major merger simulations by \citet{Burkert2008}, as a function of their mass ratios (red, orange, purple, and magenta squares for 1:1, 2:1, 3:1, and 4:1, respectively). \emph{Right panel}: comparison with the ellipticals remnants that result from the 1:1 and 3:1 major merger simulations by \citet{Burkert2008}, considering different gas ratios, star formation, and feedback procedures. The symbols and colors used here for the \citet{Burkert2008} models are the same as used in their Fig.~3 (except for their blue triangles, which are plotted here in orange to avoid confusion with our models). For more information, we refer to that paper. \emph{Lines}: different linear fits performed to the real and simulated ellipticals (consult the legend in each panel). In the left and right panels, we have overplotted the theoretical $\delta$ -- $\epsilon_\mathrm{e,intr}$ relations expected for oblate axisymmetric systems with constant values of $V/\sigma=$ 0, 0.25, 0.5, 0.75, and 1.0 \citep[see eqs.\,\ref{eq:deltaBinney} -- \ref{eq:omega} from][]{Binney2005}.} \label{fig:anisotropy}

\end{figure*}

\subsection{Relation between the bulge triaxiality and the global rotational support in the remnants}
\label{sec:relation}
In Sect.\,\ref{sec:morphology}, we have reported that the bulges of our remnants show a great diversity of triaxial shapes and that this parameter is very sensitive to small differences in the semi-axes of nearly spherical systems. However, because this parameter has been widely used as intrinsic shape indicator \citep[e.g.][]{Franx1991,Tremblay1996,Alam2002,Mendez-Abreu2010}, we analysed whether the global rotational support of the remnants is related to the shape of their central bulges as provided by $T$ or not.
In Fig.~\ref{fig:tlambda}, we show the relation between the bulge triaxiality $T$ (see Sect.~\,\ref{sec:morphology}) and the global rotational support of the galaxy (as provided by $\lambda_\mathrm{e}$, see Sect.\,\ref{sec:fastslow}) for our minor merger remnants. We also represent the location of the elliptical remnants that result from the major merger simulations by \citet{Jesseit2009} in this figure. We compared the $\lambda_\mathrm{e}$  and $T$ values of the spheroidal components that result from a merger. Noted that $T$ is estimated from the stellar material within $r=r_\mathrm{eff,bulge}$, whereas $\lambda_\mathrm{e}$ is computed for all stars within $r=R_\mathrm{eff,glx}$ in all the plotted cases. 
The region over which $\lambda_\mathrm{e}$ is estimated is $\sim 4$-10 times larger than the effective radius of the bulge in our S0 remnants, as also occurs in real S0 galaxies (see Table\,\ref{tab:kinematics}). In the case of the elliptical remnants studied by Jesseit~and~collaborators, both parameters are derived from the \emph{same} stellar material (because the spheroidal component is the whole galaxy).
The figure indicates that in both sets of simulations, oblate central spheroidal structures (i.e., with low $T$) tend to be hosted by galaxies with a higher global rotational support (i.e., higher $\lambda_\mathrm{e}$ values). This is obvious in the case of ellipticals, because $T$ and $\lambda_\mathrm{e}$ are estimated for the same stellar material. However, it is striking in the case of our S0 remnants, because both parameters are computed from different galaxy regions. The figure shows that, because the S0 remnant exhibits a higher global rotational support, its bulge becomes more oblate, too. This might be because the encounters transfer part of the rotational support to the inner regions (see EM06; EM11), which contributes to the flattening of the material at the galaxy centre. Higher global rotational support thus implies stronger flattening of the bulge, which in turn implies a more oblate bulge. 
The trends of the two simulation sets are also offset in the $\lambda_\mathrm{e}$ -- $T$ diagram. For a similar rotational support of the whole galaxy, our remnant bulges exhibit higher $T$ values than the ellipticals that result in major mergers. This means that the bulk of the rotational support indicated by $\lambda_\mathrm{e}$ in our S0 remnants mostly comes from the disc contribution within one half-light radius. This amount of rotation does not contribute to shape the central bulge \emph{directly}, but it is \emph{indirectly} related to its structure, because high global rotational support in the galaxy body must imply high rotational support in the central region as well. 
In conclusion, the 3D structure of the bulge that results from our minor merger experiments is indirectly related to the global rotational support of the whole remnant (as given by $\lambda_\mathrm{e}$ parameter).
\section{Model limitations}
\label{sec:limitations}
We have analysed S0 primary galaxies with spherical bulges. Real bulges in ETGs tend to be triaxial \citep{MendezAbreu2012}. It might be assumed that this initial condition biases the remnants towards low apparent $\epse$ values in the cases of progenitors with massive (i.e., luminous) bulges, because central mass concentrations tend to make the discs more stable against distortions (see references in EM11). But, instead, we have found that all remnants tend to exhibit higher percentages of intermediate \epse\ values at random projections than initially (Sect.\,\ref{sec:compare}). According to the definition in eq.\,\ref{eq:epsilon}, \epse\  ``departs from a simple luminosity-weighted average of the ellipticity profile, which is more strongly biased towards the central values'' \citep{Cappellari2011}. Considering that the characteristic radial scales of the remnant bulges are $\sim 4$-10 times lower than those used to estimate $\epse$ in each remnant (see Col.\,2 in Table\,\ref{tab:kinematics}), we expect the results to not be biased because of the initial shape of the bulges.  In fact, we have shown that the structure of the bulge is not directly related to the global rotational support of the galaxy (see Sect.\,\ref{sec:relation}), therefore we do not expect the shape of the original primary bulges to be necessarily determining the intrinsic values of $\epse$ and $\lambda_\mathrm{e}$ of the remnants. The effect of a non-spherical original bulge on the outcome of the mergers needs to be studied in detail.
 \begin{figure*}[ht]
\centering
   \includegraphics[width=4.5cm]{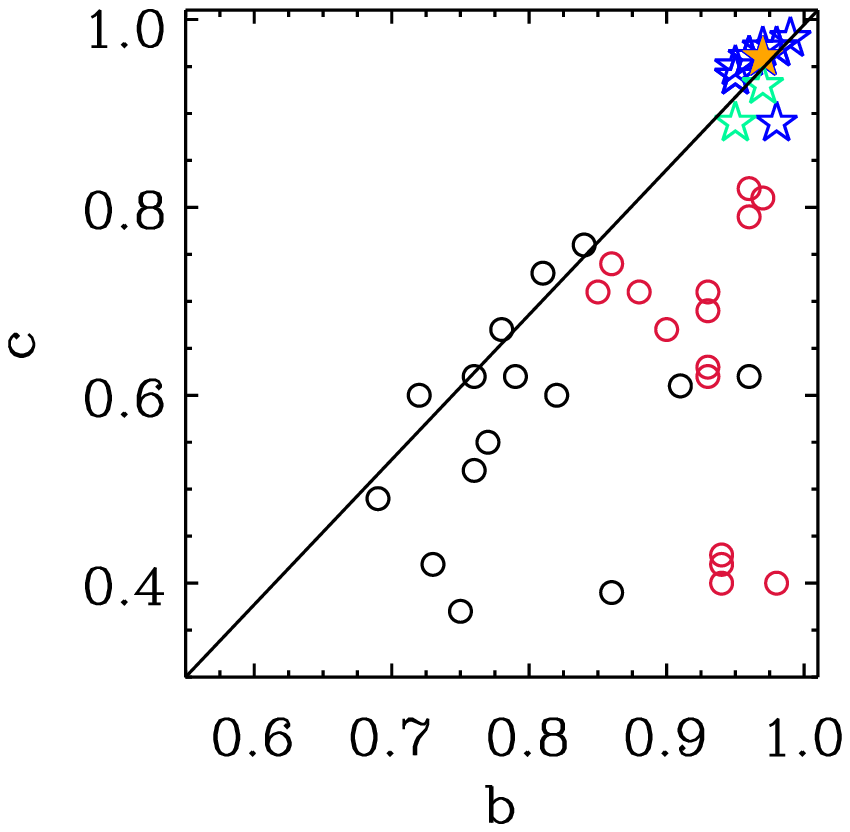}
   \includegraphics[width=4.5cm]{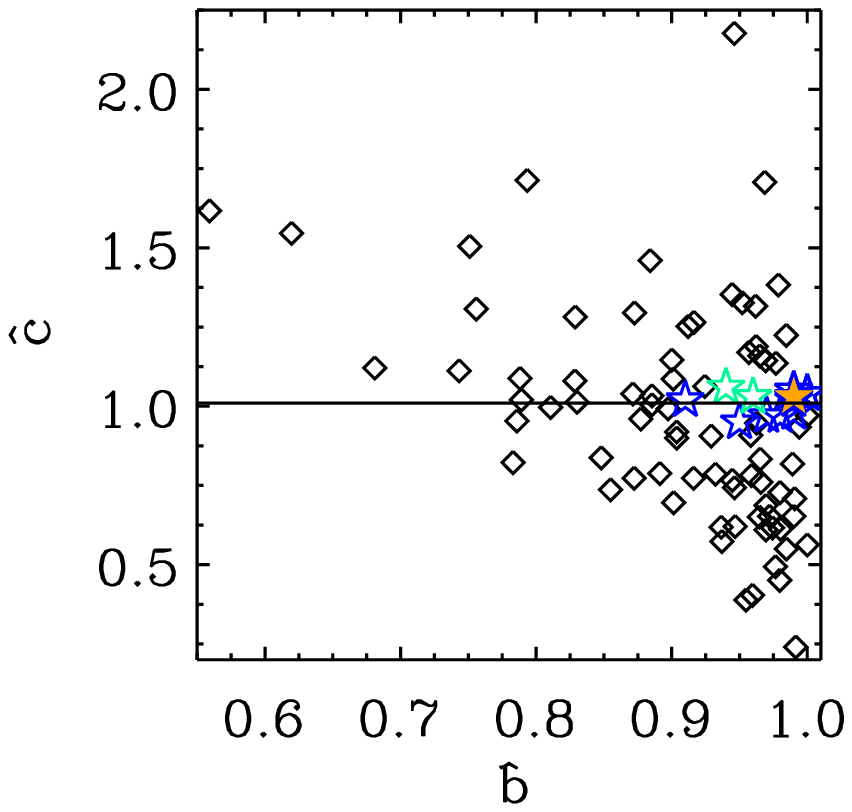}
   \includegraphics[width=4.5cm]{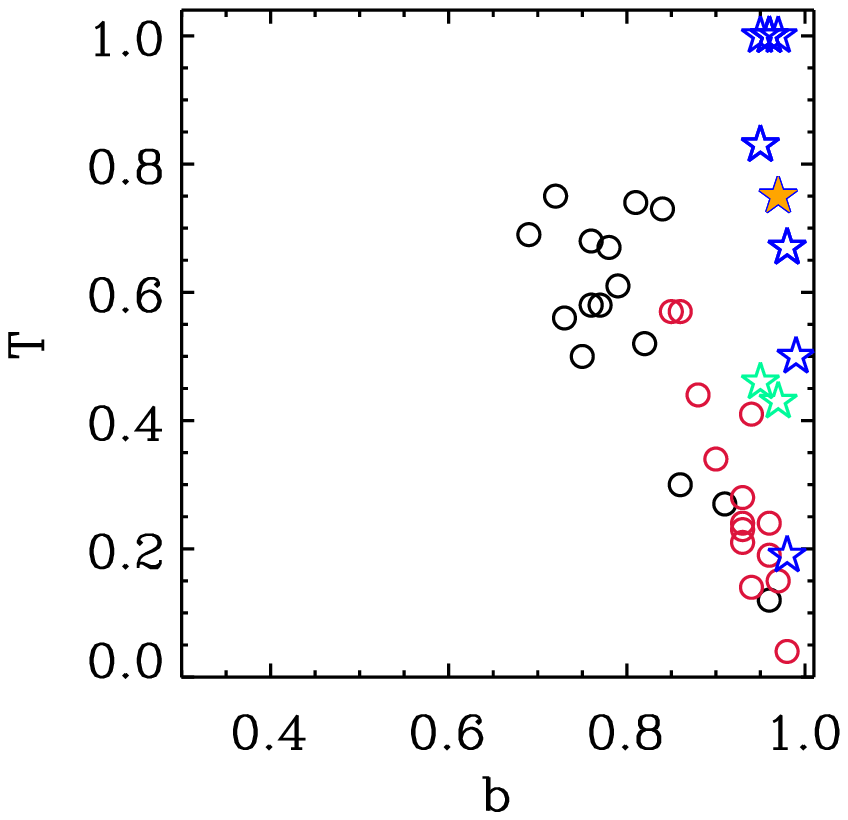}
   \includegraphics[width=4.5cm]{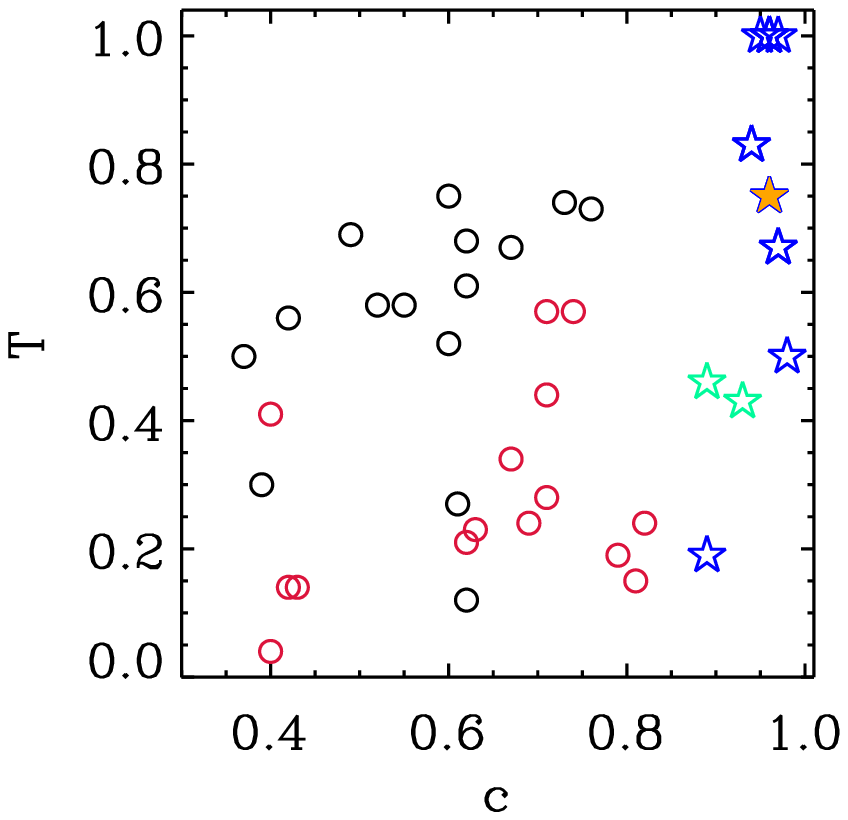}
      \caption{Shape diagrams of our remnant bulges, computed considering all stars at $r \leq r_{\mathrm{eff,bulge}}$ in the remnants. The blue and green stars in all panels represent the models with a primary S0b and S0c galaxy, respectively. The location of the original primary bulges is shown with orange stars. \emph{First panel}: minor versus major axial ratios of our remnant bulges computed according to the \citet{Cox2006} criterion ($A>B>C$), compared with their simulated remnants of equal-mass major mergers of galaxies with pure exponential stellar discs (black circles) and containing 40\% of gas (red circles). The line represents the location of a prolate spheroid with $T=1$. \emph{Second panel}: axial ratios of our remnant bulges computed according to the \citet{Mendez-Abreu2010} criterion ($A>B$ lying in the equatorial plane, and $C$ perpendicular to it), compared with the observed distribution of their sample of 115 S0--Sb bulges (black diamonds). The straight line marks $\hat{c}=1$. \emph{Third and fourth panels}: triaxiality as a function of the major and minor axial ratios of our remnant bulges, compared with the major merger experiments of \citet{Cox2006}. The legend is the same as in the left panel.}
 \label{fig:axialratios}
 \end{figure*}  
The fact that our remnants tend to exhibit a lower percentage of random projections with lower $\epse\ $ values than initially is an even more robust result if we also consider that no dissipative effects are included in the simulations. Many theoretical studies have shown that gas tends to destabilize discs, inducing the formation of non-axisymmetric distortions \citep{Barnes1996}, therefore we expect that the gas would make the stellar remnants even less axisymmetric at intermediate radial locations in the galaxy \citep[although this behaviour is not strictly general and depends on the gas content and feedback efficiency, see][]{Hopkins2009d}. 
S0s have low gas amounts in general \citep{Crocker2011,Saintongue2011}. A galaxy that becomes S0 type after a minor merger event must thus start with very low gas amount indeed \citep[see][]{Bournaud2007a,Moster2010}. Therefore, the gas amount that can be inserted in this type of simulations (binary minor mergers) must be low to obtain an S0 remnant, unless a sequence of minor merger events onto an initial gas-rich galaxy is investigated \citep[as those resulting in cosmological merging trees, see][]{Naab2013}. We can accordingly expect that the inclusion of (realistic) low gas contents in these simulations only weakly affect the final results concerning the global dynamical state of the whole galaxy. Gas would probably contribute to increase the global rotational support of the remnant in the central region, because it tends to accumulate in highly rotating co-planar orbits \citep[see][]{Cox2006,Cox2006a}, which modifies the structure in the central region. This effect could flatten the galaxy in the centre, in addition to inducing bar-like distortions at intermediate radial positions. The former effect might slightly increase the rotational support in the central region. But, as discussed above, the material within one effective radius \emph{of the bulge} does not necessarily noticeably determine or affect the intrinsic values of $\epse$ and $\lambda_\mathrm{e}$ \emph{in the galaxy as a whole}. Moreover, previous studies have also shown that gas tends to produce more round and axisymmetric remnants in major merger simulations as reported by \citet{Jesseit2007} and \citet{DiMatteo2008}, so gas does not necessarily flattens the resulting remnant. Additionally, these authors also demonstrated that the global structural properties of the remnants in dissipative models are nearly the same as those obtained in analog collisionless merger simulations. All these results indicate that the main conclusion of our study must be robust against the inclusion of gas in the simulations as well.
We also studied the effects of the disc numerical thickening in the location of our models in the  $\lambda_\mathrm{e}$ -- $\epse$ diagram. In Fig.~\,\ref{fig:thickening}, we investigate the dependence of the location of some of our models in the $\lambda_\mathrm{e}$ -- $\epse$ plane on the number of particles of the simulation. The models reproduce the initial conditions of models a, b, g, and h from Table~\,\ref{tab:kinematics} with a factor $\times 3$ and $\times 10$ more particles  \citep[][]{Tapia2010a,Tapia2010b}. We do not find significant changes or any trends with the number of particles, meaning that the number of particles used here is enough to describe the global dynamical state and evolution of these systems. The models with original S0c galaxies have twice as many  particles as the models with S0b primary galaxies, so the numerical resolution is even better in them. Therefore, the results are robust against an improved numerical resolution of the simulations.

Finally, our set of models is limited by the space of initial conditions considered for the mergers, especially in mass ratios and orbital configurations. However, the trends found are the same in all the simulated encounters. Thus, we consider that the main conclusions derived in this study can be generally applied.

\section{Discussion} \label{sec:discussion}
The studies by Kormendy and collaborators and by Emsellem and coworkers~have shown that the dynamical state of ETGs strongly differ from one system to another (more noticeably in S0s), contrary to the classical view of ETGs as pressure-supported objects. There is a general agreement about the merger-related origin of ellipticals \citep{Springel2005b,Naab2006,Naab2007,Duc2011}, as well as that the wide spread of structural and dynamical properties of S0s are evidence of different evolutionary paths for them \citep[see][]{Laurikainen2010,Cappellari2011,Erwin2012,Kormendy2012}. Recent studies tend to support secular evolutionary scenarios for S0s in which mergers play a secondary role \citep[][]{VanDenBosch1998,Bekki2002,Aragon2006,Buta2010,Silchenko2012,Laurikainen2010,Laurikainen2013}. But many S0s exhibit clear signs of having experienced recent mergers \citep{Falcon-Barroso2004,Chillingarian2011,Kim2012}. However, this does not imply that the current dynamical status of these galaxies has been set by these events either, because merger relics are also frequent in spirals  \citep[see][]{Haynes2000,Knapen2004,Mazzuca2006,Silchenko2006,Martinez-Delgado2010}. Therefore, the question of how relevant merging has been in determining the dynamical properties of the present-day S0 population remains unsettled.

The dynamical status observed in ellipticals strongly supports a major-merger formation scenario for these systems. \citet{Burkert2008} showed that major mergers force the elliptical remnants to obey a well-defined linear relation in the $\delta$ -- $\epsilon_\mathrm{e,intr}$ diagram for high ellipticities. A similar relation is fulfilled by the ellipticals with $\epsilon_\mathrm{e,intr} >0.3$ in the SAURON sample (see Fig.~\,\ref{fig:anisotropy}). Real ellipticals with $\epsilon_\mathrm{e,intr} <0.3$ exhibit a lower slope of this relation, which these authors also reproduced using a more complex cosmological context, in which the mass is accreted in subsequent mergers. The difference in slopes seems to be related to secondary aspects of the major mass accretion \citep[as the relevance of dissipative processes or the way the mass is accreted, see][]{Burkert2008}. But these results clearly indicate that the formation physics that underlies the $\delta$ -- $\epse$ linear relation obeyed by ellipticals is \emph{major} merging.
The SAURON sample used by Burkert~and~collaborators only contained six S0s. Three out of these six S0s seem to obey the same $\delta$ -- $\epsilon_\mathrm{e,intr}$ relation fitted to ellipticals, whereas the other three lie far below this relation (see the left panel of Fig.\,\ref{fig:anisotropy}). When the observational sample is extended using recent data from ATLAS$^\mathrm{3D}$ \citep[which basically contain S0s at $0.4<\epsilon_\mathrm{e,intr}<0.9$, see][]{Cappellari2013}, it is more evident that the majority of S0s are outliers of this linear correlation drawn by the ellipticals, although some S0s follow it. Should we then interpret this as a sign of the major merger-related evolution of these S0s? According to the fact that our minor merger simulations with S0b progenitors seem to obey a similar correlation (see the same figure), the answer should be ``not necessarily''. The remnants of major mergers give rise to this linear correlation in the $\delta$ -- $\epsilon_\mathrm{e,intr}$ diagram, but not all galaxies obeying it necessarily derive from major encounters , at least according to our simulations. Therefore, the widespread distribution of S0s in the $\delta$ -- $\epsilon_\mathrm{e,intr}$ plane shows that many of these galaxies may come from processes different from major mergers (as minor mergers). 

The evolutionary path proposed here (through minor mergers) requires an already gas-poor progenitor with both high intrinsic ellipticity and rotational support. Although it may be difficult to form these type of galaxies through major mergers (B11), there are many other processes that have contributed to the gas exhaustion in spirals with a minor effect on the shape and kinematics of the progenitor galaxy, such as simple fading, strangulation, and gas stripping \citep[see, e.g.,][]{Smith2010,Yagi2010}. Therefore, an evolutionary scenario for some S0s in which these processes combine with minor mergers is quite probable. In fact, gas stripping and tidal interactions and mergers are known to be taking place at the same time in many galaxies as they infall into the center of a cluster \citep[][]{Sofue1994,2012A&A...544A..99W,Vollmer2013}. Consequently, the dispersion of the S0s in the  $\delta$ -- $\epsilon_\mathrm{e,intr}$ diagram is compatible with an evolution induced by minor mergers.

The observational sample plotted in Fig.\,\ref{fig:anisotropy} is still small for a robust estimate of an upper limit to the percentage of present-day S0s that fulfill this $\delta$ -- $\epse$ linear correlation and therefore that might derive from a major merger. Moreover, errors in the $\delta$ values derived from the JAM models are very large and these estimates are based on the assumption of axisymmetry for the galaxy, which clearly does not apply for a large number of galaxies and affects the accuracy of the $\delta$ values \citep[see][]{Lablanche2012}. However, a first estimate would be that $\sim 44$\% of S0 galaxies lie within $1\sigma$ of this linear relation drawn by ellipticals and major merger remnants (8 out of a total of 18 S0s, see the first panel of the figure).  If this estimate were robust, it would mean that major merging may have been relevant for establishing the current dynamical state of $\sim 40$--50\% of present-day S0s \emph{at maximum}.

Although still uncertain, this estimate agrees quite well with the complex formation scenario for S0s that has surfaced during the past years. As commented in Sect.\,\ref{sec:introduction}, the different properties found for S0s in different environments probably indicate different formation mechanisms. Spirals in clusters seem to have turned into S0s through environmental-related processes during the last $\sim 7$-8\,Gyr \citep{Poggianti2001,Poggianti2009}, whereas this transformation seems to have been led by mergers and tidal interactions in groups during the same period \citep{Wilman2009,Bekki2011}. Considering that $\sim 50$\% of S0s reside in groups or in groups that are falling into a cluster \citep{Huchra1982,Berlind2006,Crook2007,Wilman2009}, the percentage of S0s that may have experienced major encounters is therefore similar to the percentage of S0s located onto the linear correlation in the $\delta$ -- $\epse$ diagram estimated above. However, the estimate above is not statistically significative and that the fact that an S0 galaxy follow this linear relation in the diagram does not necessarily imply that it was formed in a major merger. More robust diagnostics are required to reliably quantify how relevant merging has been for setting the dynamical structure of present-day S0s.

\begin{figure}[ht]
\centering
  \includegraphics[width=8cm]{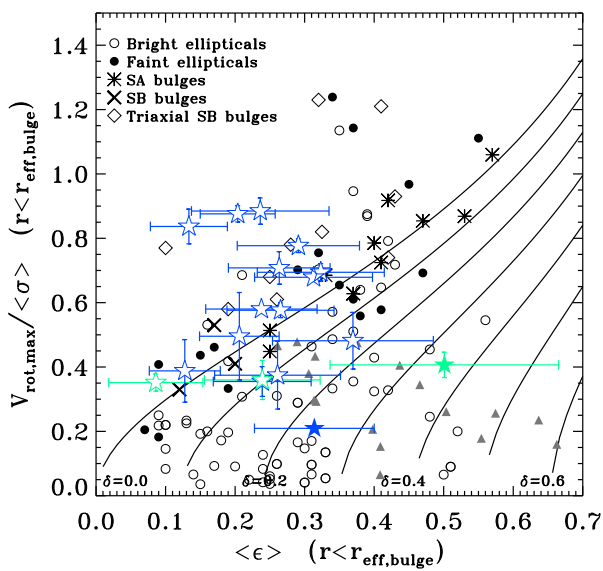}
  \caption{$V_\mathrm{rot,max}/\langle\sigma\rangle$ versus $\langle\epsilon\rangle$ at $r\leq r_\mathrm{eff}$ in our remnant bulges (\emph{blue and green empty stars} for the bulges of the remnants with S0b and S0c progenitor, respectively) and in the bulges of the original S0 galaxies (\emph{blue and green filled stars} for S0b and S0c), compared with observational data of bulges and elliptical galaxies \citep[][]{Kormendy1982,Davies1983,Davies1983a,Bender1988,Kormendy1993}. Data are labelled as shown in the legend. \emph{Solid lines}: theoretical relation of $V/\sigma$-$\epsilon$ for rotationally flattened oblate spheroids with different velocity anisotropies $\delta$ \citep[see][]{Binney2005}. \emph{Filled grey triangles}: location of the elliptical remnants of the binary mergers of disc galaxies with mass ratios from 1:1 to 3:1 by \citet{Gonzalez-Garcia2005}.}
\label{fig:vsigmak}
\end{figure}

\begin{figure}[th]
\centering
  \includegraphics[width=8cm]{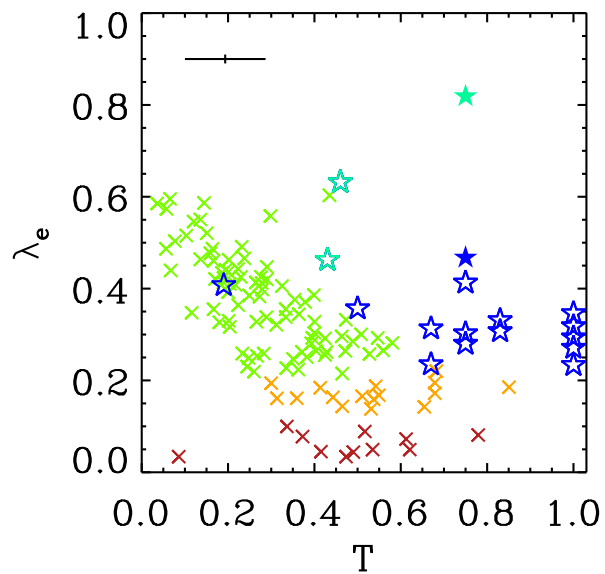}
  \caption{Relation between the bulge triaxiality and the $\lambda_\mathrm{e}$ kinematic parameter computed for the galaxy as a single component, in major and minor merger simulations. \emph{Stars}: for the remnants of our minor merger models with original S0b and S0c galaxies (empty blue and green ones, respectively) and our original S0 galaxies (filled blue and green ones). \emph{Crosses}: for the elliptical remnants of \citet{Jesseit2009} that are fast rotators in all projections (light green), that are fast rotators but that are identified as slow rotators in some projections (orange), and those that are slow rotators (red).  Median error bars in both axes are plotted in the upper left corner of the frame.} 
\label{fig:tlambda}
\end{figure}

Many studies claim that minor mergers have led the evolution of galaxies over the past $\sim 7$\,Gyr, especially of the most massive ones, whereas major mergers were more relevant at earlier epochs \citep[e.g.][]{Naab2009,Lotz2011,Oser2012,McLure2013}. Therefore, an evolutionary scenario in which major mergers at earlier epochs give rise to a population of initial S0s, that later minor mergers can evolve into different dynamical systems is quite plausible \citep[e.g.,\,the scenario proposed by][to explain the age and metallicity properties of S0 galaxies]{Silchenko2012}. A mixture of both types of mergers and of environmental and secular processes must have shaped galaxies in general (especially S0s) in a complex scenario \citep{Kannappan2009,Huertas-Company2010,Aguerri2012}. 

The relation between the possible evolutionary mechanisms of S0s and their role depending on the environment and the mass are still debated. Numerical simulations of minor mergers show that these processes typically induce secular evolution in the progenitors, even in the absence of strong bars and dissipative processes \citep[EM06;][]{Eliche-Moral2012,Eliche-Moral2013}. Environmental processes (such as strangulation) can also be efficient in groups, not just in clusters, as is commonly considered \citep[see][]{Kawata2008}. This means that it must be difficult to quantify the effects of each mechanism separately. Additionally, recent observations have reported the existence of some S0s that clearly derive from major merger events in which the disc-rebuilding has been successful, so this mechanism cannot be ignored even though the properties of present-day S0s seem too relaxed and smooth to have been formed through a mechanism as violent as a major merger \citep[][]{Peirani2009,Yang2009,Hammer2009,Hammer2009b,Hammer2010}. In fact, many studies find that major mergers may have played a key role in the definitive build-up of the bulk of the massive ETG population (both ellipticals and S0s) in low-to-intermediate density environments at $0.6<z<1$, in agreement with standard hierarchical models of galaxy formation \citep[see][]{Eliche-Moral2010,Eliche-Moral2010b,Bernardi2011,Bernardi2011b,Prieto2013}. This is why it is so relevant to develop methods for quantifying the cumulative effects of mergers in the settlement of the dynamical status of current galaxies (and in particular, of S0s). Fortunately, some studies have started to directly face this question \citep[see, e.g.,][]{Gu2013,Privon2013}, but much more work needs to be done in this direction.

\section{Conclusions} 
\label{sec:conclusions}
We have investigated whether minor mergers can explain the existence of S0s with kinematic properties intermediate between fast and slow rotators which major merger and cosmological simulations find difficult to reproduce. We analysed the properties of the remnants that result from dry mergers with mass ratios ranging 6:1 -- 18:1 onto original S0s that initially are fast rotators with high intrinsic ellipticities (they might in turn derive from gas stripping).  

We found that the minor mergers decrease the intrinsic ellipticity of the whole galaxy, but the remnants do not exhibit a higher percentage of random projections with low \epse\ values. Instead, they increase the fraction of projections with intermediate apparent ellipticities ($0.4<\epse<0.7$) due to the formation of non-axisymmetric distortions and to disc thickening. This means that the remnants become more triaxial. Minor mergers also induce a lower decrease of the rotational support  in the remnants than major mergers. These combined effects produce S0 remnants that extend over the limiting region between the distributions of fast and slow rotators in the \lambdae\ -- \epse\ diagram, spanning the whole range of apparent ellipticities up to $\epse \sim 0.8$.  Therefore, minor mergers are a plausible mechanism to generate S0s with hybrid kinematics and shape properties.

Considering the intrinsic properties of the remnant bulges, we find that the simulated mergers tend to decrease the velocity anisotropy of this sub-component (increasing the rotational support of the bulge or decreasing its intrinsic ellipticity). The remnant bulges remain nearly spherical, but exhibit a wide range of triaxialities ($0.20<T<1.0$). In addition, we showed that the triaxiality of the bulge is only indirectly related with the global rotational support of the whole remnant. 

In the plane of global anisotropy of velocities ($\delta$) vs.\,intrinsic ellipticity ($\epsilon_\mathrm{e,intr}$), some of our models extend the linear trend found in previous major merger simulations towards higher $\epsilon_\mathrm{e,intr}$ values, while others depart from it, depending on the progenitor. This contributes to increase the dispersion in the diagram. We compared these trends with those exhibited by elliptical and S0 galaxies from the SAURON and ATLAS$^\mathrm{3D}$ projects. While most real ellipticals closely follow the linear trend (consistent with a major merger origin, as already known), S0s are widepreadly in the $\delta$ -- $\epsilon_\mathrm{e,intr}$ diagram. In fact, less than $\sim 40$--50\% of real S0s are located within 1$\sigma$ of the linear trend drawn by major merger simulations.  Our simulations show that minor mergers can explain the dispersion exhibited by real S0s in this diagram. Therefore, the different trends exhibited by ellipticals and S0 galaxies in the $\delta$ -- $\epsilon_\mathrm{e}$ diagram probably point to the different role played by major mergers in the build-up of each galaxy type.

\begin{figure*}[th]
\centering
  \includegraphics*[width=4.5cm]{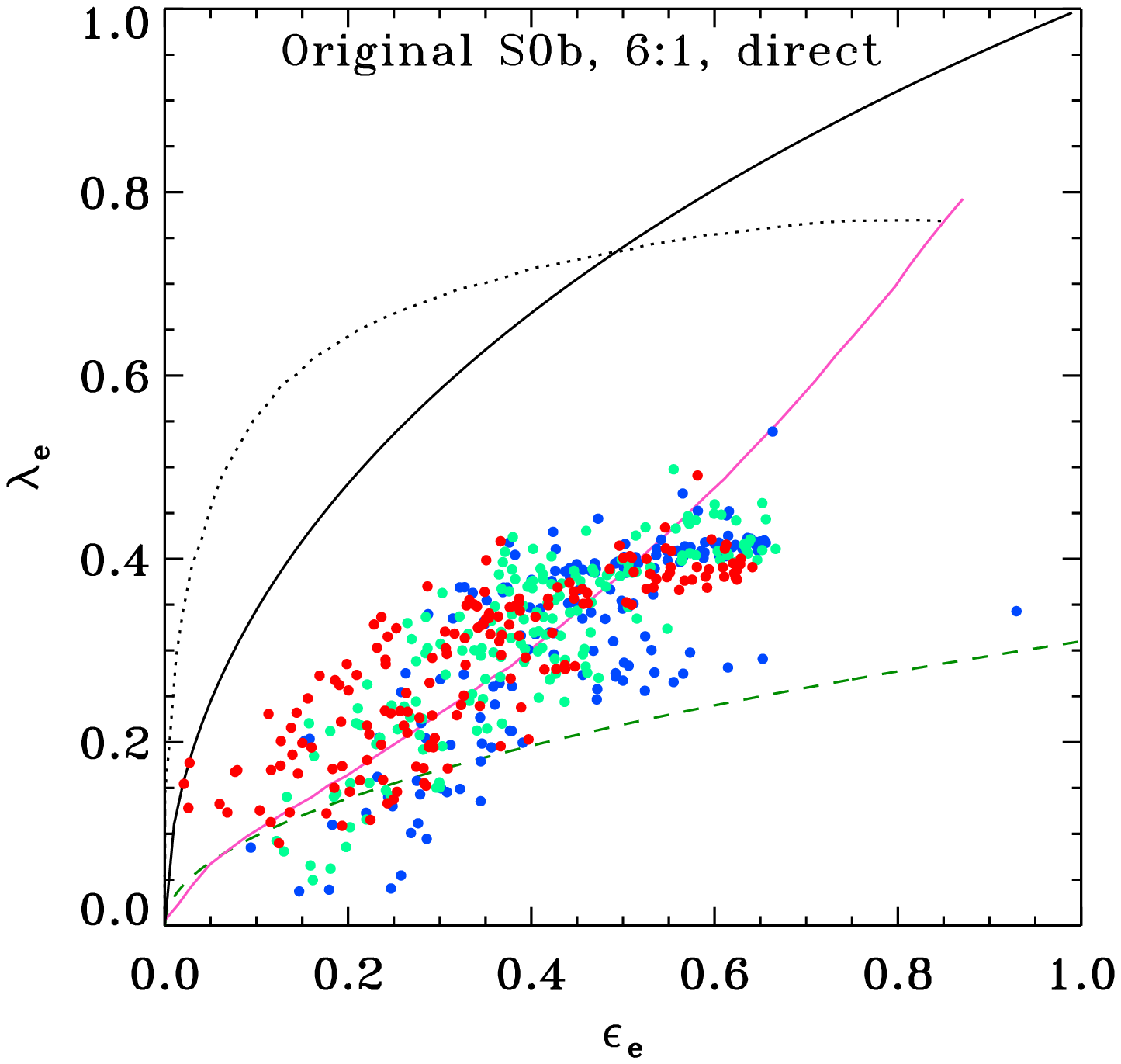}
  \includegraphics*[width=4.5cm]{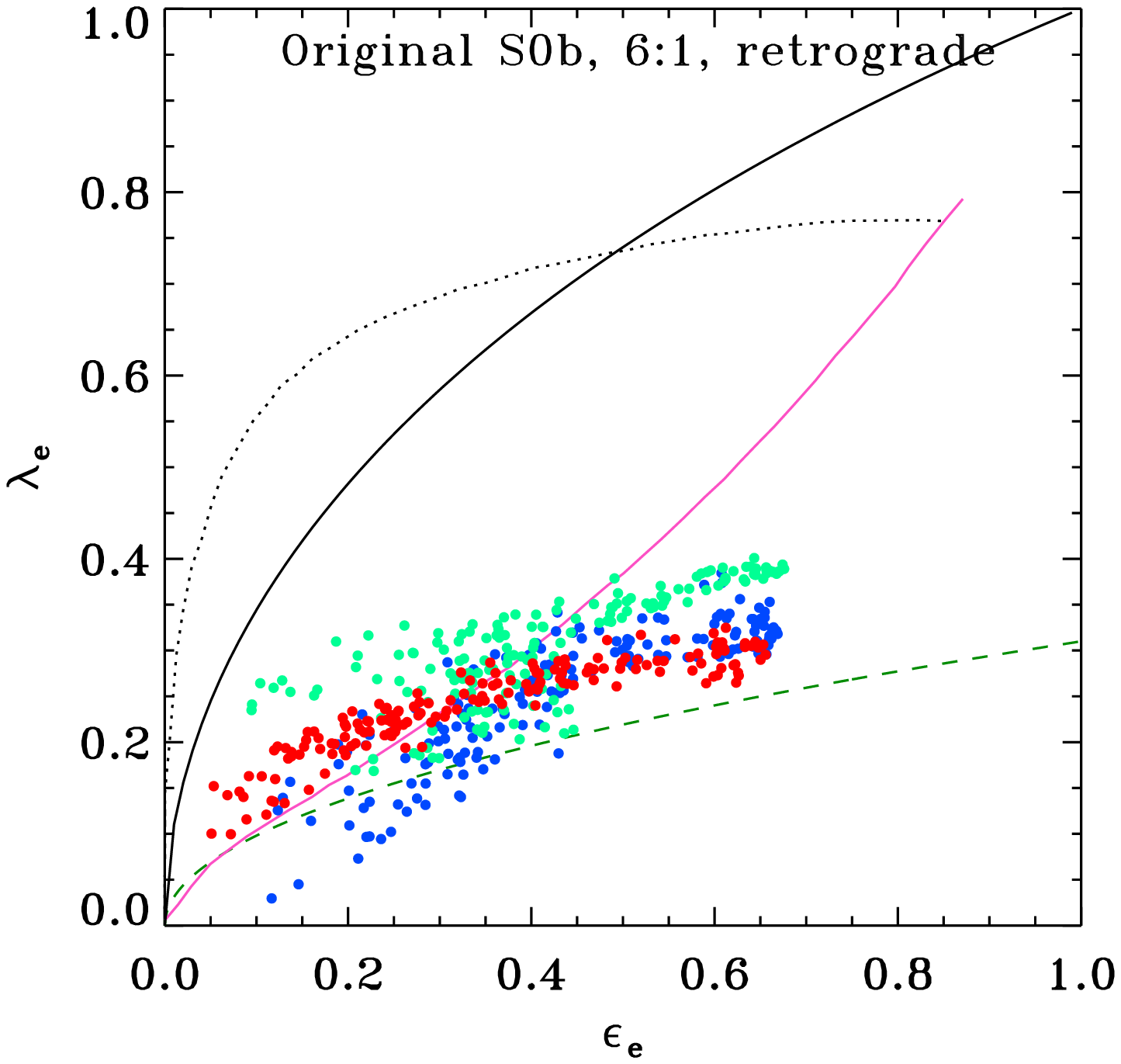}
  \includegraphics*[width=4.5cm]{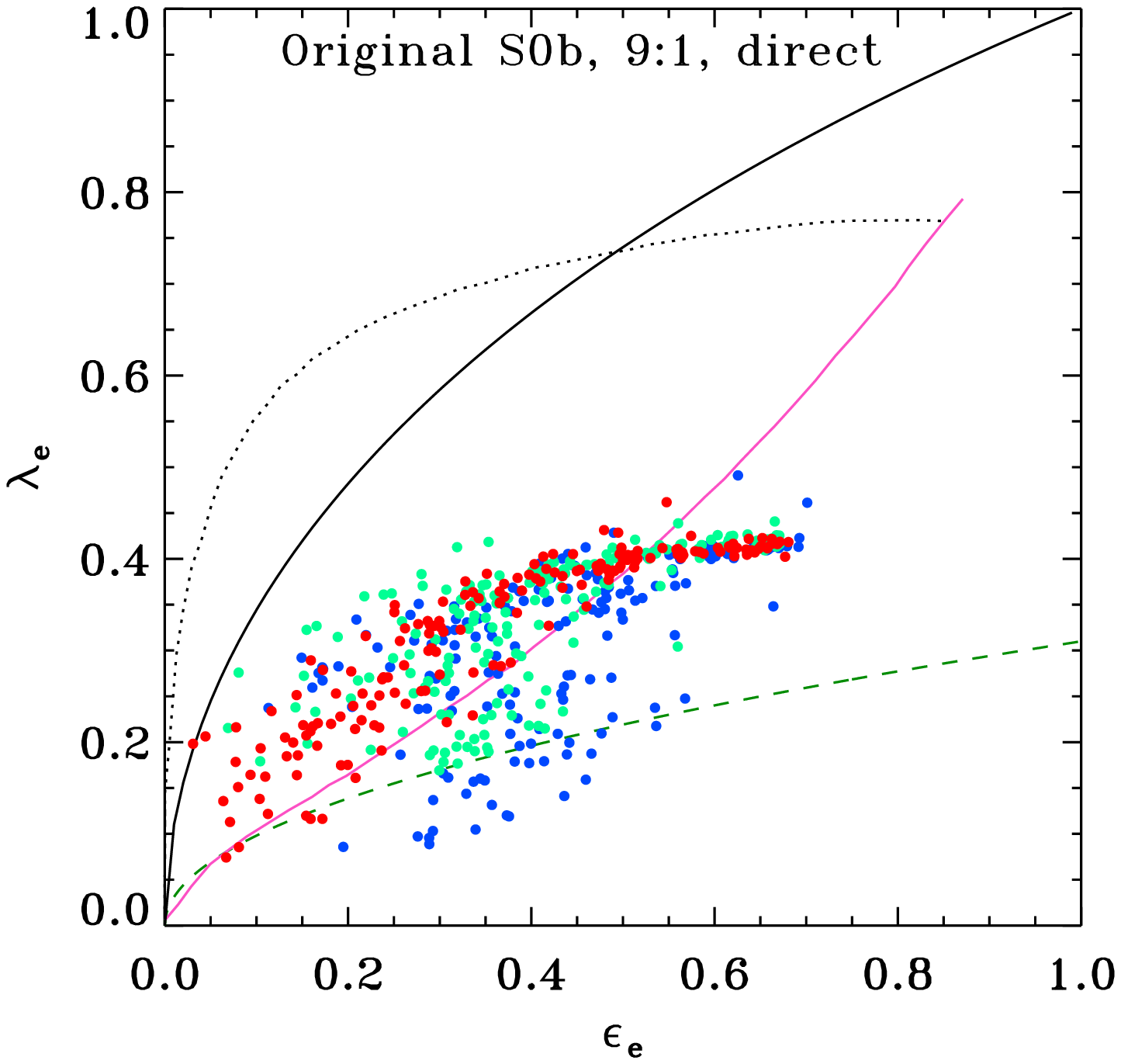}
  \includegraphics*[width=4.5cm]{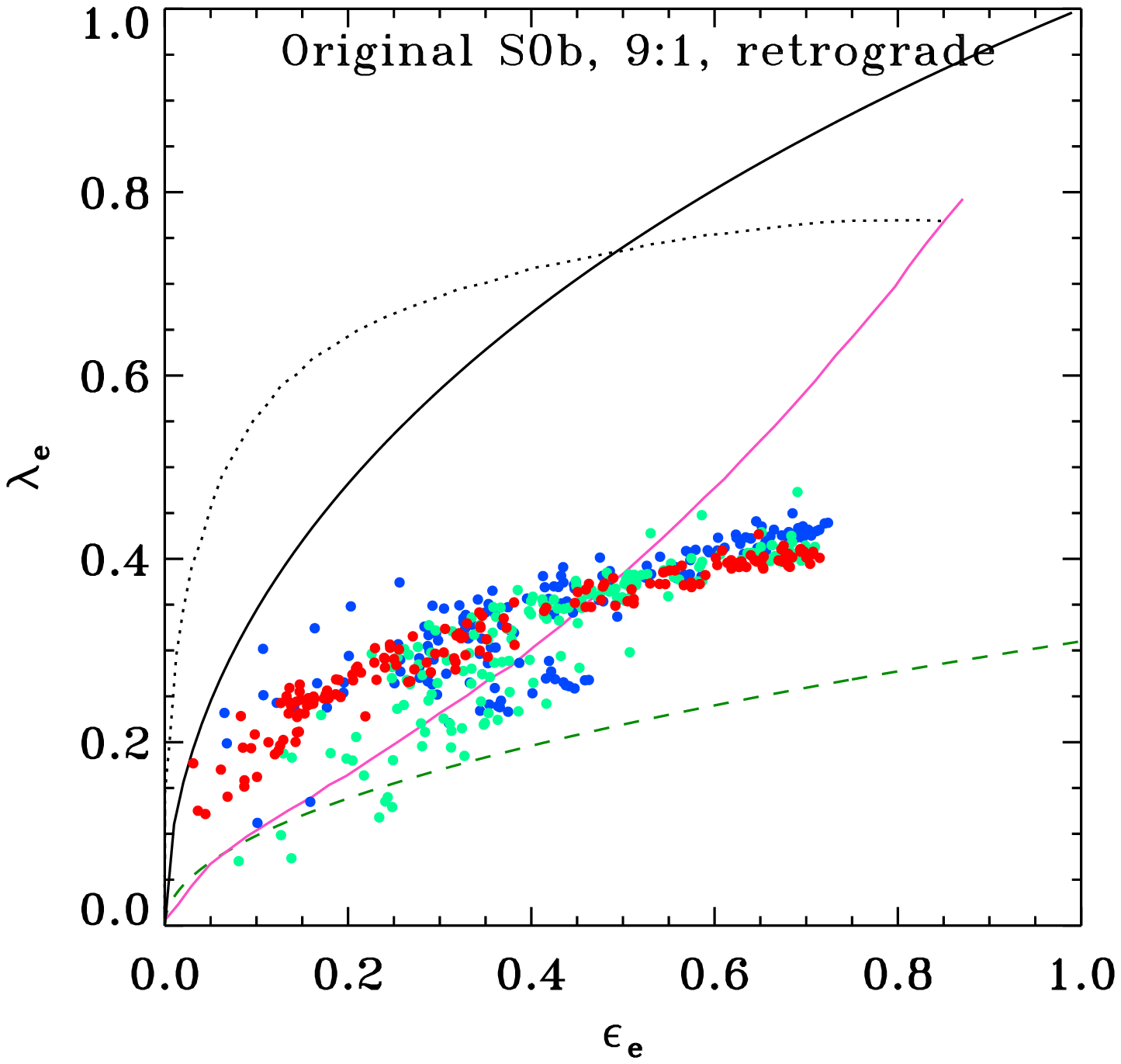}
  \caption{Dependence on the number of particles used in the simulation of the location in the $\lambda_\mathrm{e}$ -- $\epse$ diagram of the 200 random projections of our models for an identical set of initial conditions (indicated in each frame). \emph{Red dots}: models with $N = 185$K particles. \emph{Green dots}: models with $N = 555$K particles. \emph{Blue dots}: models with $N = 1\,850$K particles. The legend for the lines is the same as in Fig.\,\ref{fig:lambdaobs}.}
\label{fig:thickening}
\end{figure*}

\small  
%
\begin{acknowledgements}   
The authors thank the anonymous referee for the provided input that helped to improve this publication significantly, and to Frederic Bournaud for his interesting and useful comments. Supported by the Spanish Ministry of Economy and Competitiveness (MINECO) under projects AYA2006-12955, AYA2009-10368, AYA2009-11137, AYA2012-30717, AYA2010-21322-C03-02, AYA2010-21887-C04-04, and by the Madrid Regional Government through the AstroMadrid Project (CAM S2009/ESP-1496, http://www.laeff.cab.inta-csic.es/projects/astromadrid/main/index.php). Funded by the Spanish MICINN under the Consolider-Ingenio 2010 Program grant CSD2006-0070: "First Science with the GTC" (http://www.iac.es/consolider-ingenio-gtc/), and by the Spanish programme of International Campus of Excellence Moncloa (CEI). ACGG is a Ram\'on y Cajal Fellow of the Spanish MINECO. TT gratefully acknowledged the computer resources, technical expertise and assistance provided by the LaPalma (IAC/RES, Spain) supercomputer installations. 

\end{acknowledgements}

\bibliographystyle{aa}
\bibliography{massiveellipticals}


\begin{table*}
\centering 
\caption{Structural and kinematic properties of the progenitors and the remnants when considered as single-component systems.}
\vspace{0.1cm}
\label{tab:kinematics}
{\small
\centering
\begin{tabular}{lcccccc}
\hline\hline\vspace{-0.2cm}\\
\multicolumn{1}{c}{Model}& \multicolumn{1}{c}{$R_\mathrm{eff,glx}/r_\mathrm{eff,bulge}$} & \multicolumn{1}{c}{$\epsilon_\mathrm{e}$} & \multicolumn{1}{c}{$\mathbf{(V/\sigma)_{\mathrm{e}}}$}& \multicolumn{1}{c}{$\lambda_{\mathrm{e}}$ [from eq.~\ref{eq:lambda}]}& \multicolumn{1}{c}{$\lambda_{\mathrm{e}}$ [from eq.~\ref{eq:lambdaB1}]} & \multicolumn{1}{c}{$\delta_\mathrm{glx}$ [from eq.~\ref{eq:delta}]} \\
\multicolumn{1}{c}{(1)} & \multicolumn{1}{c}{(2)} & \multicolumn{1}{c}{(3)} & \multicolumn{1}{c}{(4)}& \multicolumn{1}{c}{(5)} & \multicolumn{1}{c}{(6)} & \multicolumn{1}{c}{(7)}\\\hline\vspace{-0.3cm}\\
Original S0b, $B/D=0.5$   & 4.59 & 0.875 $\pm$ 0.017 & 0.466 $\pm$ 0.002 & 0.468 $\pm$ 0.002 & 0.456 $\pm$ 0.002  &0.3959 $\pm$ 0.008 \\
Original S0c, $B/D=0.08$ & 8.94 & 0.869 $\pm$ 0.003 & 1.469 $\pm$ 0.023     & 0.819 $\pm$ 0.014 & 0.850 $\pm$ 0.018  &0.9126 $\pm$ 0.0008 \vspace{0.1cm}\\\hline\vspace{-0.3cm}\\
(a)\,\,\, M6 Ps Db                & 5.19 & 0.683 $\pm$ 0.002 &  0.439 $\pm$ 0.001 &  0.413 $\pm$ 0.001 & 0.435 $\pm$ 0.001 & 0.37 $\pm$ 0.04 \\
(a2) M6 Ps Db TF3             & 5.06  & 0.694 $\pm$  0.004 & 0.299 $\pm$ 0.015 & 0.303 $\pm$ 0.019 & 0.312 $\pm$ 0.017 & 0.36 $\pm$ 0.08\\
(a3) M6 Ps Db TF4             & 4.83  & 0.651 $\pm$ 0.002 & 0.284 $\pm$ 0.007 & 0.294 $\pm$ 0.010 & 0.298 $\pm$ 0.008  &0.34 $\pm$ 0.06\\
(b)\,\,\, M6 Ps Rb                 & 4.93 & 0.681 $\pm$ 0.008 & 0.224 $\pm$ 0.013 &  0.271 $\pm$ 0.018 & 0.239 $\pm$ 0.014 & 0.37 $\pm$ 0.08 \\
(c)\,\,\, M6 Pl Db                  & 4.57 & 0.638 $\pm$ 0.009  & 0.337 $\pm$ 0.020 &  0.331 $\pm$ 0.025 & 0.348 $\pm$ 0.022  &0.471 $\pm$ 0.010  \\
(d)\,\,\, M6 Pl Rb                   & 4.71 & 0.719 $\pm$ 0.004  & 0.231 $\pm$ 0.001 &  0.234 $\pm$ 0.001 & 0.246 $\pm$ 0.001  &0.45 $\pm$ 0.16  \\
(e)\,\,\, M6 Ps Ds                  & 8.52  & 0.755 $\pm$ 0.011  & 0.735 $\pm$ 0.011 & 0.632 $\pm$ 0.012 & 0.629 $\pm$ 0.010 &0.735 $\pm$ 0.002\\
(f)\,\,\, M6 Ps Rs                   &10.36 & 0.778 $\pm$ 0.004  & 0.494 $\pm$ 0.016 & 0.463 $\pm$ 0.019 & 0.478 $\pm$ 0.017  &0.701 $\pm$ 0.005  \vspace{0.05cm}\\\hline\vspace{-0.3cm}\\
(g)\,\,\, M9 Ps Db                 & 5.09  & 0.697 $\pm$ 0.008 & 0.298$\pm$ 0.023 & 0.318 $\pm$ 0.030 & 0.312 $\pm$ 0.025  &0.36 $\pm$ 0.05 \\
(g2) M9 Ps Db TF3              & 5.26 & 0.727 $\pm$ 0.006 & 0.292 $\pm$ 0.022 & 0.313 $\pm$ 0.028 & 0.306 $\pm$ 0.024 & 0.36 $\pm$ 0.01\\
(g3) M9 Ps Db TF4              & 4.61 & 0.671 $\pm$ 0.002 & 0.285 $\pm$ 0.014 & 0.307 $\pm$ 0.019 & 0.299 $\pm$ 0.016 & 0.34 $\pm$ 0.08\\
(h)\,\,\, M9 Ps Rb                 & 5.35 & 0.713 $\pm$ 0.005 & 0.222 $\pm$0.005 & 0.235 $\pm$ 0.006 & 0.237 $\pm$ 0.005 & 0.393 $\pm$ 0.002   \vspace{0.05cm}\\\hline\vspace{-0.3cm}\\
(i)\,\,\, M18 Ps Db               & 5.05 & 0.769 $\pm$ 0.004 & 0.265 $\pm$ 0.009 & 0.280 $\pm$ 0.012 & 0.280 $\pm$ 0.010 & 0.365 $\pm$ 0.023\\
(j)\,\,\, M18 Ps Rb               & 5.27 & 0.802 $\pm$ 0.005 & 0.328 $\pm$ 0.009 & 0.357 $\pm$ 0.012 & 0.339 $\pm$ 0.010 & 0.38 $\pm$ 0.07\\
(k)\,\,\,M18 Pl Db               & 4.77 & 0.704 $\pm$ 0.006 & 0.311 $\pm$ 0.017 & 0.346 $\pm$ 0.023 & 0.324 $\pm$ 0.019 & 0.442 $\pm$ 0.003 \\
(l)\,\,\, M18 Pl Rb               & 5.10 & 0.743 $\pm$ 0.004 & 0.381 $\pm$ 0.020 & 0.408 $\pm$ 0.027 & 0.387 $\pm$ 0.022 & 0.454 $\pm$ 0.010 \\\hline\\
\end{tabular}
\begin{minipage}[t]{\textwidth}{\small
\emph{Columns}: (1) Model code. (2) Ratio between the half-mass radius of each galaxy and the effective radius of its bulge. The bulge effective radii are available in Table\,2 of \citet{Eliche-Moral2013}. (3,4) Ellipticity within $r=R_\mathrm{eff,glx}$ and the $(\mathrm{V}/\sigma)_\mathrm{e}$ value computed from $N$-body data following the procedure by E11, for edge-on views of the galaxies. (5) The $\lambda_\mathrm{e}$ kinematic parameter defined by \citet{Emsellem2007} directly measured from $N$-body data (using the definition in eq.~\ref{eq:lambda}), for an edge-on view of each galaxy. (6) The $\lambda_\mathrm{e}$ kinematic parameter derived using eq.~\ref{eq:lambdaB1}, using the measured values of $(\mathrm{V}/\sigma)_\mathrm{e}$ in Col.~4, assuming $\kappa=1.1$ (see E11). (7) Anisotropy of velocities in the models, computed directly from the data using eq.~\ref{eq:delta}}.
\end{minipage}
}
\end{table*}

\begin{table*}
\centering 
\caption{Percentage of random projections in different bins of apparent ellipticity for each model.}
\vspace{0.1cm}
\label{tab:projections}
{\small
\centering
\begin{tabular}{lccccc}
\hline\hline\vspace{-0.2cm}\\
\multicolumn{1}{c}{Model} & \multicolumn{1}{c}{$\%(\epse \leq 0.2)$} & \multicolumn{1}{c}{$\%(0.2< \epse \leq 0.4)$} & \multicolumn{1}{c}{$\%(0.4 < \epse \leq 0.6)$} & \multicolumn{1}{c}{$\%(0.6 < \epse \leq 0.8)$} & \multicolumn{1}{c}{$\%(0.8 < \epse \leq 1.0)$}\\
\multicolumn{1}{c}{(1)} & \multicolumn{1}{c}{(2)} & \multicolumn{1}{c}{(3)} & \multicolumn{1}{c}{(4)}& \multicolumn{1}{c}{(5)} & \multicolumn{1}{c}{(6)}\\\hline\vspace{-0.3cm}\\
Original S0b, $B/D=0.5$  & 43.5   &  24.0  &  11.5  &  14.0  &  7.0 \\
Original S0c, $B/D=0.08$ & 38.5   &  22.5   &  19.5  &  12.0  &  7.5\vspace{0.1cm}\\\hline\vspace{-0.3cm}\\
(a)\,\,\, M6 Ps Db & 16.0   &  49.0   &  27.5  &  7.5 &    0\\
(a2) M6 Ps Db TF3 & 32.5   &  31.0   &  28.0  &  8.5 &    0\\
(a3) M6 Ps Db TF4 & 19.5   &  43.0   &  28.0  &  9.5 &    0\\
(b)\,\,\, M6 Ps Rb & 32.0   &  26.5   &  27.0  &   14.5 &    0\\
(c)\,\,\, M6 Pl Db & 43.0   &  24.0   &  25.5  &  7.5 &    0\\
(d)\,\,\, M6 Pl Rb& 19.0   &  25.0   &  31.5  &   24.5 &      0\\
(e)\,\,\, M6 Ps Ds  &39.0  &   24.5  &   13.5  &   23.0   &    0\\
(f)\,\,\, M6 Ps Rs & 6.0  &   42.0  &   29.5 &    22.5  &     0\vspace{0.05cm}\\\hline\vspace{-0.3cm}\\
(g)\,\,\, M9 Ps Db & 27.0   &  37.0   &  23.5  &   12.5   &    0\\
(g2) M9 Ps Db TF3 & 36.5   &  25.0   &  22.0  &   16.5   &    0\\
(g3) M9 Ps Db TF4 & 19.5   &  46.5  &   21.0   &  13.0   &    0\\
(h)\,\,\, M9 Ps Rb  & 24.0   &  39.5   &  20.5  &   16.0   &    0\vspace{0.05cm}\\\hline\vspace{-0.3cm}\\
(i)\,\,\, M18 Ps Db & 35.0  &   30.5   &  19.5  &   15.0   &    0\\
(j)\,\,\, M18 Ps Rb & 34.0  &   22.0   &  21.5   &  22.0 &  0.5\\
(k)\,\,\,M18 Pl Db & 49.0  &   19.5   &  17.0   &  14.5   &    0\\
(l)\,\,\, M18 Pl Rb & 27.5  &   31.5   &  21.5   &  19.5    &   0 \\\hline\\
\end{tabular}
\begin{minipage}[t]{\textwidth}{\small
\emph{Columns}: (1) Model code. (2 -- 6) Percentages of the 200 random projections considered for each model in different bins of apparent ellipticity values: $\epse \leq 0.2$, $0.2<\epse \leq 0.4$, $0.4<\epse \leq 0.6$, $0.6<\epse\leq 0.8$, and $\epse>0.8$, respectively. 
}\end{minipage}
}
\end{table*}

\begin{table*}[!t]
\centering 
\caption{Structural and kinematic properties of the bulges in the primary S0s and remnants of our merger experiments}
\vspace{0.1cm}
\label{tab:bulges}
{\small
\centering
\begin{tabular}{lccccccc}
\hline\hline\vspace{-0.2cm}\\
\multicolumn{1}{c}{Model}& \multicolumn{1}{c}{$b$} & \multicolumn{1}{c}{$c$} &  \multicolumn{1}{c}{$\hat{b}$} & \multicolumn{1}{c}{$\hat{c}$} & \multicolumn{1}{c}{T} & \multicolumn{1}{c}{$\langle\epsilon\rangle$ $(r \leq r_\mathrm{eff,bulge})$} &\multicolumn{1}{c}{$V_\mathrm{rot,max}/\langle\sigma\rangle$ $(r \leq r_\mathrm{eff,bulge})$}\\
\multicolumn{1}{c}{(1)} & \multicolumn{1}{c}{(2)} & \multicolumn{1}{c}{(3)} & \multicolumn{1}{c}{(4)}& \multicolumn{1}{c}{(5)}& \multicolumn{1}{c}{(6)} & \multicolumn{1}{c}{(7)} & \multicolumn{1}{c}{(8)} \\\hline\vspace{-0.3cm}\\
Original S0b, $B/D=0.5$   & 0.97 & 0.96 & 0.99 & 1.03 & 0.75 & 0.31 $\pm$ 0.09 & 0.21 $\pm$ 0.01   \\
Original S0c, $B/D=0.08$  & 0.97 & 0.96 & 0.99 & 1.03 & 0.75 & 0.50 $\pm$ 0.16 & 0.41 $\pm$ 0.04   \vspace{0.1cm}\\\hline\vspace{-0.3cm}\\
(a) M6 Ps Db              & 0.97 & 0.96 & 0.99 & 1.03 & 0.75 & 0.37 $\pm$ 0.12 & 0.48 $\pm$ 0.09   \\
(a2) M6 Ps Db TF3     & 0.97 & 0.96 & 0.99 & 1.03 & 0.75 & 0.26 $\pm$ 0.07 & 0.71 $\pm$ 0.05     \\
(a3) M6 Ps Db TF4     & 0.96 & 0.96 & 1.00 & 1.04 & 1.00 & 0.27 $\pm$ 0.08 & 0.58 $\pm$ 0.02    \\
(b) M6 Ps Rb              & 0.97 & 0.97 & 0.97 & 0.97 & 1.00 & 0.24 $\pm$ 0.08 & 0.58 $\pm$ 0.01   \\
(c) M6 Pl Db              & 0.95 & 0.94 & 0.99 & 1.05 & 0.83 & 0.32 $\pm$ 0.09 & 0.69 $\pm$ 0.03     \\
(d) M6 Pl Rb              & 0.95 & 0.95 & 0.95 & 0.95 & 1.00 & 0.31 $\pm$ 0.08 & 0.68 $\pm$ 0.01    \\
(e) M6 Ps Ds              & 0.95 & 0.89 & 0.94 & 1.06 & 0.46 & 0.24 $\pm$ 0.08 & 0.36 $\pm$ 0.06     \\
(f) M6 Ps Rs              & 0.97 & 0.93 & 0.96 & 1.03 & 0.43 & 0.09 $\pm$ 0.07 & 0.35 $\pm$ 0.03    \vspace{0.1cm}\\\hline\vspace{-0.3cm}\\
(g) M9 Ps Db              & 0.96 & 0.96 & 1.00 & 1.04 & 1.00 & 0.24 $\pm$ 0.10 & 0.88 $\pm$ 0.04    \\
(g2) M9 Ps Db TF3         & 0.98 & 0.97 & 0.99 & 1.02 & 0.67 & 0.13 $\pm$ 0.05 & 0.39 $\pm$ 0.10      \\
(g3) M9 Ps Db TF4         & 0.95 & 0.94 & 0.99 & 1.05 & 0.83 & 0.24 $\pm$ 0.07 & 0.36 $\pm$ 0.05    \\
(h) M9 Ps Rb              & 0.98 & 0.97 & 0.98 & 0.97 & 0.67 & 0.21 $\pm$ 0.06 & 0.50 $\pm$ 0.14     \vspace{0.05cm}\\\hline\vspace{-0.3cm}\\
(i) M18 Ps Db             & 0.97 & 0.96 & 0.99 & 1.03 & 0.75 & 0.29 $\pm$ 0.09 & 0.78 $\pm$ 0.02     \\
(j) M18 Ps Rb             & 0.99 & 0.98 & 0.99 & 0.98 & 0.50 & 0.26 $\pm$ 0.09 & 0.38 $\pm$ 0.10     \\
(k) M18 Pl Db             & 0.96 & 0.96 & 1.00 & 1.04 & 1.00 & 0.20 $\pm$ 0.05 & 0.88 $\pm$ 0.02 	\\
(l) M18 Pl Rb             & 0.98 & 0.89 & 0.91 & 1.02 & 0.19 & 0.13 $\pm$ 0.05 & 0.84 $\pm$ 0.05    \\\hline\\
\end{tabular}
\begin{minipage}[t]{\textwidth}{\small
\emph{Columns}: (1) Model code as described in Sect.~2 of \citet{Eliche-Moral2012}. (2,3) Major and minor axial ratios of the bulges according to the \citet{Cox2006} criterion ($A>B>C$). (4,5) Major and minor axial ratios of the bulges, derived considering the \citet{Mendez-Abreu2010} criterion ($A>B$ located in the equatorial plane and $C$ perpendicular to it). (6) Bulge triaxiality. (7,8) Intrinsic average ellipticity and $V_\mathrm{rot,max}/\langle\sigma\rangle$ in the bulge according to the \citet{Kormendy1983} method, corrected for the effects of numerical thickening (see Sect.\,\ref{sec:rotation}). Errors in the values listed in Cols.\,2-6 are $\sim 10$\% of the estimated values on average. 
}\end{minipage}
}
\end{table*}

\end{document}